\documentclass[%
 reprint,
superscriptaddress,
]{revtex4-2}

\usepackage{graphicx}
\usepackage{dcolumn}
\usepackage{bm}
\usepackage{amsmath, amssymb}
\usepackage{dsfont}
\usepackage{xcolor}
\usepackage{siunitx}
\usepackage{parskip}
\usepackage{subcaption}
\usepackage{soul}
\usepackage{ragged2e}
\usepackage{mleftright}
\usepackage[hidelinks]{hyperref}
\usepackage{algorithm}
\usepackage{algpseudocode}
\algrenewcommand\algorithmicrequire{\textbf{Input:}}
\algrenewcommand\algorithmicensure{\textbf{Output:}}

\newcommand\mb\mathbf

\makeatletter\def\p@subsection{} \def\p@subsubsection{}\makeatother

\newcommand\dc{d_\text{c}}

\begin{document}

\newcommand{\papertitle}{Parallel Omnidirectional Atom Rearrangement with Dynamic Holograms}

\title{\papertitle}

\author{Samuel A. Engel}
\thanks{These authors contributed equally to this work.}
\affiliation{Department of Physics, Harvard University, Cambridge, Massachusetts 02138, USA}
\affiliation{Department of Chemistry and Chemical Biology, Harvard University, Cambridge, Massachusetts 02138, USA}
\affiliation{Harvard-MIT Center for Ultracold Atoms, Cambridge, Massachusetts 02138, USA}
\author{Dewei~Gong}
\thanks{These authors contributed equally to this work.}
\affiliation{Department of Chemistry and Chemical Biology, Harvard University, Cambridge, Massachusetts 02138, USA}
\affiliation{Department of Physics, Harvard University, Cambridge, Massachusetts 02138, USA}
\affiliation{Harvard-MIT Center for Ultracold Atoms, Cambridge, Massachusetts 02138, USA}
\author{Ziguang~Lin}
\thanks{These authors contributed equally to this work.}
\author{Lysander~Christakis}
\thanks{These authors contributed equally to this work.}
\affiliation{Department of Physics, Harvard University, Cambridge, Massachusetts 02138, USA}
\affiliation{Department of Chemistry and Chemical Biology, Harvard University, Cambridge, Massachusetts 02138, USA}
\affiliation{Harvard-MIT Center for Ultracold Atoms, Cambridge, Massachusetts 02138, USA}
\author{Kang-Kuen Ni}
\email{ni@chemistry.harvard.edu}
\affiliation{Department of Chemistry and Chemical Biology, Harvard University, Cambridge, Massachusetts 02138, USA}
\affiliation{Department of Physics, Harvard University, Cambridge, Massachusetts 02138, USA}
\affiliation{Harvard-MIT Center for Ultracold Atoms, Cambridge, Massachusetts 02138, USA}
\author{Norman Y. Yao}
\email{nyao@fas.harvard.edu}
\affiliation{Department of Physics, Harvard University, Cambridge, Massachusetts 02138, USA}
\affiliation{Department of Chemistry and Chemical Biology, Harvard University, Cambridge, Massachusetts 02138, USA}
\affiliation{Harvard-MIT Center for Ultracold Atoms, Cambridge, Massachusetts 02138, USA}

\begin{abstract}

The ability to assemble and reconfigure individual atoms is central to performing quantum science using optical tweezer arrays. 
Here, we demonstrate parallel omnidirectional atom transport and rearrangement using dynamic holograms generated by a single kilohertz-rate spatial light modulator. 
By carefully characterizing survival and heating, we show that atoms remain confined even when successive holograms are commanded before the liquid crystal has settled, enabling significantly faster transport. 
To generate  holograms in real time, we develop and implement a warm-start, weighted Gerchberg--Saxton algorithm that achieves performance comparable to neural-network methods without requiring any training. 
Combining these techniques, we assemble ytterbium atoms into a 2078-site kagome lattice with 99.0\% mean filling. 
In addition, we demonstrate parallel transport between array layers, combining axial and transverse motion to reconfigure atoms in three dimensions. 
Our approach is readily adaptable to existing  tweezer platforms and provides a method for large-scale SLM-based atom assembly as well as programmable three-dimensional control.

\end{abstract}

\begingroup
\setlength{\parskip}{0pt}
\maketitle
\endgroup

Neutral atoms in optical tweezer arrays provide a powerful combination of scale, geometric flexibility, and site-resolved control for quantum computing, simulation, and metrology~\cite{kaufman2021quantum}. 
Central to this platform is the ability to assemble and dynamically reconfigure arrays by moving individual atoms \cite{barredo2016atom,endres2016atom,lee2016three,barredo2018synthetic, schlosser2023scalable,lin2025ai}.
Beyond the initial assembly of defect-free arrays, this motion can spatially separate atoms for mid-circuit measurement \cite{bluvstein2024logical,bluvstein2026fault}, program time-dependent interactions through changes in array geometry \cite{you2025control,yu2026high}, and enable optical-clock protocols with locally-controlled phase accumulation~\cite{shaw2024multi}.

The scope and efficiency of atomic reconfiguration depend on two key properties of the optical control: the geometric flexibility of the trap pattern and the speed with which that pattern can be changed.
Liquid-crystal spatial light modulators (SLMs) enable flexibility by generating arbitrary holographic arrays \cite{di2007computer,nogrette2014single}, but are conventionally limited in transport speed \cite{kim2016situ,lee2016three,lee2017defect, knottnerus2025parallel}. 
Acousto-optic deflectors (AODs) invert this trade-off. 
Megahertz-rate diffraction steers tweezers quickly, but row-column addressing constrains atom trajectories~\cite{barredo2016atom,endres2016atom,schymik2020enhanced,sheng2021efficient,tian2023parallel,wang2023accelerating,picard2026three,lu2026astigmatism}.
Pioneering recent  work has demonstrated the potential to combine these strengths using kilohertz-rate SLMs~\cite{knottnerus2025parallel} and neural-network-generated holograms to rapidly rearrange large-scale atom arrays~\cite{lin2025ai}.

While these demonstrations highlight the promise of AI-enabled SLM rearrangement, a multitude of open questions remain regarding the fundamental limitations and  capabilities of the approach. 
Among these questions, three stand out: (i) what sets the speed limit for atom transport, (ii) can holograms be generated in real time without pretrained neural networks, and (iii) can this approach enable scalable transport in three dimensions?

 \begin{figure*}[t]
    \centering    \includegraphics[width=\linewidth]{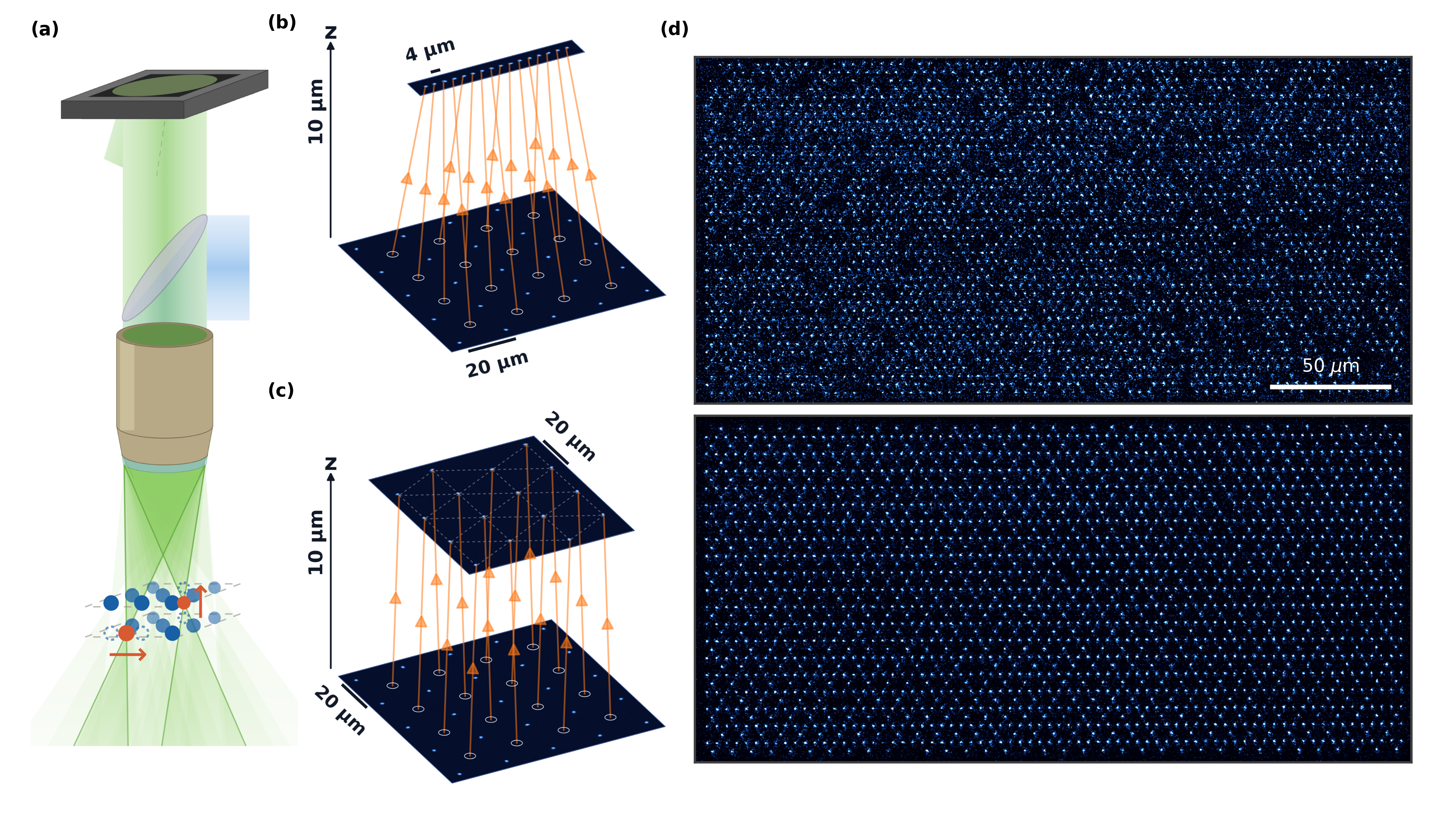}
    \caption{(a) Schematic of the experimental setup. Individual $^{174}\mathrm{Yb}$ atoms are trapped in a tweezer array generated by a kHz-rate SLM and a high-NA objective; orange atoms are undergoing radial and axial transport. 
    Fluorescence is collected through the same objective and split off by a dichroic mirror. 
    (b,c) Single-shot fluorescence images, rendered in 3D, of one representative region of a $536$-site bilayer $^{174}\mathrm{Yb}$ array formed by omnidirectional rearrangement. 
    The array is partitioned into $41$-site checkerboard regions; in each, $16$ atoms are lifted \SI{10}{\micro\meter} into a second plane and compacted into (b) a \SI{4}{\micro\meter}-spacing one-dimensional chain or (c) a triangular lattice with \SI{20}{\micro\meter} spacing. Rings mark the vacated sites and arrows indicate the site-resolved path.
    (d) Single-shot fluorescence images. Top: stochastic loading of a 3013-site triangular lattice. Bottom: large-scale rearrangement of $2071$ $^{174}\mathrm{Yb}$ atoms  into a $\SI{4}{\micro\meter}$-spacing kagome lattice with $99.7\%$ filling.    }
    \label{fig:1}
    
\end{figure*}

\begin{figure*}[t]
    \centering
    \includegraphics[width=\linewidth]{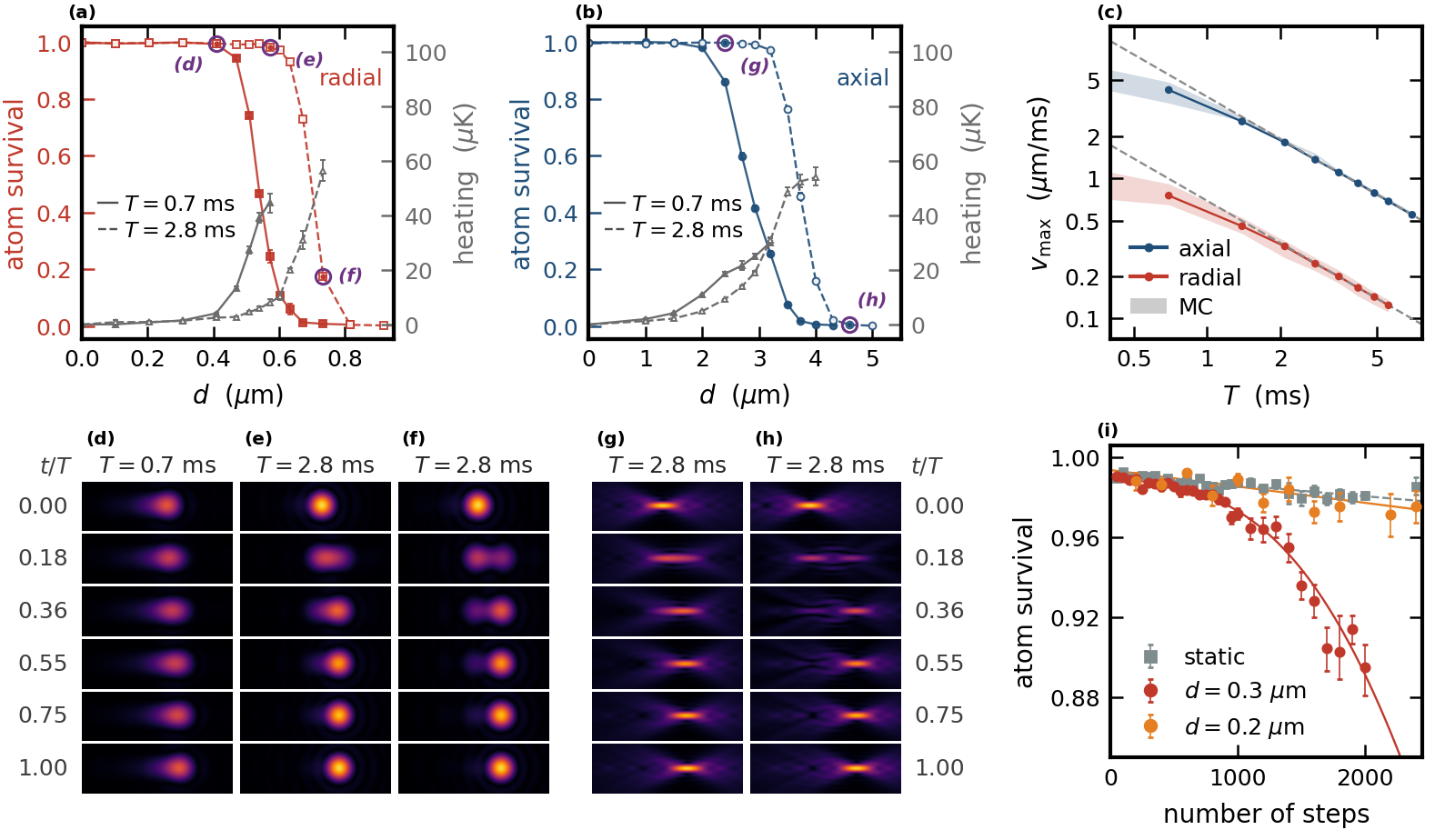}
    \caption{Atom loss and heating during transport. 
    (a) Atom survival (left y-axis) and heating (right y-axis) over $50$ radial transport steps versus step size, $d$.
    The atom temperature is measured via release-recapture and data are shown for two different refresh periods, $T$.
    (b) Analogous measurements for axial transport with 50 steps under the ping-pong protocol, where atoms are moved back-and-forth~\cite{supp}.
    (c) $v_\text{max}$ versus refresh period for both radial and axial transport~\cite{supp}. Dashed lines show $v_\text{max} = d_0/T$ with $d_0$ fixed at its long-period value.
    Shaded bands indicate the predicted $v_\text{max}$ from Monte Carlo simulations of the liquid crystal settling dynamics.
    (d)--(f) Simulations of the atom-plane light field  during radial transport  for different periods and
     step sizes corresponding to those indicated in panel (a)~\cite{supp}. 
    The left vertical axis shows time in units of $T$.
    (g,h) Analogous simulations of the atom-plane light field  during axial transport. 
    (i) Atom survival versus number of radial transport steps at small step size and $T = 0.7$\,\SI{}{\milli\second}, with a static control in gray. 
    The control and $d=\SI{0.2}{\micro\meter}$ data are fit by a single exponential while the $d=\SI{0.3}{\micro\meter}$ data are fit with $\exp(-A n^b)$, $b \approx 3$ (super-exponential).
    }
    \label{fig:2}
    
\end{figure*}

In this Letter, we take important steps toward answering all three questions.
First, intuition suggests allowing the liquid crystal (LC) to equilibrate before updating the tweezer pattern, since faster updates could compromise atom confinement.
By carefully characterizing atom survival and heating, we show that confinement persists even when the SLM pattern is refreshed on timescales much shorter than the liquid crystal settle time, enabling substantially faster atom transport and rearrangement.
Second, we demonstrate that neural networks are not required for real-time hologram generation: a warm-start weighted Gerchberg–Saxton (WGS) algorithm \cite{kim2019gerchberg} achieves comparable performance without training. 
By eliminating the need for training, this approach simplifies adaptation to different array configurations.
Finally, despite encoding the optical field in a two-dimensional hologram, we demonstrate parallel atom transport in all three spatial directions, enabling transport between array layers and scalable omnidirectional rearrangement. 
Combining these insights and techniques, we assemble nearly defect-free 2D arrays of more than 2000 ytterbium atoms and demonstrate inter-layer 3D rearrangement (Fig.~\ref{fig:1}).

\emph{Optimizing holographic atom transport dynamics}---In SLM-based rearrangement, atom transport proceeds through successive hologram updates that incrementally shift the tweezer positions.
The experimentally achievable atom transport speed is then set by the ratio $v=d/T$, where $d$ is the distance moved per hologram update and $T$ is the time between updates. 
Maximizing this speed while preserving atom survival requires understanding the physical constraints on both quantities.
For the refresh period, the natural time-scale is given by the liquid-crystal settle time, $\tau_0 \approx\SI{1.6}{\milli\second}$~\cite{supp}; for $T\lesssim\tau_0$, each new hologram is commanded before the preceding pattern has fully formed.

For the step size, a natural scale follows from the transient optical potential during a hologram update. Assuming small differences between consecutive holograms, the atom-plane light field is an exponentially-weighted superposition of the initial and final fields~\cite{lin2025ai}. 
For two Gaussian tweezers of matched depth and phase, this potential smoothly interpolates between their positions below a critical separation, $\dc$, and bifurcates above it.
Analytically, this threshold is given by $\dc=\sqrt{2}w_0$ where $w_0$ characterizes the waist of the tweezers~\cite{supp}.
Well below this threshold, atoms can adiabatically follow the optical potential as it moves; above it, they tend to remain in the fading initial potential while the new potential forms, leading to both heating and loss (Fig.~\ref{fig:2}d-h)~\cite{supp}.

To investigate these constraints as well as their interplay, we load atoms into a 1068-site square tweezer array and apply a sequence of global phase gratings that translate all tweezers uniformly in a radial direction.
We independently vary $d$ and $T$, and for each set of parameters,  we
measure the fraction of atoms remaining after 50 transport steps, as well as the temperature of the surviving atoms~\cite{supp}.
As a metric, we define $v_\textrm{max}=d_0/T$, where $d_{0}$  is the step size corresponding to $99\%$ survival per step.

As depicted in Fig.~\ref{fig:2}a (dashed red curve), we begin by exploring update periods longer than the liquid-crystal settle time, $T=\SI{2.8}{\milli\second}>\tau_0$.
Atom survival remains near unity up to $d\approx\SI{0.6}{\micro\meter}$ and then rapidly falls to zero by $\SI{0.8}{\micro\meter}$, with $d_{0}=\SI{0.69(1)}{\micro\meter}$ and $v_\text{max}=\SI{0.25(1)}{\micro\meter/\milli\second}$.
This is consistent with our bifurcation picture, which predicts a critical radial step size, $\dc^r \approx \SI{0.78}{\micro\meter}$.
As expected, the onset of heating closely coincides with that of atom loss (dashed gray curve, Fig.~\ref{fig:2}a).

Next, we consider the opposite regime where the update period is significantly shorter than the 
settle time.
Two competing expectations arise: the Gaussian-overlap picture predicts a critical step size $\dc$ that is purely geometric and independent of $T$, suggesting that faster updates should directly increase transport speed~\cite{supp}. 
On the other hand, incomplete liquid-crystal settling may distort the transient potential and reduce the tolerable step size~\cite{supp}. 

This latter expectation is indeed borne out by our data.
For the minimum update period supported by our SLM electronics, $T_\text{min} \approx \SI{0.7}{\milli\second}<\tau_0$ (solid curves, Fig.~\ref{fig:2}a), we observe the same qualitative behavior with the onset of heating and loss shifted to a smaller step-size, corresponding to $d_{0}=\SI{0.53(1)}{\micro\meter}$. 
Crucially, despite the smaller tolerable step-size, $v_\text{max}$ nevertheless increases threefold, from $\SI{0.25(1)}{\micro\meter/\milli\second}$ to $\SI{0.75(1)}{\micro\meter/\milli\second}$ (Fig.~\ref{fig:2}c).

Two remarks are in order. 
First, as illustrated in Fig.~\ref{fig:2}c, for update periods longer than $\tau_0$, the critical step size \emph{does} remain constant leading to a  $1/T$ scaling of $v_\text{max}$.
Second, the measured survival curves in Fig.~\ref{fig:2}a demonstrate the onset of loss for varying step sizes, but are insufficient to accurately characterize the  survival probability at small step sizes. 
Therefore, we repeat our radial transport experiments (for $T=\SI{0.7}{\milli\second}$) for up to ${\sim}2000$ update steps, with a zero-movement control (Fig.~\ref{fig:2}i).
For $d = \SI{0.2}{\micro\meter}$ we observe no discernible transport loss, while for $d = \SI{0.3}{\micro\meter}$, we observe a loss of around $9\%$.
Atom survival decays super-exponentially, suggesting that accumulated heating is a significant source of loss in this small step-size regime.

We now turn to axial ($\hat{z}$) atom transport.
Since our tweezers are significantly more extended along $\hat{z}$, the corresponding bifurcation threshold is larger, $\dc^z \approx \SI{3.7}{\micro\meter}$.
We perform an analogous set of axial survival and heating measurements by adding global defocus kernels to the hologram, with compensation for the Gouy phase~\cite{supp}.

The behavior of axial atom survival closely mirrors the radial case (blue curves, Fig.~\ref{fig:2}b).
For $T=\SI{2.8}{\milli\second}$ (dashed blue curve), we observe a survival threshold $d_0 = \SI{3.62(4)}{\micro\meter}$ (consistent with the estimated bifurcation threshold) and $v_\text{max} = \SI{1.30(2)}{\micro\meter/\milli\second}$.
Meanwhile for $T_\text{min}=\SI{0.7}{\milli\second}$, loss sets in at smaller step sizes, with $d_0 = \SI{2.69(10)}{\micro\meter}$; nevertheless, faster updating again yields a roughly threefold increase to $v_\text{max} = \SI{3.86(15)}{\micro\meter/\milli\second}$.
We note that the full axial period sweep (blue, Fig.~\ref{fig:2}c) exhibits the same two regimes: $1/T$ scaling for $T \gtrsim \tau_0$, and a departure from this scaling at the shortest periods.

Interestingly, in contrast to the radial case, we find that atom heating does not track atom loss in the axial case. 
For example, at $T = \SI{2.8}{\milli\second}$, heating onsets at approximately $\SI{2.0}{\micro\meter}$, well below the onset of atom loss (Fig.~\ref{fig:2}b).
Moreover, unlike the radial case, reducing the update period to $T_\textrm{min}=\SI{0.7}{\milli\second}$ does not produce a comparable shift of the axial heating curve toward smaller step sizes.

\begin{figure}[t]
    \centering

    \includegraphics[width=1\linewidth]{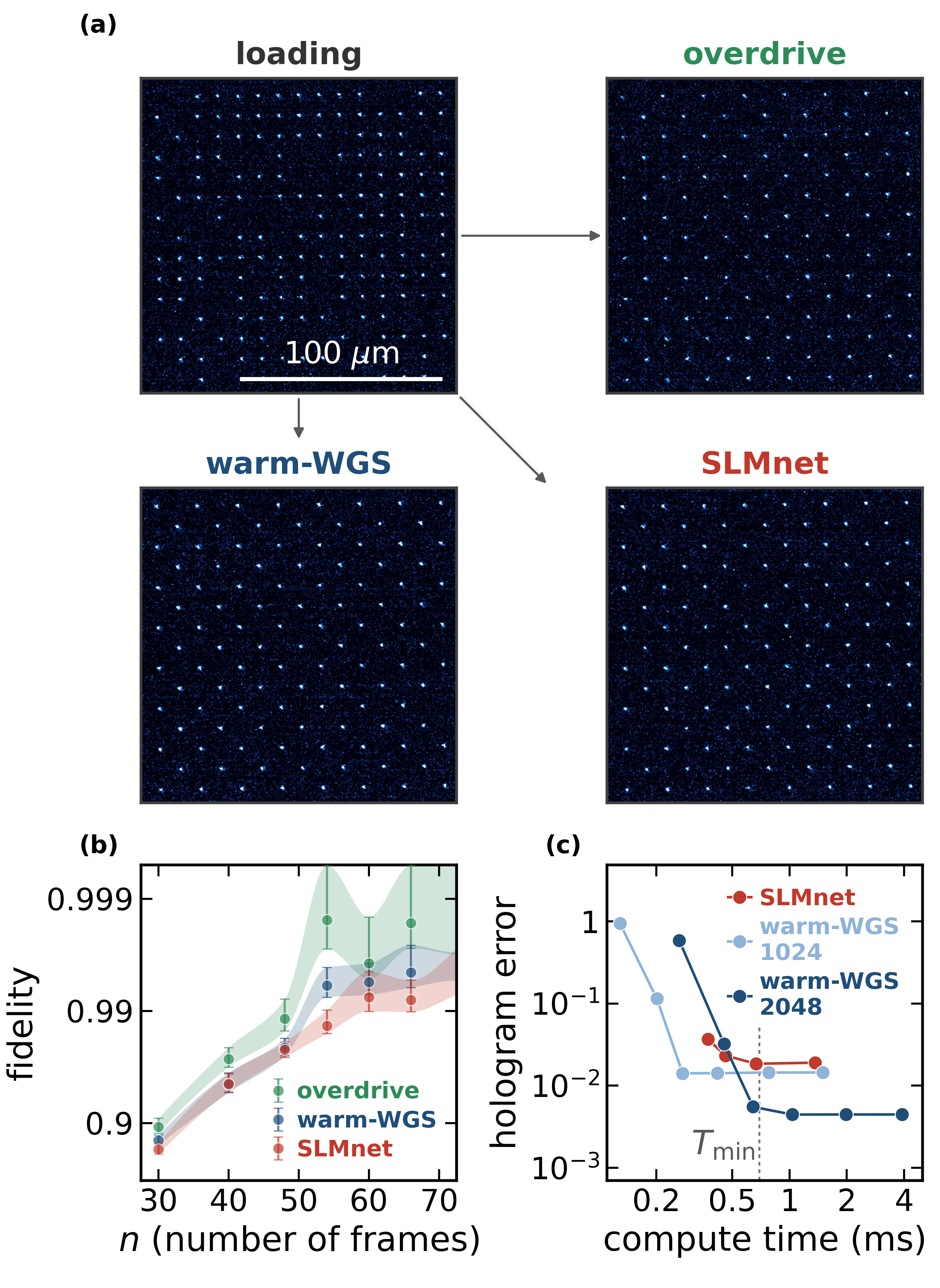}
        \caption{Hologram generation for rearrangement. (a) Single-shot fluorescence image of  stochastic loading into our $1068$-site, $\SI{10}{\micro\meter}$-spacing square lattice.
        Subsequent rearrangement into a checkerboard pattern is shown using hologram generation via SLMnet, warm-WGS and warm-WGS with overdrive.
        Note that the fluorescence images  have been cropped to one quadrant to enhance visibility.
     (b) Performance of each hologram generation method versus $n$, the number of frames for checkerboard rearrangement. 
     The fidelity is normalized to atom survival with no-movement. (c) Simulated hologram error versus compute time for various methods and parameters~\cite{supp}. 
     }
    \label{fig:3}
\end{figure}

\emph{Real-time hologram generation for atom rearrangement}---Having established speed limits for uniform, unidirectional transport, we now consider hologram generation for \emph{arbitrary} atom rearrangement.
Conventionally, SLM hologram generation is performed via the  weighted Gerchberg--Saxton algorithm, which achieves excellent positional control and intensity uniformity~\cite{chew2024ultraprecise}. However, our characterization of transport suggests two other requirements on the hologram sequence: (i) tweezer phases must be matched between consecutive frames~\cite{supp,knottnerus2025parallel,lin2025ai} and (ii) each frame should be computed within $T_\text{min}$ since faster updates yield faster transport (Fig.~\ref{fig:2}c). WGS leaves tweezer phases uncontrolled, causing bifurcation-induced atom loss at smaller step sizes, and typically takes tens of iterations to converge.

Previous work elegantly addressed both issues
with a 19k-parameter convolutional neural network (SLMnet) that approximates phase-constrained WGS in a single forward pass~\cite{lin2025ai}.
Building on recent work~\cite{huang2026phase,zhang2026algorithm}, we propose and implement a modified algorithm, which we dub warm-WGS, aimed at meeting both requirements without needing a neural network. 
Specifically, warm-WGS imposes a phase constraint at each iteration and  initializes each frame from the previous frame's hologram; this enables convergence in around three iterations at $2048\times 2048$ resolution (Fig.~\ref{fig:3}c).

We compare the performance of SLMnet and warm-WGS using the rearrangement of a 1068-site square lattice (\SI{10}{\micro\meter} spacing) into a checkerboard as a benchmark (Fig.~\ref{fig:3}a).
After an initial image of the stochastically loaded array, we assign atoms to target sites~\cite{lee2017defect,schymik2020enhanced} and split the overall transport into $n$ intermediate frames; each frame is converted into a hologram by either SLMnet or warm-WGS and sent to the SLM at intervals of $T_\text{min}$~\cite{supp}.

We find that both methods achieve near-defect-free rearrangement, reaching $99\%$ normalized fidelity by $n=53(3)$ (warm-WGS) and $n = 59(4)$ (SLMnet) frames (Fig.~\ref{fig:3}a,b).
Further improvements to the rearrangement speed require the ability to perform larger transport steps at our minimum update period, $T_\textrm{min}$. 
To this end, we utilize ``overdrive'', a standard technique for accelerating the liquid-crystal's response \cite{thalhammer2013overdrive}: rather than commanding the next frame's phase directly, one commands a phase that overshoots along the direction of liquid-crystal relaxation~\cite{supp}.
Overdrive adds negligible computational overhead and reduces the number of warm-WGS frames required for the checkerboard rearrangement to $n = 47(2)$ (Fig.~\ref{fig:3}b).

Thus far, we have focused on SLM rearrangement within a single plane. 
Extending this control to arbitrary three-dimensional trajectories requires encoding traps at different depths in a single two-dimensional hologram~\cite{kim2016situ,barredo2018synthetic,lin2025ai}. 
Standard 3D WGS addresses this problem by propagating the optical field to each distinct layer at every iteration, introducing a computational cost that grows with the number of layers.
We obviate this cost by working entirely in a single focal plane, where each out-of-plane tweezer is described by a precomputed defocus pattern~\cite{supp}. 
This representation allows us to apply warm-WGS directly to generate holograms for three-dimensional rearrangement. 
Our resulting algorithm outperforms both standard multi-layer WGS and a three-dimensional version of SLMnet, achieving comparable or higher quality simulated holograms in less time~\cite{supp}.

We conclude with two remarks highlighting the tradeoffs between warm-WGS and SLMnet.
On the one hand, warm-WGS requires no training and applies directly to arbitrary array sizes, spacings, and dimensionality, whereas SLMnet must be trained over the full range of configurations it will encounter~\cite{supp}.
On the other hand,  our implementation of warm-WGS computes each frame from the previous one, while SLMnet frames are generated independently and  can  therefore be parallelized. 
In principle, one can also envision parallelizing warm-WGS by independently initializing each frame from a computed, interpolated target~\cite{supp}.

\emph{Large-scale assembly and three-dimensional rearrangement}---Combining these techniques, we demonstrate two capabilities: (i) assembly of the largest rearranged ytterbium array to date~\cite{norcia2024iterative,zhu2026high} and (ii) scalable kHz-rate three-dimensional rearrangement.
The former is depicted in Fig.~\ref{fig:1}d, where $2071$ $^{174}\mathrm{Yb}$ atoms have been rearranged into a $2078$-site kagome lattice with a filling fraction of 99.7\% (mean 99.0\% over 195 shots)~\cite{supp}.
The total time for the rearrangement sequence is $\sim$\SI{56}{\milli \second} (median), which compares favorably to experiments utilizing AOD-based rearrangement~\cite{pichard2024rearrangement,lim2026defect,chiu2025continuous}.
We note that our current array size is limited only by the available laser power and not by the SLM-rearrangement technique~\cite{supp}.

For three-dimensional transport, we demonstrate two rearrangement sequences. 
First, we extract selected atoms from a rearranged 2D array into a separate plane in order to form a 1D chain (Fig.~\ref{fig:1}b).
Second, starting from a rearranged 2D checkerboard, we transport atoms from one of the two checkerboard sublattices into a triangular lattice $\sim \SI{10}{\micro\meter}$ away in $\hat{z}$; for this square--triangular bilayer, we achieve an average filling of $96.7\%$ across $536$ sites (Fig.~\ref{fig:1}c)~\cite{supp}. 
In both sequences, atom trajectories combine axial and transverse motion,  demonstrating three-dimensional reconfiguration beyond a simple axial displacement.
To image the atoms in different layers, we sequentially translate each layer into the focal plane of the objective.
Alternatively, one could, in principle, compress the 3D array into a single focal plane via a prescribed mapping and then perform imaging, allowing for single-shot reconstruction of the full three-dimensional array while avoiding the repeated exposure of individual layers.

\emph{Discussion and outlook}.---Our results demonstrate the potential for utilizing kHz-rate SLMs for large-scale atom rearrangement with three-dimensional control. 
We emphasize that this approach should be readily adaptable to nearly all existing SLM-based tweezer platforms, since the hardware requirements are modest:  it requires no optics beyond those already needed to generate the array and the holograms can be computed in real time on a single commercial GPU without requiring neural network training. 

Looking forward, our work opens the door to a number of intriguing future directions.
First, interlayer transport could enable quantum computing architectures to stack certain ``zones'' such as the storage or reservoir zone~\cite{manetsch2025tweezer,bluvstein2024logical,chiu2025continuous} along the optical axis.
Second, assembling atoms across multiple layers also brings genuinely three-dimensional lattices within reach of Rydberg quantum simulators \cite{barredo2018synthetic,song2021quantum}, such as pyrochlore arrays proposed to host a U(1) quantum spin liquid \cite{shah2025quantum}. 
Finally, the combination of large-scale assembly and flexible control over array geometry paves the way for studies of quantum magnetism on frustrated lattices, where dipolar interactions are predicted to stabilize quantum spin liquid phases \cite{bintz2024dirac,bornet2026dirac,machado2026dipolar,mao2026chiral}.

\emph{Acknowledgments}---We gratefully acknowledge the insights of and discussions with J.~Deng, C.~Laumann, R.~Lin, T.~Ono, G.~Semeghini, and J.~Zeiher.
We thank Meadowlark and AVR Optics for providing us with a high-speed SLM for high-power 532 nm laser light. 
This work was supported in part by multiple awards from the NSF including the Center for Ultracold Atoms (an NSF Physics Frontiers Center), PHY-2409713, PHY-2608240 and 
 QLCI Award 2553619 for Fault-Tolerant Quantum Systems, Architectures and Applications. 
 N. Y. Yao 
acknowledges support from a Brown Investigator award.
AI tools were used for assistance in operating the experimental control software, analyzing data, preparing figures, and editing the manuscript; all results and text have been verified by the authors.

\nocite{saskin2019narrow,wei202610,bytyqi2026device,deters2026ultrafast,zhang2024scaled}

\bibliography{slm}

\begin{thebibliography}{46}%
\makeatletter
\providecommand \@ifxundefined [1]{%
 \@ifx{#1\undefined}
}%
\providecommand \@ifnum [1]{%
 \ifnum #1\expandafter \@firstoftwo
 \else \expandafter \@secondoftwo
 \fi
}%
\providecommand \@ifx [1]{%
 \ifx #1\expandafter \@firstoftwo
 \else \expandafter \@secondoftwo
 \fi
}%
\providecommand \natexlab [1]{#1}%
\providecommand \enquote  [1]{``#1''}%
\providecommand \bibnamefont  [1]{#1}%
\providecommand \bibfnamefont [1]{#1}%
\providecommand \citenamefont [1]{#1}%
\providecommand \href@noop [0]{\@secondoftwo}%
\providecommand \href [0]{\begingroup \@sanitize@url \@href}%
\providecommand \@href[1]{\@@startlink{#1}\@@href}%
\providecommand \@@href[1]{\endgroup#1\@@endlink}%
\providecommand \@sanitize@url [0]{\catcode `\\12\catcode `\$12\catcode `\&12\catcode `\#12\catcode `\^12\catcode `\_12\catcode `\%12\relax}%
\providecommand \@@startlink[1]{}%
\providecommand \@@endlink[0]{}%
\providecommand \url  [0]{\begingroup\@sanitize@url \@url }%
\providecommand \@url [1]{\endgroup\@href {#1}{\urlprefix }}%
\providecommand \urlprefix  [0]{URL }%
\providecommand \Eprint [0]{\href }%
\providecommand \doibase [0]{https://doi.org/}%
\providecommand \selectlanguage [0]{\@gobble}%
\providecommand \bibinfo  [0]{\@secondoftwo}%
\providecommand \bibfield  [0]{\@secondoftwo}%
\providecommand \translation [1]{[#1]}%
\providecommand \BibitemOpen [0]{}%
\providecommand \bibitemStop [0]{}%
\providecommand \bibitemNoStop [0]{.\EOS\space}%
\providecommand \EOS [0]{\spacefactor3000\relax}%
\providecommand \BibitemShut  [1]{\csname bibitem#1\endcsname}%
\let\auto@bib@innerbib\@empty
\bibitem [{\citenamefont {Kaufman}\ and\ \citenamefont {Ni}(2021)}]{kaufman2021quantum}%
  \BibitemOpen
  \bibfield  {author} {\bibinfo {author} {\bibfnamefont {A.~M.}\ \bibnamefont {Kaufman}}\ and\ \bibinfo {author} {\bibfnamefont {K.-K.}\ \bibnamefont {Ni}},\ }\bibfield  {title} {\bibinfo {title} {Quantum science with optical tweezer arrays of ultracold atoms and molecules},\ }\href@noop {} {\bibfield  {journal} {\bibinfo  {journal} {Nature Physics}\ }\textbf {\bibinfo {volume} {17}},\ \bibinfo {pages} {1324} (\bibinfo {year} {2021})}\BibitemShut {NoStop}%
\bibitem [{\citenamefont {Barredo}\ \emph {et~al.}(2016)\citenamefont {Barredo}, \citenamefont {De~L{\'e}s{\'e}leuc}, \citenamefont {Lienhard}, \citenamefont {Lahaye},\ and\ \citenamefont {Browaeys}}]{barredo2016atom}%
  \BibitemOpen
  \bibfield  {author} {\bibinfo {author} {\bibfnamefont {D.}~\bibnamefont {Barredo}}, \bibinfo {author} {\bibfnamefont {S.}~\bibnamefont {De~L{\'e}s{\'e}leuc}}, \bibinfo {author} {\bibfnamefont {V.}~\bibnamefont {Lienhard}}, \bibinfo {author} {\bibfnamefont {T.}~\bibnamefont {Lahaye}},\ and\ \bibinfo {author} {\bibfnamefont {A.}~\bibnamefont {Browaeys}},\ }\bibfield  {title} {\bibinfo {title} {An atom-by-atom assembler of defect-free arbitrary two-dimensional atomic arrays},\ }\href@noop {} {\bibfield  {journal} {\bibinfo  {journal} {Science}\ }\textbf {\bibinfo {volume} {354}},\ \bibinfo {pages} {1021} (\bibinfo {year} {2016})}\BibitemShut {NoStop}%
\bibitem [{\citenamefont {Endres}\ \emph {et~al.}(2016)\citenamefont {Endres}, \citenamefont {Bernien}, \citenamefont {Keesling}, \citenamefont {Levine}, \citenamefont {Anschuetz}, \citenamefont {Krajenbrink}, \citenamefont {Senko}, \citenamefont {Vuletic}, \citenamefont {Greiner},\ and\ \citenamefont {Lukin}}]{endres2016atom}%
  \BibitemOpen
  \bibfield  {author} {\bibinfo {author} {\bibfnamefont {M.}~\bibnamefont {Endres}}, \bibinfo {author} {\bibfnamefont {H.}~\bibnamefont {Bernien}}, \bibinfo {author} {\bibfnamefont {A.}~\bibnamefont {Keesling}}, \bibinfo {author} {\bibfnamefont {H.}~\bibnamefont {Levine}}, \bibinfo {author} {\bibfnamefont {E.~R.}\ \bibnamefont {Anschuetz}}, \bibinfo {author} {\bibfnamefont {A.}~\bibnamefont {Krajenbrink}}, \bibinfo {author} {\bibfnamefont {C.}~\bibnamefont {Senko}}, \bibinfo {author} {\bibfnamefont {V.}~\bibnamefont {Vuletic}}, \bibinfo {author} {\bibfnamefont {M.}~\bibnamefont {Greiner}},\ and\ \bibinfo {author} {\bibfnamefont {M.~D.}\ \bibnamefont {Lukin}},\ }\bibfield  {title} {\bibinfo {title} {Atom-by-atom assembly of defect-free one-dimensional cold atom arrays},\ }\href@noop {} {\bibfield  {journal} {\bibinfo  {journal} {Science}\ }\textbf {\bibinfo {volume} {354}},\ \bibinfo {pages} {1024} (\bibinfo {year} {2016})}\BibitemShut {NoStop}%
\bibitem [{\citenamefont {Lee}\ \emph {et~al.}(2016)\citenamefont {Lee}, \citenamefont {Kim},\ and\ \citenamefont {Ahn}}]{lee2016three}%
  \BibitemOpen
  \bibfield  {author} {\bibinfo {author} {\bibfnamefont {W.}~\bibnamefont {Lee}}, \bibinfo {author} {\bibfnamefont {H.}~\bibnamefont {Kim}},\ and\ \bibinfo {author} {\bibfnamefont {J.}~\bibnamefont {Ahn}},\ }\bibfield  {title} {\bibinfo {title} {Three-dimensional rearrangement of single atoms using actively controlled optical microtraps},\ }\href@noop {} {\bibfield  {journal} {\bibinfo  {journal} {Optics express}\ }\textbf {\bibinfo {volume} {24}},\ \bibinfo {pages} {9816} (\bibinfo {year} {2016})}\BibitemShut {NoStop}%
\bibitem [{\citenamefont {Barredo}\ \emph {et~al.}(2018)\citenamefont {Barredo}, \citenamefont {Lienhard}, \citenamefont {De~Leseleuc}, \citenamefont {Lahaye},\ and\ \citenamefont {Browaeys}}]{barredo2018synthetic}%
  \BibitemOpen
  \bibfield  {author} {\bibinfo {author} {\bibfnamefont {D.}~\bibnamefont {Barredo}}, \bibinfo {author} {\bibfnamefont {V.}~\bibnamefont {Lienhard}}, \bibinfo {author} {\bibfnamefont {S.}~\bibnamefont {De~Leseleuc}}, \bibinfo {author} {\bibfnamefont {T.}~\bibnamefont {Lahaye}},\ and\ \bibinfo {author} {\bibfnamefont {A.}~\bibnamefont {Browaeys}},\ }\bibfield  {title} {\bibinfo {title} {Synthetic three-dimensional atomic structures assembled atom by atom},\ }\href@noop {} {\bibfield  {journal} {\bibinfo  {journal} {Nature}\ }\textbf {\bibinfo {volume} {561}},\ \bibinfo {pages} {79} (\bibinfo {year} {2018})}\BibitemShut {NoStop}%
\bibitem [{\citenamefont {Schlosser}\ \emph {et~al.}(2023)\citenamefont {Schlosser}, \citenamefont {Tichelmann}, \citenamefont {Sch{\"a}ffner}, \citenamefont {de~Mello}, \citenamefont {Hambach}, \citenamefont {Sch{\"u}tz},\ and\ \citenamefont {Birkl}}]{schlosser2023scalable}%
  \BibitemOpen
  \bibfield  {author} {\bibinfo {author} {\bibfnamefont {M.}~\bibnamefont {Schlosser}}, \bibinfo {author} {\bibfnamefont {S.}~\bibnamefont {Tichelmann}}, \bibinfo {author} {\bibfnamefont {D.}~\bibnamefont {Sch{\"a}ffner}}, \bibinfo {author} {\bibfnamefont {D.~O.}\ \bibnamefont {de~Mello}}, \bibinfo {author} {\bibfnamefont {M.}~\bibnamefont {Hambach}}, \bibinfo {author} {\bibfnamefont {J.}~\bibnamefont {Sch{\"u}tz}},\ and\ \bibinfo {author} {\bibfnamefont {G.}~\bibnamefont {Birkl}},\ }\bibfield  {title} {\bibinfo {title} {Scalable multilayer architecture of assembled single-atom qubit arrays in a three-dimensional talbot tweezer lattice},\ }\href {https://doi.org/10.1103/PhysRevLett.130.180601} {\bibfield  {journal} {\bibinfo  {journal} {Phys. Rev. Lett.}\ }\textbf {\bibinfo {volume} {130}},\ \bibinfo {pages} {180601} (\bibinfo {year} {2023})}\BibitemShut {NoStop}%
\bibitem [{\citenamefont {Lin}\ \emph {et~al.}(2025)\citenamefont {Lin}, \citenamefont {Zhong}, \citenamefont {Li}, \citenamefont {Zhao}, \citenamefont {Zheng}, \citenamefont {Hu}, \citenamefont {Wu}, \citenamefont {Wu}, \citenamefont {Ma}, \citenamefont {Gao} \emph {et~al.}}]{lin2025ai}%
  \BibitemOpen
  \bibfield  {author} {\bibinfo {author} {\bibfnamefont {R.}~\bibnamefont {Lin}}, \bibinfo {author} {\bibfnamefont {H.-S.}\ \bibnamefont {Zhong}}, \bibinfo {author} {\bibfnamefont {Y.}~\bibnamefont {Li}}, \bibinfo {author} {\bibfnamefont {Z.-R.}\ \bibnamefont {Zhao}}, \bibinfo {author} {\bibfnamefont {L.-T.}\ \bibnamefont {Zheng}}, \bibinfo {author} {\bibfnamefont {T.-R.}\ \bibnamefont {Hu}}, \bibinfo {author} {\bibfnamefont {H.-M.}\ \bibnamefont {Wu}}, \bibinfo {author} {\bibfnamefont {Z.}~\bibnamefont {Wu}}, \bibinfo {author} {\bibfnamefont {W.-J.}\ \bibnamefont {Ma}}, \bibinfo {author} {\bibfnamefont {Y.}~\bibnamefont {Gao}}, \emph {et~al.},\ }\bibfield  {title} {\bibinfo {title} {Ai-enabled parallel assembly of thousands of defect-free neutral atom arrays},\ }\href@noop {} {\bibfield  {journal} {\bibinfo  {journal} {Physical Review Letters}\ }\textbf {\bibinfo {volume} {135}},\ \bibinfo {pages} {060602} (\bibinfo {year} {2025})}\BibitemShut {NoStop}%
\bibitem [{\citenamefont {Bluvstein}\ \emph {et~al.}(2024)\citenamefont {Bluvstein}, \citenamefont {Evered}, \citenamefont {Geim}, \citenamefont {Li}, \citenamefont {Zhou}, \citenamefont {Manovitz}, \citenamefont {Ebadi}, \citenamefont {Cain}, \citenamefont {Kalinowski}, \citenamefont {Hangleiter} \emph {et~al.}}]{bluvstein2024logical}%
  \BibitemOpen
  \bibfield  {author} {\bibinfo {author} {\bibfnamefont {D.}~\bibnamefont {Bluvstein}}, \bibinfo {author} {\bibfnamefont {S.~J.}\ \bibnamefont {Evered}}, \bibinfo {author} {\bibfnamefont {A.~A.}\ \bibnamefont {Geim}}, \bibinfo {author} {\bibfnamefont {S.~H.}\ \bibnamefont {Li}}, \bibinfo {author} {\bibfnamefont {H.}~\bibnamefont {Zhou}}, \bibinfo {author} {\bibfnamefont {T.}~\bibnamefont {Manovitz}}, \bibinfo {author} {\bibfnamefont {S.}~\bibnamefont {Ebadi}}, \bibinfo {author} {\bibfnamefont {M.}~\bibnamefont {Cain}}, \bibinfo {author} {\bibfnamefont {M.}~\bibnamefont {Kalinowski}}, \bibinfo {author} {\bibfnamefont {D.}~\bibnamefont {Hangleiter}}, \emph {et~al.},\ }\bibfield  {title} {\bibinfo {title} {Logical quantum processor based on reconfigurable atom arrays},\ }\href@noop {} {\bibfield  {journal} {\bibinfo  {journal} {Nature}\ }\textbf {\bibinfo {volume} {626}},\ \bibinfo {pages} {58} (\bibinfo {year} {2024})}\BibitemShut {NoStop}%
\bibitem [{\citenamefont {Bluvstein}\ \emph {et~al.}(2026)\citenamefont {Bluvstein}, \citenamefont {Geim}, \citenamefont {Li}, \citenamefont {Evered}, \citenamefont {Bonilla~Ataides}, \citenamefont {Baranes}, \citenamefont {Gu}, \citenamefont {Manovitz}, \citenamefont {Xu}, \citenamefont {Kalinowski} \emph {et~al.}}]{bluvstein2026fault}%
  \BibitemOpen
  \bibfield  {author} {\bibinfo {author} {\bibfnamefont {D.}~\bibnamefont {Bluvstein}}, \bibinfo {author} {\bibfnamefont {A.~A.}\ \bibnamefont {Geim}}, \bibinfo {author} {\bibfnamefont {S.~H.}\ \bibnamefont {Li}}, \bibinfo {author} {\bibfnamefont {S.~J.}\ \bibnamefont {Evered}}, \bibinfo {author} {\bibfnamefont {J.~P.}\ \bibnamefont {Bonilla~Ataides}}, \bibinfo {author} {\bibfnamefont {G.}~\bibnamefont {Baranes}}, \bibinfo {author} {\bibfnamefont {A.}~\bibnamefont {Gu}}, \bibinfo {author} {\bibfnamefont {T.}~\bibnamefont {Manovitz}}, \bibinfo {author} {\bibfnamefont {M.}~\bibnamefont {Xu}}, \bibinfo {author} {\bibfnamefont {M.}~\bibnamefont {Kalinowski}}, \emph {et~al.},\ }\bibfield  {title} {\bibinfo {title} {A fault-tolerant neutral-atom architecture for universal quantum computation},\ }\href@noop {} {\bibfield  {journal} {\bibinfo  {journal} {Nature}\ }\textbf {\bibinfo {volume} {649}},\ \bibinfo {pages} {39} (\bibinfo {year} {2026})}\BibitemShut {NoStop}%
\bibitem [{\citenamefont {You}\ \emph {et~al.}(2025)\citenamefont {You}, \citenamefont {Doyle}, \citenamefont {Liu}, \citenamefont {Yu},\ and\ \citenamefont {Periwal}}]{you2025control}%
  \BibitemOpen
  \bibfield  {author} {\bibinfo {author} {\bibfnamefont {J.}~\bibnamefont {You}}, \bibinfo {author} {\bibfnamefont {J.~M.}\ \bibnamefont {Doyle}}, \bibinfo {author} {\bibfnamefont {Z.}~\bibnamefont {Liu}}, \bibinfo {author} {\bibfnamefont {S.~S.}\ \bibnamefont {Yu}},\ and\ \bibinfo {author} {\bibfnamefont {A.}~\bibnamefont {Periwal}},\ }\bibfield  {title} {\bibinfo {title} {Control of dipolar dynamics by geometrical programming},\ }\href {https://doi.org/10.1103/k22t-xgcy} {\bibfield  {journal} {\bibinfo  {journal} {Phys. Rev. Lett.}\ }\textbf {\bibinfo {volume} {135}},\ \bibinfo {pages} {253002} (\bibinfo {year} {2025})}\BibitemShut {NoStop}%
\bibitem [{\citenamefont {Yu}\ \emph {et~al.}(2026)\citenamefont {Yu}, \citenamefont {Periwal}, \citenamefont {You}, \citenamefont {Liu}, \citenamefont {Lyu}, \citenamefont {Cho}, \citenamefont {Anderegg}, \citenamefont {Chae},\ and\ \citenamefont {Doyle}}]{yu2026high}%
  \BibitemOpen
  \bibfield  {author} {\bibinfo {author} {\bibfnamefont {S.~S.}\ \bibnamefont {Yu}}, \bibinfo {author} {\bibfnamefont {A.}~\bibnamefont {Periwal}}, \bibinfo {author} {\bibfnamefont {J.}~\bibnamefont {You}}, \bibinfo {author} {\bibfnamefont {Z.}~\bibnamefont {Liu}}, \bibinfo {author} {\bibfnamefont {Q.}~\bibnamefont {Lyu}}, \bibinfo {author} {\bibfnamefont {Y.}~\bibnamefont {Cho}}, \bibinfo {author} {\bibfnamefont {L.}~\bibnamefont {Anderegg}}, \bibinfo {author} {\bibfnamefont {E.}~\bibnamefont {Chae}},\ and\ \bibinfo {author} {\bibfnamefont {J.~M.}\ \bibnamefont {Doyle}},\ }\href@noop {} {\bibinfo {title} {High-fidelity entanglement of polar molecules by dynamic geometric control}} (\bibinfo {year} {2026}),\ \Eprint {https://arxiv.org/abs/2607.13008} {arXiv:2607.13008} \BibitemShut {NoStop}%
\bibitem [{\citenamefont {Shaw}\ \emph {et~al.}(2024)\citenamefont {Shaw}, \citenamefont {Finkelstein}, \citenamefont {Tsai}, \citenamefont {Scholl}, \citenamefont {Yoon}, \citenamefont {Choi},\ and\ \citenamefont {Endres}}]{shaw2024multi}%
  \BibitemOpen
  \bibfield  {author} {\bibinfo {author} {\bibfnamefont {A.~L.}\ \bibnamefont {Shaw}}, \bibinfo {author} {\bibfnamefont {R.}~\bibnamefont {Finkelstein}}, \bibinfo {author} {\bibfnamefont {R.~B.-S.}\ \bibnamefont {Tsai}}, \bibinfo {author} {\bibfnamefont {P.}~\bibnamefont {Scholl}}, \bibinfo {author} {\bibfnamefont {T.~H.}\ \bibnamefont {Yoon}}, \bibinfo {author} {\bibfnamefont {J.}~\bibnamefont {Choi}},\ and\ \bibinfo {author} {\bibfnamefont {M.}~\bibnamefont {Endres}},\ }\bibfield  {title} {\bibinfo {title} {Multi-ensemble metrology by programming local rotations with atom movements},\ }\href@noop {} {\bibfield  {journal} {\bibinfo  {journal} {Nature Physics}\ }\textbf {\bibinfo {volume} {20}},\ \bibinfo {pages} {195} (\bibinfo {year} {2024})}\BibitemShut {NoStop}%
\bibitem [{\citenamefont {Di~Leonardo}\ \emph {et~al.}(2007)\citenamefont {Di~Leonardo}, \citenamefont {Ianni},\ and\ \citenamefont {Ruocco}}]{di2007computer}%
  \BibitemOpen
  \bibfield  {author} {\bibinfo {author} {\bibfnamefont {R.}~\bibnamefont {Di~Leonardo}}, \bibinfo {author} {\bibfnamefont {F.}~\bibnamefont {Ianni}},\ and\ \bibinfo {author} {\bibfnamefont {G.}~\bibnamefont {Ruocco}},\ }\bibfield  {title} {\bibinfo {title} {Computer generation of optimal holograms for optical trap arrays},\ }\href@noop {} {\bibfield  {journal} {\bibinfo  {journal} {Optics express}\ }\textbf {\bibinfo {volume} {15}},\ \bibinfo {pages} {1913} (\bibinfo {year} {2007})}\BibitemShut {NoStop}%
\bibitem [{\citenamefont {Nogrette}\ \emph {et~al.}(2014)\citenamefont {Nogrette}, \citenamefont {Labuhn}, \citenamefont {Ravets}, \citenamefont {Barredo}, \citenamefont {B{\'e}guin}, \citenamefont {Vernier}, \citenamefont {Lahaye},\ and\ \citenamefont {Browaeys}}]{nogrette2014single}%
  \BibitemOpen
  \bibfield  {author} {\bibinfo {author} {\bibfnamefont {F.}~\bibnamefont {Nogrette}}, \bibinfo {author} {\bibfnamefont {H.}~\bibnamefont {Labuhn}}, \bibinfo {author} {\bibfnamefont {S.}~\bibnamefont {Ravets}}, \bibinfo {author} {\bibfnamefont {D.}~\bibnamefont {Barredo}}, \bibinfo {author} {\bibfnamefont {L.}~\bibnamefont {B{\'e}guin}}, \bibinfo {author} {\bibfnamefont {A.}~\bibnamefont {Vernier}}, \bibinfo {author} {\bibfnamefont {T.}~\bibnamefont {Lahaye}},\ and\ \bibinfo {author} {\bibfnamefont {A.}~\bibnamefont {Browaeys}},\ }\bibfield  {title} {\bibinfo {title} {Single-atom trapping in holographic 2d arrays of microtraps with arbitrary geometries},\ }\href@noop {} {\bibfield  {journal} {\bibinfo  {journal} {Physical Review X}\ }\textbf {\bibinfo {volume} {4}},\ \bibinfo {pages} {021034} (\bibinfo {year} {2014})}\BibitemShut {NoStop}%
\bibitem [{\citenamefont {Kim}\ \emph {et~al.}(2016)\citenamefont {Kim}, \citenamefont {Lee}, \citenamefont {Lee}, \citenamefont {Jo}, \citenamefont {Song},\ and\ \citenamefont {Ahn}}]{kim2016situ}%
  \BibitemOpen
  \bibfield  {author} {\bibinfo {author} {\bibfnamefont {H.}~\bibnamefont {Kim}}, \bibinfo {author} {\bibfnamefont {W.}~\bibnamefont {Lee}}, \bibinfo {author} {\bibfnamefont {H.-g.}\ \bibnamefont {Lee}}, \bibinfo {author} {\bibfnamefont {H.}~\bibnamefont {Jo}}, \bibinfo {author} {\bibfnamefont {Y.}~\bibnamefont {Song}},\ and\ \bibinfo {author} {\bibfnamefont {J.}~\bibnamefont {Ahn}},\ }\bibfield  {title} {\bibinfo {title} {In situ single-atom array synthesis using dynamic holographic optical tweezers},\ }\href@noop {} {\bibfield  {journal} {\bibinfo  {journal} {Nature communications}\ }\textbf {\bibinfo {volume} {7}},\ \bibinfo {pages} {13317} (\bibinfo {year} {2016})}\BibitemShut {NoStop}%
\bibitem [{\citenamefont {Lee}\ \emph {et~al.}(2017)\citenamefont {Lee}, \citenamefont {Kim},\ and\ \citenamefont {Ahn}}]{lee2017defect}%
  \BibitemOpen
  \bibfield  {author} {\bibinfo {author} {\bibfnamefont {W.}~\bibnamefont {Lee}}, \bibinfo {author} {\bibfnamefont {H.}~\bibnamefont {Kim}},\ and\ \bibinfo {author} {\bibfnamefont {J.}~\bibnamefont {Ahn}},\ }\bibfield  {title} {\bibinfo {title} {Defect-free atomic array formation using the hungarian matching algorithm},\ }\href@noop {} {\bibfield  {journal} {\bibinfo  {journal} {Physical Review A}\ }\textbf {\bibinfo {volume} {95}},\ \bibinfo {pages} {053424} (\bibinfo {year} {2017})}\BibitemShut {NoStop}%
\bibitem [{\citenamefont {Knottnerus}\ \emph {et~al.}(2025)\citenamefont {Knottnerus}, \citenamefont {Tseng}, \citenamefont {Urech}, \citenamefont {Spreeuw},\ and\ \citenamefont {Schreck}}]{knottnerus2025parallel}%
  \BibitemOpen
  \bibfield  {author} {\bibinfo {author} {\bibfnamefont {I.}~\bibnamefont {Knottnerus}}, \bibinfo {author} {\bibfnamefont {Y.~C.}\ \bibnamefont {Tseng}}, \bibinfo {author} {\bibfnamefont {A.}~\bibnamefont {Urech}}, \bibinfo {author} {\bibfnamefont {R.}~\bibnamefont {Spreeuw}},\ and\ \bibinfo {author} {\bibfnamefont {F.}~\bibnamefont {Schreck}},\ }\bibfield  {title} {\bibinfo {title} {Parallel assembly of neutral atom arrays with an slm using linear phase interpolation},\ }\href {https://doi.org/10.21468/SciPostPhys.19.4.118} {\bibfield  {journal} {\bibinfo  {journal} {SciPost Phys.}\ }\textbf {\bibinfo {volume} {19}},\ \bibinfo {pages} {118} (\bibinfo {year} {2025})}\BibitemShut {NoStop}%
\bibitem [{\citenamefont {Schymik}\ \emph {et~al.}(2020)\citenamefont {Schymik}, \citenamefont {Lienhard}, \citenamefont {Barredo}, \citenamefont {Scholl}, \citenamefont {Williams}, \citenamefont {Browaeys},\ and\ \citenamefont {Lahaye}}]{schymik2020enhanced}%
  \BibitemOpen
  \bibfield  {author} {\bibinfo {author} {\bibfnamefont {K.-N.}\ \bibnamefont {Schymik}}, \bibinfo {author} {\bibfnamefont {V.}~\bibnamefont {Lienhard}}, \bibinfo {author} {\bibfnamefont {D.}~\bibnamefont {Barredo}}, \bibinfo {author} {\bibfnamefont {P.}~\bibnamefont {Scholl}}, \bibinfo {author} {\bibfnamefont {H.}~\bibnamefont {Williams}}, \bibinfo {author} {\bibfnamefont {A.}~\bibnamefont {Browaeys}},\ and\ \bibinfo {author} {\bibfnamefont {T.}~\bibnamefont {Lahaye}},\ }\bibfield  {title} {\bibinfo {title} {Enhanced atom-by-atom assembly of arbitrary tweezer arrays},\ }\href@noop {} {\bibfield  {journal} {\bibinfo  {journal} {Physical Review A}\ }\textbf {\bibinfo {volume} {102}},\ \bibinfo {pages} {063107} (\bibinfo {year} {2020})}\BibitemShut {NoStop}%
\bibitem [{\citenamefont {Sheng}\ \emph {et~al.}(2021)\citenamefont {Sheng}, \citenamefont {Hou}, \citenamefont {He}, \citenamefont {Xu}, \citenamefont {Wang}, \citenamefont {Zhuang}, \citenamefont {Li}, \citenamefont {Liu}, \citenamefont {Wang},\ and\ \citenamefont {Zhan}}]{sheng2021efficient}%
  \BibitemOpen
  \bibfield  {author} {\bibinfo {author} {\bibfnamefont {C.}~\bibnamefont {Sheng}}, \bibinfo {author} {\bibfnamefont {J.}~\bibnamefont {Hou}}, \bibinfo {author} {\bibfnamefont {X.}~\bibnamefont {He}}, \bibinfo {author} {\bibfnamefont {P.}~\bibnamefont {Xu}}, \bibinfo {author} {\bibfnamefont {K.}~\bibnamefont {Wang}}, \bibinfo {author} {\bibfnamefont {J.}~\bibnamefont {Zhuang}}, \bibinfo {author} {\bibfnamefont {X.}~\bibnamefont {Li}}, \bibinfo {author} {\bibfnamefont {M.}~\bibnamefont {Liu}}, \bibinfo {author} {\bibfnamefont {J.}~\bibnamefont {Wang}},\ and\ \bibinfo {author} {\bibfnamefont {M.}~\bibnamefont {Zhan}},\ }\bibfield  {title} {\bibinfo {title} {Efficient preparation of two-dimensional defect-free atom arrays with near-fewest sorting-atom moves},\ }\href@noop {} {\bibfield  {journal} {\bibinfo  {journal} {Physical Review Research}\ }\textbf {\bibinfo {volume} {3}},\ \bibinfo {pages} {023008} (\bibinfo {year} {2021})}\BibitemShut {NoStop}%
\bibitem [{\citenamefont {Tian}\ \emph {et~al.}(2023)\citenamefont {Tian}, \citenamefont {Wee}, \citenamefont {Qu}, \citenamefont {Lim}, \citenamefont {Datla}, \citenamefont {Koh},\ and\ \citenamefont {Loh}}]{tian2023parallel}%
  \BibitemOpen
  \bibfield  {author} {\bibinfo {author} {\bibfnamefont {W.}~\bibnamefont {Tian}}, \bibinfo {author} {\bibfnamefont {W.~J.}\ \bibnamefont {Wee}}, \bibinfo {author} {\bibfnamefont {A.}~\bibnamefont {Qu}}, \bibinfo {author} {\bibfnamefont {B.~J.~M.}\ \bibnamefont {Lim}}, \bibinfo {author} {\bibfnamefont {P.~R.}\ \bibnamefont {Datla}}, \bibinfo {author} {\bibfnamefont {V.~P.~W.}\ \bibnamefont {Koh}},\ and\ \bibinfo {author} {\bibfnamefont {H.}~\bibnamefont {Loh}},\ }\bibfield  {title} {\bibinfo {title} {Parallel assembly of arbitrary defect-free atom arrays with a multitweezer algorithm},\ }\href@noop {} {\bibfield  {journal} {\bibinfo  {journal} {Physical Review Applied}\ }\textbf {\bibinfo {volume} {19}},\ \bibinfo {pages} {034048} (\bibinfo {year} {2023})}\BibitemShut {NoStop}%
\bibitem [{\citenamefont {Wang}\ \emph {et~al.}(2023)\citenamefont {Wang}, \citenamefont {Zhang}, \citenamefont {Zhang}, \citenamefont {Mei}, \citenamefont {Wang}, \citenamefont {Hu},\ and\ \citenamefont {Chen}}]{wang2023accelerating}%
  \BibitemOpen
  \bibfield  {author} {\bibinfo {author} {\bibfnamefont {S.}~\bibnamefont {Wang}}, \bibinfo {author} {\bibfnamefont {W.}~\bibnamefont {Zhang}}, \bibinfo {author} {\bibfnamefont {T.}~\bibnamefont {Zhang}}, \bibinfo {author} {\bibfnamefont {S.}~\bibnamefont {Mei}}, \bibinfo {author} {\bibfnamefont {Y.}~\bibnamefont {Wang}}, \bibinfo {author} {\bibfnamefont {J.}~\bibnamefont {Hu}},\ and\ \bibinfo {author} {\bibfnamefont {W.}~\bibnamefont {Chen}},\ }\bibfield  {title} {\bibinfo {title} {Accelerating the assembly of defect-free atomic arrays with maximum parallelisms},\ }\href@noop {} {\bibfield  {journal} {\bibinfo  {journal} {Physical Review Applied}\ }\textbf {\bibinfo {volume} {19}},\ \bibinfo {pages} {054032} (\bibinfo {year} {2023})}\BibitemShut {NoStop}%
\bibitem [{\citenamefont {Picard}\ and\ \citenamefont {Endres}(2026)}]{picard2026three}%
  \BibitemOpen
  \bibfield  {author} {\bibinfo {author} {\bibfnamefont {L.~R.}\ \bibnamefont {Picard}}\ and\ \bibinfo {author} {\bibfnamefont {M.}~\bibnamefont {Endres}},\ }\bibfield  {title} {\bibinfo {title} {A three-dimensional acousto-optic deflector},\ }\href@noop {} {\bibfield  {journal} {\bibinfo  {journal} {Device}\ } (\bibinfo {year} {2026})}\BibitemShut {NoStop}%
\bibitem [{\citenamefont {Lu}\ \emph {et~al.}(2026)\citenamefont {Lu}, \citenamefont {Song}, \citenamefont {Xiang}, \citenamefont {Ho}, \citenamefont {Lee}, \citenamefont {Yan},\ and\ \citenamefont {Stamper-Kurn}}]{lu2026astigmatism}%
  \BibitemOpen
  \bibfield  {author} {\bibinfo {author} {\bibfnamefont {Y.-H.}\ \bibnamefont {Lu}}, \bibinfo {author} {\bibfnamefont {N.}~\bibnamefont {Song}}, \bibinfo {author} {\bibfnamefont {T.}~\bibnamefont {Xiang}}, \bibinfo {author} {\bibfnamefont {J.}~\bibnamefont {Ho}}, \bibinfo {author} {\bibfnamefont {T.-C.}\ \bibnamefont {Lee}}, \bibinfo {author} {\bibfnamefont {Z.}~\bibnamefont {Yan}},\ and\ \bibinfo {author} {\bibfnamefont {D.~M.}\ \bibnamefont {Stamper-Kurn}},\ }\bibfield  {title} {\bibinfo {title} {Astigmatism-free 3d optical tweezer control for rapid atom rearrangement},\ }\href@noop {} {\bibfield  {journal} {\bibinfo  {journal} {Optica Quantum}\ }\textbf {\bibinfo {volume} {4}},\ \bibinfo {pages} {241} (\bibinfo {year} {2026})}\BibitemShut {NoStop}%
\bibitem [{sup()}]{supp}%
  \BibitemOpen
  \href@noop {} {}\bibinfo {note} {See Supplemental Material at [URL] for additional details, which includes Refs. [42-46].}\BibitemShut {Stop}%
\bibitem [{\citenamefont {Kim}\ \emph {et~al.}(2019)\citenamefont {Kim}, \citenamefont {Kim}, \citenamefont {Lee},\ and\ \citenamefont {Ahn}}]{kim2019gerchberg}%
  \BibitemOpen
  \bibfield  {author} {\bibinfo {author} {\bibfnamefont {H.}~\bibnamefont {Kim}}, \bibinfo {author} {\bibfnamefont {M.}~\bibnamefont {Kim}}, \bibinfo {author} {\bibfnamefont {W.}~\bibnamefont {Lee}},\ and\ \bibinfo {author} {\bibfnamefont {J.}~\bibnamefont {Ahn}},\ }\bibfield  {title} {\bibinfo {title} {Gerchberg-saxton algorithm for fast and efficient atom rearrangement in optical tweezer traps},\ }\href@noop {} {\bibfield  {journal} {\bibinfo  {journal} {Optics express}\ }\textbf {\bibinfo {volume} {27}},\ \bibinfo {pages} {2184} (\bibinfo {year} {2019})}\BibitemShut {NoStop}%
\bibitem [{\citenamefont {Chew}\ \emph {et~al.}(2024)\citenamefont {Chew}, \citenamefont {Poitrinal}, \citenamefont {Tomita}, \citenamefont {Kitade}, \citenamefont {Mauricio}, \citenamefont {Ohmori},\ and\ \citenamefont {de~L\'es\'eleuc}}]{chew2024ultraprecise}%
  \BibitemOpen
  \bibfield  {author} {\bibinfo {author} {\bibfnamefont {Y.~T.}\ \bibnamefont {Chew}}, \bibinfo {author} {\bibfnamefont {M.}~\bibnamefont {Poitrinal}}, \bibinfo {author} {\bibfnamefont {T.}~\bibnamefont {Tomita}}, \bibinfo {author} {\bibfnamefont {S.}~\bibnamefont {Kitade}}, \bibinfo {author} {\bibfnamefont {J.}~\bibnamefont {Mauricio}}, \bibinfo {author} {\bibfnamefont {K.}~\bibnamefont {Ohmori}},\ and\ \bibinfo {author} {\bibfnamefont {S.}~\bibnamefont {de~L\'es\'eleuc}},\ }\bibfield  {title} {\bibinfo {title} {Ultraprecise holographic optical tweezer array},\ }\href {https://doi.org/10.1103/PhysRevA.110.053518} {\bibfield  {journal} {\bibinfo  {journal} {Phys. Rev. A}\ }\textbf {\bibinfo {volume} {110}},\ \bibinfo {pages} {053518} (\bibinfo {year} {2024})}\BibitemShut {NoStop}%
\bibitem [{\citenamefont {Huang}\ \emph {et~al.}(2026)\citenamefont {Huang}, \citenamefont {Huang}, \citenamefont {Yao}, \citenamefont {Wang},\ and\ \citenamefont {Liu}}]{huang2026phase}%
  \BibitemOpen
  \bibfield  {author} {\bibinfo {author} {\bibfnamefont {E.}~\bibnamefont {Huang}}, \bibinfo {author} {\bibfnamefont {J.}~\bibnamefont {Huang}}, \bibinfo {author} {\bibfnamefont {H.}~\bibnamefont {Yao}}, \bibinfo {author} {\bibfnamefont {X.}~\bibnamefont {Wang}},\ and\ \bibinfo {author} {\bibfnamefont {J.-G.}\ \bibnamefont {Liu}},\ }\href@noop {} {\bibinfo {title} {Phase-stable hologram updates for large-scale neutral-atom array reconfiguration}} (\bibinfo {year} {2026}),\ \Eprint {https://arxiv.org/abs/2604.04600} {arXiv:2604.04600} \BibitemShut {NoStop}%
\bibitem [{\citenamefont {Zhang}\ \emph {et~al.}(2026)\citenamefont {Zhang}, \citenamefont {Li}, \citenamefont {Zhai},\ and\ \citenamefont {Chen}}]{zhang2026algorithm}%
  \BibitemOpen
  \bibfield  {author} {\bibinfo {author} {\bibfnamefont {T.}~\bibnamefont {Zhang}}, \bibinfo {author} {\bibfnamefont {X.}~\bibnamefont {Li}}, \bibinfo {author} {\bibfnamefont {H.}~\bibnamefont {Zhai}},\ and\ \bibinfo {author} {\bibfnamefont {L.}~\bibnamefont {Chen}},\ }\href@noop {} {\bibinfo {title} {An algorithm for fast assembling large-scale defect-free atom arrays}} (\bibinfo {year} {2026}),\ \Eprint {https://arxiv.org/abs/2604.08669} {arXiv:2604.08669} \BibitemShut {NoStop}%
\bibitem [{\citenamefont {Thalhammer}\ \emph {et~al.}(2013)\citenamefont {Thalhammer}, \citenamefont {Bowman}, \citenamefont {Love}, \citenamefont {Padgett},\ and\ \citenamefont {Ritsch-Marte}}]{thalhammer2013overdrive}%
  \BibitemOpen
  \bibfield  {author} {\bibinfo {author} {\bibfnamefont {G.}~\bibnamefont {Thalhammer}}, \bibinfo {author} {\bibfnamefont {R.~W.}\ \bibnamefont {Bowman}}, \bibinfo {author} {\bibfnamefont {G.~D.}\ \bibnamefont {Love}}, \bibinfo {author} {\bibfnamefont {M.~J.}\ \bibnamefont {Padgett}},\ and\ \bibinfo {author} {\bibfnamefont {M.}~\bibnamefont {Ritsch-Marte}},\ }\bibfield  {title} {\bibinfo {title} {Speeding up liquid crystal slms using overdrive with phase change reduction},\ }\href {https://doi.org/10.1364/OE.21.001779} {\bibfield  {journal} {\bibinfo  {journal} {Opt. Express}\ }\textbf {\bibinfo {volume} {21}},\ \bibinfo {pages} {1779} (\bibinfo {year} {2013})}\BibitemShut {NoStop}%
\bibitem [{\citenamefont {Norcia}\ \emph {et~al.}(2024)\citenamefont {Norcia}, \citenamefont {Kim}, \citenamefont {Cairncross}, \citenamefont {Stone}, \citenamefont {Ryou}, \citenamefont {Jaffe}, \citenamefont {Brown}, \citenamefont {Barnes}, \citenamefont {Battaglino}, \citenamefont {Bohdanowicz}, \citenamefont {Brown}, \citenamefont {Cassella}, \citenamefont {Chen}, \citenamefont {Coxe}, \citenamefont {Crow}, \citenamefont {Epstein}, \citenamefont {Griger}, \citenamefont {Halperin}, \citenamefont {Hummel}, \citenamefont {Jones}, \citenamefont {Kindem}, \citenamefont {King}, \citenamefont {Kotru}, \citenamefont {Lauigan}, \citenamefont {Li}, \citenamefont {Lu}, \citenamefont {Megidish}, \citenamefont {Marjanovic}, \citenamefont {McDonald}, \citenamefont {Mittiga}, \citenamefont {Muniz}, \citenamefont {Narayanaswami}, \citenamefont {Nishiguchi}, \citenamefont {Paule}, \citenamefont {Pawlak}, \citenamefont {Peng}, \citenamefont {Pudenz}, \citenamefont {Rodr{\'{\i}}guez~P{\'e}rez}, \citenamefont {Smull},
  \citenamefont {Stack}, \citenamefont {Urbanek}, \citenamefont {van~de Veerdonk}, \citenamefont {Vendeiro}, \citenamefont {Wadleigh}, \citenamefont {Wilkason}, \citenamefont {Wu}, \citenamefont {Xie}, \citenamefont {Zalys-Geller}, \citenamefont {Zhang},\ and\ \citenamefont {Bloom}}]{norcia2024iterative}%
  \BibitemOpen
  \bibfield  {author} {\bibinfo {author} {\bibfnamefont {M.~A.}\ \bibnamefont {Norcia}}, \bibinfo {author} {\bibfnamefont {H.}~\bibnamefont {Kim}}, \bibinfo {author} {\bibfnamefont {W.~B.}\ \bibnamefont {Cairncross}}, \bibinfo {author} {\bibfnamefont {M.}~\bibnamefont {Stone}}, \bibinfo {author} {\bibfnamefont {A.}~\bibnamefont {Ryou}}, \bibinfo {author} {\bibfnamefont {M.}~\bibnamefont {Jaffe}}, \bibinfo {author} {\bibfnamefont {M.~O.}\ \bibnamefont {Brown}}, \bibinfo {author} {\bibfnamefont {K.}~\bibnamefont {Barnes}}, \bibinfo {author} {\bibfnamefont {P.}~\bibnamefont {Battaglino}}, \bibinfo {author} {\bibfnamefont {T.~C.}\ \bibnamefont {Bohdanowicz}}, \bibinfo {author} {\bibfnamefont {A.}~\bibnamefont {Brown}}, \bibinfo {author} {\bibfnamefont {K.}~\bibnamefont {Cassella}}, \bibinfo {author} {\bibfnamefont {C.-A.}\ \bibnamefont {Chen}}, \bibinfo {author} {\bibfnamefont {R.}~\bibnamefont {Coxe}}, \bibinfo {author} {\bibfnamefont {D.}~\bibnamefont {Crow}}, \bibinfo {author} {\bibfnamefont {J.}~\bibnamefont
  {Epstein}}, \bibinfo {author} {\bibfnamefont {C.}~\bibnamefont {Griger}}, \bibinfo {author} {\bibfnamefont {E.}~\bibnamefont {Halperin}}, \bibinfo {author} {\bibfnamefont {F.}~\bibnamefont {Hummel}}, \bibinfo {author} {\bibfnamefont {A.~M.~W.}\ \bibnamefont {Jones}}, \bibinfo {author} {\bibfnamefont {J.~M.}\ \bibnamefont {Kindem}}, \bibinfo {author} {\bibfnamefont {J.}~\bibnamefont {King}}, \bibinfo {author} {\bibfnamefont {K.}~\bibnamefont {Kotru}}, \bibinfo {author} {\bibfnamefont {J.}~\bibnamefont {Lauigan}}, \bibinfo {author} {\bibfnamefont {M.}~\bibnamefont {Li}}, \bibinfo {author} {\bibfnamefont {M.}~\bibnamefont {Lu}}, \bibinfo {author} {\bibfnamefont {E.}~\bibnamefont {Megidish}}, \bibinfo {author} {\bibfnamefont {J.}~\bibnamefont {Marjanovic}}, \bibinfo {author} {\bibfnamefont {M.}~\bibnamefont {McDonald}}, \bibinfo {author} {\bibfnamefont {T.}~\bibnamefont {Mittiga}}, \bibinfo {author} {\bibfnamefont {J.~A.}\ \bibnamefont {Muniz}}, \bibinfo {author} {\bibfnamefont {S.}~\bibnamefont
  {Narayanaswami}}, \bibinfo {author} {\bibfnamefont {C.}~\bibnamefont {Nishiguchi}}, \bibinfo {author} {\bibfnamefont {T.}~\bibnamefont {Paule}}, \bibinfo {author} {\bibfnamefont {K.~A.}\ \bibnamefont {Pawlak}}, \bibinfo {author} {\bibfnamefont {L.~S.}\ \bibnamefont {Peng}}, \bibinfo {author} {\bibfnamefont {K.~L.}\ \bibnamefont {Pudenz}}, \bibinfo {author} {\bibfnamefont {D.}~\bibnamefont {Rodr{\'{\i}}guez~P{\'e}rez}}, \bibinfo {author} {\bibfnamefont {A.}~\bibnamefont {Smull}}, \bibinfo {author} {\bibfnamefont {D.}~\bibnamefont {Stack}}, \bibinfo {author} {\bibfnamefont {M.}~\bibnamefont {Urbanek}}, \bibinfo {author} {\bibfnamefont {R.~J.~M.}\ \bibnamefont {van~de Veerdonk}}, \bibinfo {author} {\bibfnamefont {Z.}~\bibnamefont {Vendeiro}}, \bibinfo {author} {\bibfnamefont {L.}~\bibnamefont {Wadleigh}}, \bibinfo {author} {\bibfnamefont {T.}~\bibnamefont {Wilkason}}, \bibinfo {author} {\bibfnamefont {T.-Y.}\ \bibnamefont {Wu}}, \bibinfo {author} {\bibfnamefont {X.}~\bibnamefont {Xie}}, \bibinfo {author}
  {\bibfnamefont {E.}~\bibnamefont {Zalys-Geller}}, \bibinfo {author} {\bibfnamefont {X.}~\bibnamefont {Zhang}},\ and\ \bibinfo {author} {\bibfnamefont {B.~J.}\ \bibnamefont {Bloom}},\ }\bibfield  {title} {\bibinfo {title} {Iterative assembly of ${}^{171}$$\mathrm{Yb}$ atom arrays with cavity-enhanced optical lattices},\ }\href {https://doi.org/10.1103/PRXQuantum.5.030316} {\bibfield  {journal} {\bibinfo  {journal} {PRX Quantum}\ }\textbf {\bibinfo {volume} {5}},\ \bibinfo {pages} {030316} (\bibinfo {year} {2024})}\BibitemShut {NoStop}%
\bibitem [{\citenamefont {Zhu}\ \emph {et~al.}(2026)\citenamefont {Zhu}, \citenamefont {Chen}, \citenamefont {Zhou}, \citenamefont {Xie}, \citenamefont {Jiang}, \citenamefont {Ding}, \citenamefont {Wu}, \citenamefont {Yang}, \citenamefont {Wang}, \citenamefont {Gong}, \citenamefont {Zhang}, \citenamefont {Zhang},\ and\ \citenamefont {Peng}}]{zhu2026high}%
  \BibitemOpen
  \bibfield  {author} {\bibinfo {author} {\bibfnamefont {J.}~\bibnamefont {Zhu}}, \bibinfo {author} {\bibfnamefont {C.}~\bibnamefont {Chen}}, \bibinfo {author} {\bibfnamefont {L.}~\bibnamefont {Zhou}}, \bibinfo {author} {\bibfnamefont {X.}~\bibnamefont {Xie}}, \bibinfo {author} {\bibfnamefont {C.}~\bibnamefont {Jiang}}, \bibinfo {author} {\bibfnamefont {Z.}~\bibnamefont {Ding}}, \bibinfo {author} {\bibfnamefont {F.}~\bibnamefont {Wu}}, \bibinfo {author} {\bibfnamefont {F.}~\bibnamefont {Yang}}, \bibinfo {author} {\bibfnamefont {G.}~\bibnamefont {Wang}}, \bibinfo {author} {\bibfnamefont {Q.}~\bibnamefont {Gong}}, \bibinfo {author} {\bibfnamefont {P.}~\bibnamefont {Zhang}}, \bibinfo {author} {\bibfnamefont {S.}~\bibnamefont {Zhang}},\ and\ \bibinfo {author} {\bibfnamefont {P.}~\bibnamefont {Peng}},\ }\bibfield  {title} {\bibinfo {title} {High-efficiency loading of 2400 ytterbium atoms in optical tweezer arrays},\ }\href {https://doi.org/10.1103/bp6k-8zmd} {\bibfield  {journal} {\bibinfo  {journal} {Phys. Rev.
  Lett.}\ }\textbf {\bibinfo {volume} {137}},\ \bibinfo {pages} {063201} (\bibinfo {year} {2026})}\BibitemShut {NoStop}%
\bibitem [{\citenamefont {Pichard}\ \emph {et~al.}(2024)\citenamefont {Pichard}, \citenamefont {Lim}, \citenamefont {Bloch}, \citenamefont {Vaneecloo}, \citenamefont {Bourachot}, \citenamefont {Both}, \citenamefont {M{\'e}riaux}, \citenamefont {Dutartre}, \citenamefont {Hostein}, \citenamefont {Paris}, \citenamefont {Ximenez}, \citenamefont {Signoles}, \citenamefont {Browaeys}, \citenamefont {Lahaye},\ and\ \citenamefont {Dreon}}]{pichard2024rearrangement}%
  \BibitemOpen
  \bibfield  {author} {\bibinfo {author} {\bibfnamefont {G.}~\bibnamefont {Pichard}}, \bibinfo {author} {\bibfnamefont {D.}~\bibnamefont {Lim}}, \bibinfo {author} {\bibfnamefont {{\'E}.}~\bibnamefont {Bloch}}, \bibinfo {author} {\bibfnamefont {J.}~\bibnamefont {Vaneecloo}}, \bibinfo {author} {\bibfnamefont {L.}~\bibnamefont {Bourachot}}, \bibinfo {author} {\bibfnamefont {G.-J.}\ \bibnamefont {Both}}, \bibinfo {author} {\bibfnamefont {G.}~\bibnamefont {M{\'e}riaux}}, \bibinfo {author} {\bibfnamefont {S.}~\bibnamefont {Dutartre}}, \bibinfo {author} {\bibfnamefont {R.}~\bibnamefont {Hostein}}, \bibinfo {author} {\bibfnamefont {J.}~\bibnamefont {Paris}}, \bibinfo {author} {\bibfnamefont {B.}~\bibnamefont {Ximenez}}, \bibinfo {author} {\bibfnamefont {A.}~\bibnamefont {Signoles}}, \bibinfo {author} {\bibfnamefont {A.}~\bibnamefont {Browaeys}}, \bibinfo {author} {\bibfnamefont {T.}~\bibnamefont {Lahaye}},\ and\ \bibinfo {author} {\bibfnamefont {D.}~\bibnamefont {Dreon}},\ }\bibfield  {title} {\bibinfo {title}
  {Rearrangement of individual atoms in a 2000-site optical-tweezer array at cryogenic temperatures},\ }\href {https://doi.org/10.1103/PhysRevApplied.22.024073} {\bibfield  {journal} {\bibinfo  {journal} {Phys. Rev. Appl.}\ }\textbf {\bibinfo {volume} {22}},\ \bibinfo {pages} {024073} (\bibinfo {year} {2024})}\BibitemShut {NoStop}%
\bibitem [{\citenamefont {Lim}\ \emph {et~al.}(2026)\citenamefont {Lim}, \citenamefont {Mamann}, \citenamefont {Pichard}, \citenamefont {Bourachot}, \citenamefont {Lindberg}, \citenamefont {Hamot}, \citenamefont {Le~Bars}, \citenamefont {Fasola}, \citenamefont {Tan}, \citenamefont {Cournez}, \citenamefont {Dutartre}, \citenamefont {Cartry}, \citenamefont {Lemettre}, \citenamefont {Hostein}, \citenamefont {Paris}, \citenamefont {Ferreyrol}, \citenamefont {Collardey}, \citenamefont {Signoles}, \citenamefont {Lahaye}, \citenamefont {Monmeyran},\ and\ \citenamefont {Ximenez}}]{lim2026defect}%
  \BibitemOpen
  \bibfield  {author} {\bibinfo {author} {\bibfnamefont {D.}~\bibnamefont {Lim}}, \bibinfo {author} {\bibfnamefont {H.}~\bibnamefont {Mamann}}, \bibinfo {author} {\bibfnamefont {G.}~\bibnamefont {Pichard}}, \bibinfo {author} {\bibfnamefont {L.}~\bibnamefont {Bourachot}}, \bibinfo {author} {\bibfnamefont {A.}~\bibnamefont {Lindberg}}, \bibinfo {author} {\bibfnamefont {C.}~\bibnamefont {Hamot}}, \bibinfo {author} {\bibfnamefont {H.}~\bibnamefont {Le~Bars}}, \bibinfo {author} {\bibfnamefont {F.}~\bibnamefont {Fasola}}, \bibinfo {author} {\bibfnamefont {S.}~\bibnamefont {Tan}}, \bibinfo {author} {\bibfnamefont {G.}~\bibnamefont {Cournez}}, \bibinfo {author} {\bibfnamefont {S.}~\bibnamefont {Dutartre}}, \bibinfo {author} {\bibfnamefont {T.}~\bibnamefont {Cartry}}, \bibinfo {author} {\bibfnamefont {S.}~\bibnamefont {Lemettre}}, \bibinfo {author} {\bibfnamefont {R.}~\bibnamefont {Hostein}}, \bibinfo {author} {\bibfnamefont {J.}~\bibnamefont {Paris}}, \bibinfo {author} {\bibfnamefont {F.}~\bibnamefont {Ferreyrol}},
  \bibinfo {author} {\bibfnamefont {A.}~\bibnamefont {Collardey}}, \bibinfo {author} {\bibfnamefont {A.}~\bibnamefont {Signoles}}, \bibinfo {author} {\bibfnamefont {T.}~\bibnamefont {Lahaye}}, \bibinfo {author} {\bibfnamefont {C.}~\bibnamefont {Monmeyran}},\ and\ \bibinfo {author} {\bibfnamefont {B.}~\bibnamefont {Ximenez}},\ }\bibfield  {title} {\bibinfo {title} {Defect-free arrays at the thousand-atom scale in a 4-k cryogenic environment},\ }\href {https://doi.org/10.1103/45t1-vl5y} {\bibfield  {journal} {\bibinfo  {journal} {PRX Quantum}\ }\textbf {\bibinfo {volume} {7}},\ \bibinfo {pages} {033032} (\bibinfo {year} {2026})}\BibitemShut {NoStop}%
\bibitem [{\citenamefont {Chiu}\ \emph {et~al.}(2025)\citenamefont {Chiu}, \citenamefont {Trapp}, \citenamefont {Guo}, \citenamefont {Abobeih}, \citenamefont {Stewart}, \citenamefont {Hollerith}, \citenamefont {Stroganov}, \citenamefont {Kalinowski}, \citenamefont {Geim}, \citenamefont {Evered} \emph {et~al.}}]{chiu2025continuous}%
  \BibitemOpen
  \bibfield  {author} {\bibinfo {author} {\bibfnamefont {N.-C.}\ \bibnamefont {Chiu}}, \bibinfo {author} {\bibfnamefont {E.~C.}\ \bibnamefont {Trapp}}, \bibinfo {author} {\bibfnamefont {J.}~\bibnamefont {Guo}}, \bibinfo {author} {\bibfnamefont {M.~H.}\ \bibnamefont {Abobeih}}, \bibinfo {author} {\bibfnamefont {L.~M.}\ \bibnamefont {Stewart}}, \bibinfo {author} {\bibfnamefont {S.}~\bibnamefont {Hollerith}}, \bibinfo {author} {\bibfnamefont {P.~L.}\ \bibnamefont {Stroganov}}, \bibinfo {author} {\bibfnamefont {M.}~\bibnamefont {Kalinowski}}, \bibinfo {author} {\bibfnamefont {A.~A.}\ \bibnamefont {Geim}}, \bibinfo {author} {\bibfnamefont {S.~J.}\ \bibnamefont {Evered}}, \emph {et~al.},\ }\bibfield  {title} {\bibinfo {title} {Continuous operation of a coherent 3,000-qubit system},\ }\href@noop {} {\bibfield  {journal} {\bibinfo  {journal} {Nature}\ }\textbf {\bibinfo {volume} {646}},\ \bibinfo {pages} {1075} (\bibinfo {year} {2025})}\BibitemShut {NoStop}%
\bibitem [{\citenamefont {Manetsch}\ \emph {et~al.}(2025)\citenamefont {Manetsch}, \citenamefont {Nomura}, \citenamefont {Bataille}, \citenamefont {Lv}, \citenamefont {Leung},\ and\ \citenamefont {Endres}}]{manetsch2025tweezer}%
  \BibitemOpen
  \bibfield  {author} {\bibinfo {author} {\bibfnamefont {H.~J.}\ \bibnamefont {Manetsch}}, \bibinfo {author} {\bibfnamefont {G.}~\bibnamefont {Nomura}}, \bibinfo {author} {\bibfnamefont {E.}~\bibnamefont {Bataille}}, \bibinfo {author} {\bibfnamefont {X.}~\bibnamefont {Lv}}, \bibinfo {author} {\bibfnamefont {K.~H.}\ \bibnamefont {Leung}},\ and\ \bibinfo {author} {\bibfnamefont {M.}~\bibnamefont {Endres}},\ }\bibfield  {title} {\bibinfo {title} {A tweezer array with 6,100 highly coherent atomic qubits},\ }\href@noop {} {\bibfield  {journal} {\bibinfo  {journal} {Nature}\ }\textbf {\bibinfo {volume} {647}},\ \bibinfo {pages} {60} (\bibinfo {year} {2025})}\BibitemShut {NoStop}%
\bibitem [{\citenamefont {Song}\ \emph {et~al.}(2021)\citenamefont {Song}, \citenamefont {Kim}, \citenamefont {Hwang}, \citenamefont {Lee},\ and\ \citenamefont {Ahn}}]{song2021quantum}%
  \BibitemOpen
  \bibfield  {author} {\bibinfo {author} {\bibfnamefont {Y.}~\bibnamefont {Song}}, \bibinfo {author} {\bibfnamefont {M.}~\bibnamefont {Kim}}, \bibinfo {author} {\bibfnamefont {H.}~\bibnamefont {Hwang}}, \bibinfo {author} {\bibfnamefont {W.}~\bibnamefont {Lee}},\ and\ \bibinfo {author} {\bibfnamefont {J.}~\bibnamefont {Ahn}},\ }\bibfield  {title} {\bibinfo {title} {Quantum simulation of cayley-tree ising hamiltonians with three-dimensional rydberg atoms},\ }\href {https://doi.org/10.1103/PhysRevResearch.3.013286} {\bibfield  {journal} {\bibinfo  {journal} {Phys. Rev. Res.}\ }\textbf {\bibinfo {volume} {3}},\ \bibinfo {pages} {013286} (\bibinfo {year} {2021})}\BibitemShut {NoStop}%
\bibitem [{\citenamefont {Shah}\ \emph {et~al.}(2025)\citenamefont {Shah}, \citenamefont {Nambiar}, \citenamefont {Gorshkov},\ and\ \citenamefont {Galitski}}]{shah2025quantum}%
  \BibitemOpen
  \bibfield  {author} {\bibinfo {author} {\bibfnamefont {J.}~\bibnamefont {Shah}}, \bibinfo {author} {\bibfnamefont {G.}~\bibnamefont {Nambiar}}, \bibinfo {author} {\bibfnamefont {A.~V.}\ \bibnamefont {Gorshkov}},\ and\ \bibinfo {author} {\bibfnamefont {V.}~\bibnamefont {Galitski}},\ }\bibfield  {title} {\bibinfo {title} {Quantum spin ice in three-dimensional rydberg atom arrays},\ }\href {https://doi.org/10.1103/PhysRevX.15.011025} {\bibfield  {journal} {\bibinfo  {journal} {Phys. Rev. X}\ }\textbf {\bibinfo {volume} {15}},\ \bibinfo {pages} {011025} (\bibinfo {year} {2025})}\BibitemShut {NoStop}%
\bibitem [{\citenamefont {Bintz}\ \emph {et~al.}(2024)\citenamefont {Bintz}, \citenamefont {Liu}, \citenamefont {Hauschild}, \citenamefont {Khalifa}, \citenamefont {Chatterjee}, \citenamefont {Zaletel},\ and\ \citenamefont {Yao}}]{bintz2024dirac}%
  \BibitemOpen
  \bibfield  {author} {\bibinfo {author} {\bibfnamefont {M.}~\bibnamefont {Bintz}}, \bibinfo {author} {\bibfnamefont {V.~S.}\ \bibnamefont {Liu}}, \bibinfo {author} {\bibfnamefont {J.}~\bibnamefont {Hauschild}}, \bibinfo {author} {\bibfnamefont {A.}~\bibnamefont {Khalifa}}, \bibinfo {author} {\bibfnamefont {S.}~\bibnamefont {Chatterjee}}, \bibinfo {author} {\bibfnamefont {M.~P.}\ \bibnamefont {Zaletel}},\ and\ \bibinfo {author} {\bibfnamefont {N.~Y.}\ \bibnamefont {Yao}},\ }\href@noop {} {\bibinfo {title} {Dirac spin liquid in quantum dipole arrays}} (\bibinfo {year} {2024}),\ \Eprint {https://arxiv.org/abs/2406.00098} {arXiv:2406.00098} \BibitemShut {NoStop}%
\bibitem [{\citenamefont {Bornet}\ \emph {et~al.}(2026)\citenamefont {Bornet}, \citenamefont {Bintz}, \citenamefont {Chen}, \citenamefont {Emperauger}, \citenamefont {Qiao}, \citenamefont {Martin}, \citenamefont {Barredo}, \citenamefont {Chatterjee}, \citenamefont {Liu}, \citenamefont {Lahaye} \emph {et~al.}}]{bornet2026dirac}%
  \BibitemOpen
  \bibfield  {author} {\bibinfo {author} {\bibfnamefont {G.}~\bibnamefont {Bornet}}, \bibinfo {author} {\bibfnamefont {M.}~\bibnamefont {Bintz}}, \bibinfo {author} {\bibfnamefont {C.}~\bibnamefont {Chen}}, \bibinfo {author} {\bibfnamefont {G.}~\bibnamefont {Emperauger}}, \bibinfo {author} {\bibfnamefont {M.}~\bibnamefont {Qiao}}, \bibinfo {author} {\bibfnamefont {R.}~\bibnamefont {Martin}}, \bibinfo {author} {\bibfnamefont {D.}~\bibnamefont {Barredo}}, \bibinfo {author} {\bibfnamefont {S.}~\bibnamefont {Chatterjee}}, \bibinfo {author} {\bibfnamefont {V.~S.}\ \bibnamefont {Liu}}, \bibinfo {author} {\bibfnamefont {T.}~\bibnamefont {Lahaye}}, \emph {et~al.},\ }\href@noop {} {\bibinfo {title} {Dirac spin liquid candidate in a rydberg quantum simulator}} (\bibinfo {year} {2026}),\ \Eprint {https://arxiv.org/abs/2602.14323} {arXiv:2602.14323} \BibitemShut {NoStop}%
\bibitem [{\citenamefont {Machado}\ \emph {et~al.}(2026)\citenamefont {Machado}, \citenamefont {Chern}, \citenamefont {Zaletel},\ and\ \citenamefont {Yao}}]{machado2026dipolar}%
  \BibitemOpen
  \bibfield  {author} {\bibinfo {author} {\bibfnamefont {F.}~\bibnamefont {Machado}}, \bibinfo {author} {\bibfnamefont {S.}~\bibnamefont {Chern}}, \bibinfo {author} {\bibfnamefont {M.~P.}\ \bibnamefont {Zaletel}},\ and\ \bibinfo {author} {\bibfnamefont {N.~Y.}\ \bibnamefont {Yao}},\ }\href@noop {} {\bibinfo {title} {A dipolar chiral spin liquid on the breathed kagome lattice}} (\bibinfo {year} {2026}),\ \Eprint {https://arxiv.org/abs/2603.25784} {arXiv:2603.25784} \BibitemShut {NoStop}%
\bibitem [{\citenamefont {Mao}\ \emph {et~al.}(2026)\citenamefont {Mao}, \citenamefont {Ma},\ and\ \citenamefont {Xu}}]{mao2026chiral}%
  \BibitemOpen
  \bibfield  {author} {\bibinfo {author} {\bibfnamefont {Y.-F.}\ \bibnamefont {Mao}}, \bibinfo {author} {\bibfnamefont {S.}~\bibnamefont {Ma}},\ and\ \bibinfo {author} {\bibfnamefont {Y.}~\bibnamefont {Xu}},\ }\href@noop {} {\bibinfo {title} {Chiral spin liquid in rydberg atom arrays}} (\bibinfo {year} {2026}),\ \Eprint {https://arxiv.org/abs/2603.21147} {arXiv:2603.21147} \BibitemShut {NoStop}%
\bibitem [{\citenamefont {Saskin}\ \emph {et~al.}(2019)\citenamefont {Saskin}, \citenamefont {Wilson}, \citenamefont {Grinkemeyer},\ and\ \citenamefont {Thompson}}]{saskin2019narrow}%
  \BibitemOpen
  \bibfield  {author} {\bibinfo {author} {\bibfnamefont {S.}~\bibnamefont {Saskin}}, \bibinfo {author} {\bibfnamefont {J.~T.}\ \bibnamefont {Wilson}}, \bibinfo {author} {\bibfnamefont {B.}~\bibnamefont {Grinkemeyer}},\ and\ \bibinfo {author} {\bibfnamefont {J.~D.}\ \bibnamefont {Thompson}},\ }\bibfield  {title} {\bibinfo {title} {Narrow-line cooling and imaging of ytterbium atoms in an optical tweezer array},\ }\href {https://doi.org/10.1103/PhysRevLett.122.143002} {\bibfield  {journal} {\bibinfo  {journal} {Phys. Rev. Lett.}\ }\textbf {\bibinfo {volume} {122}},\ \bibinfo {pages} {143002} (\bibinfo {year} {2019})}\BibitemShut {NoStop}%
\bibitem [{\citenamefont {Wei}\ \emph {et~al.}(2026)\citenamefont {Wei}, \citenamefont {Li}, \citenamefont {Karve}, \citenamefont {Shaw}, \citenamefont {Schuster},\ and\ \citenamefont {Simon}}]{wei202610}%
  \BibitemOpen
  \bibfield  {author} {\bibinfo {author} {\bibfnamefont {X.}~\bibnamefont {Wei}}, \bibinfo {author} {\bibfnamefont {Z.}~\bibnamefont {Li}}, \bibinfo {author} {\bibfnamefont {A.~V.}\ \bibnamefont {Karve}}, \bibinfo {author} {\bibfnamefont {A.~L.}\ \bibnamefont {Shaw}}, \bibinfo {author} {\bibfnamefont {D.~I.}\ \bibnamefont {Schuster}},\ and\ \bibinfo {author} {\bibfnamefont {J.}~\bibnamefont {Simon}},\ }\href@noop {} {\bibinfo {title} {A 10 megahertz spatial light modulator}} (\bibinfo {year} {2026}),\ \Eprint {https://arxiv.org/abs/2601.08906} {arXiv:2601.08906} \BibitemShut {NoStop}%
\bibitem [{\citenamefont {Bytyqi}\ \emph {et~al.}(2026)\citenamefont {Bytyqi}, \citenamefont {Sinclair}, \citenamefont {Ramette},\ and\ \citenamefont {Vuleti{\'c}}}]{bytyqi2026device}%
  \BibitemOpen
  \bibfield  {author} {\bibinfo {author} {\bibfnamefont {E.}~\bibnamefont {Bytyqi}}, \bibinfo {author} {\bibfnamefont {J.}~\bibnamefont {Sinclair}}, \bibinfo {author} {\bibfnamefont {J.}~\bibnamefont {Ramette}},\ and\ \bibinfo {author} {\bibfnamefont {V.}~\bibnamefont {Vuleti{\'c}}},\ }\bibfield  {title} {\bibinfo {title} {Device for mhz-rate rastering of arbitrary 2d optical potentials},\ }\href@noop {} {\bibfield  {journal} {\bibinfo  {journal} {Optics Letters}\ }\textbf {\bibinfo {volume} {51}},\ \bibinfo {pages} {3729} (\bibinfo {year} {2026})}\BibitemShut {NoStop}%
\bibitem [{\citenamefont {Deters}\ \emph {et~al.}(2026)\citenamefont {Deters}, \citenamefont {Li}, \citenamefont {Douglas}, \citenamefont {Greiner},\ and\ \citenamefont {Young}}]{deters2026ultrafast}%
  \BibitemOpen
  \bibfield  {author} {\bibinfo {author} {\bibfnamefont {A.~D.}\ \bibnamefont {Deters}}, \bibinfo {author} {\bibfnamefont {Y.}~\bibnamefont {Li}}, \bibinfo {author} {\bibfnamefont {A.}~\bibnamefont {Douglas}}, \bibinfo {author} {\bibfnamefont {M.}~\bibnamefont {Greiner}},\ and\ \bibinfo {author} {\bibfnamefont {A.~W.}\ \bibnamefont {Young}},\ }\href@noop {} {\bibinfo {title} {Ultrafast and high resolution spatial light modulation for cold atoms}} (\bibinfo {year} {2026}),\ \Eprint {https://arxiv.org/abs/2608.18071} {arXiv:2608.18071} \BibitemShut {NoStop}%
\bibitem [{\citenamefont {Zhang}\ \emph {et~al.}(2024)\citenamefont {Zhang}, \citenamefont {Peng}, \citenamefont {Paul},\ and\ \citenamefont {Thompson}}]{zhang2024scaled}%
  \BibitemOpen
  \bibfield  {author} {\bibinfo {author} {\bibfnamefont {B.}~\bibnamefont {Zhang}}, \bibinfo {author} {\bibfnamefont {P.}~\bibnamefont {Peng}}, \bibinfo {author} {\bibfnamefont {A.}~\bibnamefont {Paul}},\ and\ \bibinfo {author} {\bibfnamefont {J.~D.}\ \bibnamefont {Thompson}},\ }\bibfield  {title} {\bibinfo {title} {Scaled local gate controller for optically addressed qubits},\ }\href@noop {} {\bibfield  {journal} {\bibinfo  {journal} {Optica}\ }\textbf {\bibinfo {volume} {11}},\ \bibinfo {pages} {227} (\bibinfo {year} {2024})}\BibitemShut {NoStop}%
\end{thebibliography}%


\begin{thebibliography}{13}%
\makeatletter
\providecommand \@ifxundefined [1]{%
 \@ifx{#1\undefined}
}%
\providecommand \@ifnum [1]{%
 \ifnum #1\expandafter \@firstoftwo
 \else \expandafter \@secondoftwo
 \fi
}%
\providecommand \@ifx [1]{%
 \ifx #1\expandafter \@firstoftwo
 \else \expandafter \@secondoftwo
 \fi
}%
\providecommand \natexlab [1]{#1}%
\providecommand \enquote  [1]{``#1''}%
\providecommand \bibnamefont  [1]{#1}%
\providecommand \bibfnamefont [1]{#1}%
\providecommand \citenamefont [1]{#1}%
\providecommand \href@noop [0]{\@secondoftwo}%
\providecommand \href [0]{\begingroup \@sanitize@url \@href}%
\providecommand \@href[1]{\@@startlink{#1}\@@href}%
\providecommand \@@href[1]{\endgroup#1\@@endlink}%
\providecommand \@sanitize@url [0]{\catcode `\\12\catcode `\$12\catcode `\&12\catcode `\#12\catcode `\^12\catcode `\_12\catcode `\%12\relax}%
\providecommand \@@startlink[1]{}%
\providecommand \@@endlink[0]{}%
\providecommand \url  [0]{\begingroup\@sanitize@url \@url }%
\providecommand \@url [1]{\endgroup\@href {#1}{\urlprefix }}%
\providecommand \urlprefix  [0]{URL }%
\providecommand \Eprint [0]{\href }%
\providecommand \doibase [0]{https://doi.org/}%
\providecommand \selectlanguage [0]{\@gobble}%
\providecommand \bibinfo  [0]{\@secondoftwo}%
\providecommand \bibfield  [0]{\@secondoftwo}%
\providecommand \translation [1]{[#1]}%
\providecommand \BibitemOpen [0]{}%
\providecommand \bibitemStop [0]{}%
\providecommand \bibitemNoStop [0]{.\EOS\space}%
\providecommand \EOS [0]{\spacefactor3000\relax}%
\providecommand \BibitemShut  [1]{\csname bibitem#1\endcsname}%
\let\auto@bib@innerbib\@empty
\bibitem [{\citenamefont {Zhu}\ \emph {et~al.}(2026)\citenamefont {Zhu}, \citenamefont {Chen}, \citenamefont {Zhou}, \citenamefont {Xie}, \citenamefont {Jiang}, \citenamefont {Ding}, \citenamefont {Wu}, \citenamefont {Yang}, \citenamefont {Wang}, \citenamefont {Gong}, \citenamefont {Zhang}, \citenamefont {Zhang},\ and\ \citenamefont {Peng}}]{zhu2026high}%
  \BibitemOpen
  \bibfield  {author} {\bibinfo {author} {\bibfnamefont {J.}~\bibnamefont {Zhu}}, \bibinfo {author} {\bibfnamefont {C.}~\bibnamefont {Chen}}, \bibinfo {author} {\bibfnamefont {L.}~\bibnamefont {Zhou}}, \bibinfo {author} {\bibfnamefont {X.}~\bibnamefont {Xie}}, \bibinfo {author} {\bibfnamefont {C.}~\bibnamefont {Jiang}}, \bibinfo {author} {\bibfnamefont {Z.}~\bibnamefont {Ding}}, \bibinfo {author} {\bibfnamefont {F.}~\bibnamefont {Wu}}, \bibinfo {author} {\bibfnamefont {F.}~\bibnamefont {Yang}}, \bibinfo {author} {\bibfnamefont {G.}~\bibnamefont {Wang}}, \bibinfo {author} {\bibfnamefont {Q.}~\bibnamefont {Gong}}, \bibinfo {author} {\bibfnamefont {P.}~\bibnamefont {Zhang}}, \bibinfo {author} {\bibfnamefont {S.}~\bibnamefont {Zhang}},\ and\ \bibinfo {author} {\bibfnamefont {P.}~\bibnamefont {Peng}},\ }\bibfield  {title} {\bibinfo {title} {High-efficiency loading of 2400 ytterbium atoms in optical tweezer arrays},\ }\href {https://doi.org/10.1103/bp6k-8zmd} {\bibfield  {journal} {\bibinfo  {journal} {Phys. Rev.
  Lett.}\ }\textbf {\bibinfo {volume} {137}},\ \bibinfo {pages} {063201} (\bibinfo {year} {2026})}\BibitemShut {NoStop}%
\bibitem [{\citenamefont {Saskin}\ \emph {et~al.}(2019)\citenamefont {Saskin}, \citenamefont {Wilson}, \citenamefont {Grinkemeyer},\ and\ \citenamefont {Thompson}}]{saskin2019narrow}%
  \BibitemOpen
  \bibfield  {author} {\bibinfo {author} {\bibfnamefont {S.}~\bibnamefont {Saskin}}, \bibinfo {author} {\bibfnamefont {J.~T.}\ \bibnamefont {Wilson}}, \bibinfo {author} {\bibfnamefont {B.}~\bibnamefont {Grinkemeyer}},\ and\ \bibinfo {author} {\bibfnamefont {J.~D.}\ \bibnamefont {Thompson}},\ }\bibfield  {title} {\bibinfo {title} {Narrow-line cooling and imaging of ytterbium atoms in an optical tweezer array},\ }\href {https://doi.org/10.1103/PhysRevLett.122.143002} {\bibfield  {journal} {\bibinfo  {journal} {Phys. Rev. Lett.}\ }\textbf {\bibinfo {volume} {122}},\ \bibinfo {pages} {143002} (\bibinfo {year} {2019})}\BibitemShut {NoStop}%
\bibitem [{\citenamefont {Kim}\ \emph {et~al.}(2019)\citenamefont {Kim}, \citenamefont {Kim}, \citenamefont {Lee},\ and\ \citenamefont {Ahn}}]{kim2019gerchberg}%
  \BibitemOpen
  \bibfield  {author} {\bibinfo {author} {\bibfnamefont {H.}~\bibnamefont {Kim}}, \bibinfo {author} {\bibfnamefont {M.}~\bibnamefont {Kim}}, \bibinfo {author} {\bibfnamefont {W.}~\bibnamefont {Lee}},\ and\ \bibinfo {author} {\bibfnamefont {J.}~\bibnamefont {Ahn}},\ }\bibfield  {title} {\bibinfo {title} {Gerchberg-saxton algorithm for fast and efficient atom rearrangement in optical tweezer traps},\ }\href@noop {} {\bibfield  {journal} {\bibinfo  {journal} {Optics express}\ }\textbf {\bibinfo {volume} {27}},\ \bibinfo {pages} {2184} (\bibinfo {year} {2019})}\BibitemShut {NoStop}%
\bibitem [{\citenamefont {Lin}\ \emph {et~al.}(2025)\citenamefont {Lin}, \citenamefont {Zhong}, \citenamefont {Li}, \citenamefont {Zhao}, \citenamefont {Zheng}, \citenamefont {Hu}, \citenamefont {Wu}, \citenamefont {Wu}, \citenamefont {Ma}, \citenamefont {Gao} \emph {et~al.}}]{lin2025ai}%
  \BibitemOpen
  \bibfield  {author} {\bibinfo {author} {\bibfnamefont {R.}~\bibnamefont {Lin}}, \bibinfo {author} {\bibfnamefont {H.-S.}\ \bibnamefont {Zhong}}, \bibinfo {author} {\bibfnamefont {Y.}~\bibnamefont {Li}}, \bibinfo {author} {\bibfnamefont {Z.-R.}\ \bibnamefont {Zhao}}, \bibinfo {author} {\bibfnamefont {L.-T.}\ \bibnamefont {Zheng}}, \bibinfo {author} {\bibfnamefont {T.-R.}\ \bibnamefont {Hu}}, \bibinfo {author} {\bibfnamefont {H.-M.}\ \bibnamefont {Wu}}, \bibinfo {author} {\bibfnamefont {Z.}~\bibnamefont {Wu}}, \bibinfo {author} {\bibfnamefont {W.-J.}\ \bibnamefont {Ma}}, \bibinfo {author} {\bibfnamefont {Y.}~\bibnamefont {Gao}}, \emph {et~al.},\ }\bibfield  {title} {\bibinfo {title} {Ai-enabled parallel assembly of thousands of defect-free neutral atom arrays},\ }\href@noop {} {\bibfield  {journal} {\bibinfo  {journal} {Physical Review Letters}\ }\textbf {\bibinfo {volume} {135}},\ \bibinfo {pages} {060602} (\bibinfo {year} {2025})}\BibitemShut {NoStop}%
\bibitem [{\citenamefont {Wei}\ \emph {et~al.}(2026)\citenamefont {Wei}, \citenamefont {Li}, \citenamefont {Karve}, \citenamefont {Shaw}, \citenamefont {Schuster},\ and\ \citenamefont {Simon}}]{wei202610}%
  \BibitemOpen
  \bibfield  {author} {\bibinfo {author} {\bibfnamefont {X.}~\bibnamefont {Wei}}, \bibinfo {author} {\bibfnamefont {Z.}~\bibnamefont {Li}}, \bibinfo {author} {\bibfnamefont {A.~V.}\ \bibnamefont {Karve}}, \bibinfo {author} {\bibfnamefont {A.~L.}\ \bibnamefont {Shaw}}, \bibinfo {author} {\bibfnamefont {D.~I.}\ \bibnamefont {Schuster}},\ and\ \bibinfo {author} {\bibfnamefont {J.}~\bibnamefont {Simon}},\ }\href@noop {} {\bibinfo {title} {A 10 megahertz spatial light modulator}} (\bibinfo {year} {2026}),\ \Eprint {https://arxiv.org/abs/2601.08906} {arXiv:2601.08906} \BibitemShut {NoStop}%
\bibitem [{\citenamefont {Bytyqi}\ \emph {et~al.}(2026)\citenamefont {Bytyqi}, \citenamefont {Sinclair}, \citenamefont {Ramette},\ and\ \citenamefont {Vuleti{\'c}}}]{bytyqi2026device}%
  \BibitemOpen
  \bibfield  {author} {\bibinfo {author} {\bibfnamefont {E.}~\bibnamefont {Bytyqi}}, \bibinfo {author} {\bibfnamefont {J.}~\bibnamefont {Sinclair}}, \bibinfo {author} {\bibfnamefont {J.}~\bibnamefont {Ramette}},\ and\ \bibinfo {author} {\bibfnamefont {V.}~\bibnamefont {Vuleti{\'c}}},\ }\bibfield  {title} {\bibinfo {title} {Device for mhz-rate rastering of arbitrary 2d optical potentials},\ }\href@noop {} {\bibfield  {journal} {\bibinfo  {journal} {Optics Letters}\ }\textbf {\bibinfo {volume} {51}},\ \bibinfo {pages} {3729} (\bibinfo {year} {2026})}\BibitemShut {NoStop}%
\bibitem [{\citenamefont {Deters}\ \emph {et~al.}(2026)\citenamefont {Deters}, \citenamefont {Li}, \citenamefont {Douglas}, \citenamefont {Greiner},\ and\ \citenamefont {Young}}]{deters2026ultrafast}%
  \BibitemOpen
  \bibfield  {author} {\bibinfo {author} {\bibfnamefont {A.~D.}\ \bibnamefont {Deters}}, \bibinfo {author} {\bibfnamefont {Y.}~\bibnamefont {Li}}, \bibinfo {author} {\bibfnamefont {A.}~\bibnamefont {Douglas}}, \bibinfo {author} {\bibfnamefont {M.}~\bibnamefont {Greiner}},\ and\ \bibinfo {author} {\bibfnamefont {A.~W.}\ \bibnamefont {Young}},\ }\href@noop {} {\bibinfo {title} {Ultrafast and high resolution spatial light modulation for cold atoms}} (\bibinfo {year} {2026}),\ \Eprint {https://arxiv.org/abs/2608.18071} {arXiv:2608.18071} \BibitemShut {NoStop}%
\bibitem [{\citenamefont {Zhang}\ \emph {et~al.}(2024)\citenamefont {Zhang}, \citenamefont {Peng}, \citenamefont {Paul},\ and\ \citenamefont {Thompson}}]{zhang2024scaled}%
  \BibitemOpen
  \bibfield  {author} {\bibinfo {author} {\bibfnamefont {B.}~\bibnamefont {Zhang}}, \bibinfo {author} {\bibfnamefont {P.}~\bibnamefont {Peng}}, \bibinfo {author} {\bibfnamefont {A.}~\bibnamefont {Paul}},\ and\ \bibinfo {author} {\bibfnamefont {J.~D.}\ \bibnamefont {Thompson}},\ }\bibfield  {title} {\bibinfo {title} {Scaled local gate controller for optically addressed qubits},\ }\href@noop {} {\bibfield  {journal} {\bibinfo  {journal} {Optica}\ }\textbf {\bibinfo {volume} {11}},\ \bibinfo {pages} {227} (\bibinfo {year} {2024})}\BibitemShut {NoStop}%
\bibitem [{\citenamefont {Pichard}\ \emph {et~al.}(2024)\citenamefont {Pichard}, \citenamefont {Lim}, \citenamefont {Bloch}, \citenamefont {Vaneecloo}, \citenamefont {Bourachot}, \citenamefont {Both}, \citenamefont {M{\'e}riaux}, \citenamefont {Dutartre}, \citenamefont {Hostein}, \citenamefont {Paris}, \citenamefont {Ximenez}, \citenamefont {Signoles}, \citenamefont {Browaeys}, \citenamefont {Lahaye},\ and\ \citenamefont {Dreon}}]{pichard2024rearrangement}%
  \BibitemOpen
  \bibfield  {author} {\bibinfo {author} {\bibfnamefont {G.}~\bibnamefont {Pichard}}, \bibinfo {author} {\bibfnamefont {D.}~\bibnamefont {Lim}}, \bibinfo {author} {\bibfnamefont {{\'E}.}~\bibnamefont {Bloch}}, \bibinfo {author} {\bibfnamefont {J.}~\bibnamefont {Vaneecloo}}, \bibinfo {author} {\bibfnamefont {L.}~\bibnamefont {Bourachot}}, \bibinfo {author} {\bibfnamefont {G.-J.}\ \bibnamefont {Both}}, \bibinfo {author} {\bibfnamefont {G.}~\bibnamefont {M{\'e}riaux}}, \bibinfo {author} {\bibfnamefont {S.}~\bibnamefont {Dutartre}}, \bibinfo {author} {\bibfnamefont {R.}~\bibnamefont {Hostein}}, \bibinfo {author} {\bibfnamefont {J.}~\bibnamefont {Paris}}, \bibinfo {author} {\bibfnamefont {B.}~\bibnamefont {Ximenez}}, \bibinfo {author} {\bibfnamefont {A.}~\bibnamefont {Signoles}}, \bibinfo {author} {\bibfnamefont {A.}~\bibnamefont {Browaeys}}, \bibinfo {author} {\bibfnamefont {T.}~\bibnamefont {Lahaye}},\ and\ \bibinfo {author} {\bibfnamefont {D.}~\bibnamefont {Dreon}},\ }\bibfield  {title} {\bibinfo {title}
  {Rearrangement of individual atoms in a 2000-site optical-tweezer array at cryogenic temperatures},\ }\href {https://doi.org/10.1103/PhysRevApplied.22.024073} {\bibfield  {journal} {\bibinfo  {journal} {Phys. Rev. Appl.}\ }\textbf {\bibinfo {volume} {22}},\ \bibinfo {pages} {024073} (\bibinfo {year} {2024})}\BibitemShut {NoStop}%
\bibitem [{\citenamefont {Chiu}\ \emph {et~al.}(2025)\citenamefont {Chiu}, \citenamefont {Trapp}, \citenamefont {Guo}, \citenamefont {Abobeih}, \citenamefont {Stewart}, \citenamefont {Hollerith}, \citenamefont {Stroganov}, \citenamefont {Kalinowski}, \citenamefont {Geim}, \citenamefont {Evered} \emph {et~al.}}]{chiu2025continuous}%
  \BibitemOpen
  \bibfield  {author} {\bibinfo {author} {\bibfnamefont {N.-C.}\ \bibnamefont {Chiu}}, \bibinfo {author} {\bibfnamefont {E.~C.}\ \bibnamefont {Trapp}}, \bibinfo {author} {\bibfnamefont {J.}~\bibnamefont {Guo}}, \bibinfo {author} {\bibfnamefont {M.~H.}\ \bibnamefont {Abobeih}}, \bibinfo {author} {\bibfnamefont {L.~M.}\ \bibnamefont {Stewart}}, \bibinfo {author} {\bibfnamefont {S.}~\bibnamefont {Hollerith}}, \bibinfo {author} {\bibfnamefont {P.~L.}\ \bibnamefont {Stroganov}}, \bibinfo {author} {\bibfnamefont {M.}~\bibnamefont {Kalinowski}}, \bibinfo {author} {\bibfnamefont {A.~A.}\ \bibnamefont {Geim}}, \bibinfo {author} {\bibfnamefont {S.~J.}\ \bibnamefont {Evered}}, \emph {et~al.},\ }\bibfield  {title} {\bibinfo {title} {Continuous operation of a coherent 3,000-qubit system},\ }\href@noop {} {\bibfield  {journal} {\bibinfo  {journal} {Nature}\ }\textbf {\bibinfo {volume} {646}},\ \bibinfo {pages} {1075} (\bibinfo {year} {2025})}\BibitemShut {NoStop}%
\bibitem [{\citenamefont {Huang}\ \emph {et~al.}(2026)\citenamefont {Huang}, \citenamefont {Huang}, \citenamefont {Yao}, \citenamefont {Wang},\ and\ \citenamefont {Liu}}]{huang2026phase}%
  \BibitemOpen
  \bibfield  {author} {\bibinfo {author} {\bibfnamefont {E.}~\bibnamefont {Huang}}, \bibinfo {author} {\bibfnamefont {J.}~\bibnamefont {Huang}}, \bibinfo {author} {\bibfnamefont {H.}~\bibnamefont {Yao}}, \bibinfo {author} {\bibfnamefont {X.}~\bibnamefont {Wang}},\ and\ \bibinfo {author} {\bibfnamefont {J.-G.}\ \bibnamefont {Liu}},\ }\href@noop {} {\bibinfo {title} {Phase-stable hologram updates for large-scale neutral-atom array reconfiguration}} (\bibinfo {year} {2026}),\ \Eprint {https://arxiv.org/abs/2604.04600} {arXiv:2604.04600} \BibitemShut {NoStop}%
\bibitem [{\citenamefont {Zhang}\ \emph {et~al.}(2026)\citenamefont {Zhang}, \citenamefont {Li}, \citenamefont {Zhai},\ and\ \citenamefont {Chen}}]{zhang2026algorithm}%
  \BibitemOpen
  \bibfield  {author} {\bibinfo {author} {\bibfnamefont {T.}~\bibnamefont {Zhang}}, \bibinfo {author} {\bibfnamefont {X.}~\bibnamefont {Li}}, \bibinfo {author} {\bibfnamefont {H.}~\bibnamefont {Zhai}},\ and\ \bibinfo {author} {\bibfnamefont {L.}~\bibnamefont {Chen}},\ }\href@noop {} {\bibinfo {title} {An algorithm for fast assembling large-scale defect-free atom arrays}} (\bibinfo {year} {2026}),\ \Eprint {https://arxiv.org/abs/2604.08669} {arXiv:2604.08669} \BibitemShut {NoStop}%
\bibitem [{\citenamefont {Thalhammer}\ \emph {et~al.}(2013)\citenamefont {Thalhammer}, \citenamefont {Bowman}, \citenamefont {Love}, \citenamefont {Padgett},\ and\ \citenamefont {Ritsch-Marte}}]{thalhammer2013overdrive}%
  \BibitemOpen
  \bibfield  {author} {\bibinfo {author} {\bibfnamefont {G.}~\bibnamefont {Thalhammer}}, \bibinfo {author} {\bibfnamefont {R.~W.}\ \bibnamefont {Bowman}}, \bibinfo {author} {\bibfnamefont {G.~D.}\ \bibnamefont {Love}}, \bibinfo {author} {\bibfnamefont {M.~J.}\ \bibnamefont {Padgett}},\ and\ \bibinfo {author} {\bibfnamefont {M.}~\bibnamefont {Ritsch-Marte}},\ }\bibfield  {title} {\bibinfo {title} {Speeding up liquid crystal slms using overdrive with phase change reduction},\ }\href {https://doi.org/10.1364/OE.21.001779} {\bibfield  {journal} {\bibinfo  {journal} {Opt. Express}\ }\textbf {\bibinfo {volume} {21}},\ \bibinfo {pages} {1779} (\bibinfo {year} {2013})}\BibitemShut {NoStop}%
\end{thebibliography}%

\end{document}


\onecolumngrid

\setcounter{section}{0}
\setcounter{subsection}{0}
\setcounter{figure}{0}
\setcounter{equation}{0}
\setcounter{table}{0}
\renewcommand{\thesection}{S\arabic{section}}
\renewcommand{\thesubsection}{\thesection.\arabic{subsection}}
\renewcommand{\thesubsubsection}{\thesubsection.\arabic{subsubsection}}
\renewcommand{\thefigure}{S\arabic{figure}}
\renewcommand{\theequation}{S\arabic{equation}}
\renewcommand{\thetable}{S\arabic{table}}

\begin{center}
  {\large\bfseries Supplemental Material for
  
  \emph{Parallel Omnidirectional Atom Rearrangement with Dynamic Holograms}}
  \label{sm}
\end{center}
\setcounter{page}{1}

\addtocontents{toc}{\protect\setcounter{tocdepth}{2}}
\tableofcontents

\section{Apparatus\label{sm:apparatus}}

The optical tweezers are generated using a \SI{532}{\nano\meter} laser (Precilasers FL-SF-532-35-CW). The tweezer beam power is stabilized using a water-cooled quartz AOM. A kHz-rate SLM (Meadowlark P1K-400-600-PC8, provided by AVR Optics) displays and updates holograms to modulate the phase of the tweezer beam. The beam is then relayed and expanded via a 4f-system to fill the back aperture of a custom-built objective (Special Optics, NA = 0.6, back aperture = \SI{24}{\milli\meter}, effective focal length = \SI{20}{\milli\meter}, field of view = \SI{800}{\micro\meter}). A home-built spatial filter, consisting of a mirror with a drilled through-hole, is placed at the intermediate image plane of the 4f-system to spatially filter out the SLM's zeroth diffraction order. In practice, we use up to \SI{30}{\watt} laser power and can generate more than 3,000 tweezers.

To accommodate high optical power at \SI{532}{\nano\meter}, where the standard high-speed liquid crystal is susceptible to optical damage, the manufacturer uses a modified LC formulation with a higher damage threshold but a slower intrinsic response compared with the standard HSP1K model, whose response time is approximately \SI{1}{\milli\second}. To compensate for this slower response, our SLM is built with a narrower cell gap and hence a thinner LC layer, giving a nominal $10$--$90\%$ switching time of \SI{2.3}{\milli\second} (manufacturer specification), at the cost of a reduced phase modulation depth of 1.2 waves ($2.4\pi$) at \SI{532}{\nano\meter} at room temperature. Because the modulation depth further decreases with increasing angle of incidence (AOI) and LC temperature, we keep the AOI below \SI{2.5}{\degree} and water-cool the LC panel to maintain its temperature below \SI{25}{\celsius} at full laser power.

\section{Atom Loading and Imaging\label{sm:loadimage}}

Atom loading is performed by spatially overlapping the optical tweezers with a magneto-optical trap (MOT) operated on the narrow  $^{1}S_{0} \rightarrow {}^{3}P_{1}$ intercombination line of $^{174}\mathrm{Yb}$ at \SI{556}{\nano\meter}. To ensure single-atom occupancy in each trap, red-detuned light-assisted collisions are induced using the same narrow-line transition, leading to  approximately 55\% stochastic loading probability. In large arrays, we precede this step with an additional blue-detuned light-assisted collision stage, boosting the loading rate to approximately 75\%. However, blue-detuned light also induces Doppler heating, which can lead to atom loss. To this end, we experimentally found a parameter regime where the interplay between single-atom loss in two-body collisions (one atom left) and Doppler heating (no atoms left) results in enhanced loading rate, similar to results in \cite{zhu2026high}. Following initial loading, we apply three-dimensional Doppler cooling to reach temperatures of approximately \SI{7}{\micro\kelvin}.

For imaging, we collect fluorescence from the atoms using the $^{1}S_{0} \rightarrow {}^{1}P_{1}$ transition at \SI{399}{\nano\meter}. The fluorescence is collected through the same objective used to form the tweezers and imaged onto a qCMOS camera (Hamamatsu ORCA-Quest C15550-20UP). During imaging, rapid photon scattering on the broad \SI{399}{\nano\meter} transition causes substantial heating. We therefore simultaneously apply narrow-line cooling at \SI{556}{\nano\meter}, yielding an atomic temperature of approximately \SI{20}{\micro\kelvin} after imaging. This combination is well suited to large arrays because detection at \SI{399}{\nano\meter} avoids the background that complicates direct imaging at \SI{556}{\nano\meter}. The background arises from Raman scattering of the high-power \SI{532}{\nano\meter} trapping light along the optical path, particularly in the objective lenses and housing \cite{saskin2019narrow}.

We characterize the atom loading and imaging in Fig.~\ref{fig:loadimage}. The measured rearrangement fidelity is limited by imaging errors and atom loss during imaging. Our atom-transport measurements suggest that the underlying rearrangement fidelity is higher. We perform only one round of rearrangement in this work. Additional rounds could further reduce the number of defects by retaining reservoir atoms for subsequent rounds, at the cost of increased total rearrangement time. We note that empty traps could also be removed by updating the holograms after the first image, allowing their optical power to be redistributed to occupied traps to increase trap depths and improve imaging fidelity and survival in subsequent rounds.

\section{Properties of Optical Tweezers\label{sm:trapcal}}

Uniform trap depths are required to ensure consistent loading, cooling, and imaging across the array. We measure trap depths spectroscopically on the \SI{556}{\nano\meter} $^{1}S_{0}\rightarrow{}^{3}P_{1}$ transition. For a trap laser wavelength of \SI{532}{\nano\meter}, the $|m_J|= 0$ state is near magic, whereas the $|m_J|= 1$ states experience a tensor light shift \cite{saskin2019narrow}. The splitting between these states therefore serves as a measurement of the local trap depth of each tweezer. To equalize depths, we start from a WGS hologram, scan the $|m_J| = 1$ resonance at every site, and feed the measured depths back into WGS, one iteration per measurement. Each iteration updates only the amplitudes near each trap, with all phases and the background amplitude held fixed, so that changes in the inter-tweezer speckle do not re-randomize the depths between cycles. For the 1068-site array with \SI{10}{\micro\meter} spacing used in Figs.~\ref{fig:2} and~\ref{fig:3}, this yields a mean depth of $U_0 = \SI{418(16)}{\micro\kelvin}$ (day-to-day global variation) with a site-to-site coefficient of variation of 1.5\%; results for the large arrays are shown in Fig.~\ref{fig:loadimage}. 

In addition to the trap depth, we also measure the trap frequencies by parametric heating. The median radial and axial trap frequencies in the 1068-site array are 85.7 kHz and 16.5 kHz. In the harmonic approximation, these frequencies and the measured depth give a waist $w_0 = \SI{0.525}{\micro\meter}$ and an effective Rayleigh range $z_R = \SI{1.93}{\micro\meter}$, longer than the Gaussian-beam value $\pi w_0^2/\lambda = \SI{1.63}{\micro\meter}$. We therefore use the calibrated spot shape rather than a Gaussian model when computing transport thresholds~\pref{sm:superposition}.

\section{SLM Electronics and LC Settle\label{sm:LC}}

The transport speeds in the main text are governed by two timescales of the SLM: the electronic refresh period, which limits how often a new hologram can be commanded, and the response time of the liquid crystal (LC), which sets how quickly the displayed phase follows. We characterize both here.

Each SLM pixel applies an rms voltage across the LC layer, which reorients the molecules and changes the local retardance. We map voltages to pixel phase with Meadowlark's factory-provided lookup table, which we measure to be linear but to effectively span only ${\sim}1.7\pi$ at \SI{532}{\nano\meter}. This reduced stroke scales wrapped phases by ${\sim}0.85$, which sends roughly $1-\text{sinc}^2(0.15) \approx 7\%$ of the light into other diffraction orders without shifting the tweezers.

To avoid damage from a sustained DC field, the drive is a square wave alternating between $\pm V_\text{rms}$ on an internal clock with period $T_\text{min} = 0.696$ ms (1437 Hz). A new hologram can be applied only at these polarity flips, so $T_\text{min}$ is the shortest frame period, and periods that are not integer multiples of $T_\text{min}$ produce aliased frame timing. Throughout this work we mostly use $T = T_\text{min}$ and $4T_\text{min} = 2.785$ ms, rounded to 0.7 ms and 2.8 ms in the main text.

To measure the LC response, we display binary gratings whose stripes alternate between pixel phases $\psi$ and $\psi'$. The first-order diffracted intensity is $I_1 \propto \sin^2(\Delta\psi/2)$ with $\Delta\psi = \psi' - \psi$, so measuring $I_1$ on a camera determines $\Delta\psi$. Sweeping the voltages of both stripes first yields a lookup table between voltage and retardance. We then switch one stripe from $\psi_1$ to $\psi_2$ and record $I_1$ at a series of delays using strobed imaging, repeating each delay with $\psi' = 0$ and $\psi' = \pi/2$ to resolve the ambiguity of $\sin^2$. This gives the time-resolved response for each pair of initial and commanded pixel phases.

An ideal pixel relaxes exponentially from its initial phase $\psi_0$ toward the commanded phase $\psi_c$,
\begin{align}  
\psi(t) = \psi_0 e^{-t/\tau} + \psi_c\left(1 - e^{-t/\tau}\right). \label{eq:pixeltaudef}
\end{align}

In practice, $\tau$ depends on the commanded pixel phases and on the direction of the change; we neglect pixel-to-pixel variation, as well as hysteresis and crosstalk. We fit Eq.~\eqref{eq:pixeltaudef} to a sum of measured responses over phase pairs between two uniform random holograms; this gives an effective settle time $\tau_0 = \SI{1.6(3)}{\milli\second}$ (\SI{1.4(3)}{\milli\second} rising voltage, decreasing pixel phase; \SI{1.8(3)}{\milli\second} falling voltage), where the uncertainty reflects statistical uncertainty in our calibrated LC dynamics, dependence on the choice of initial and final holograms, and the weighting of the measured responses across phase pairs. We are also limited in measuring the fastest responses by our strobe time resolution (${\sim}\SI{0.3}{\milli\second}$). We therefore treat $\tau_0$ as a characteristic scale only; the simulations in~\ref{sm:mc} use the calibrated dynamics.

\section{Transport Protocols\label{sm:protocol}}

The uniform transport measurements of Fig.~\ref{fig:2} use the following protocols. To test the limits of radial ($+x$) movement at some step size $d$ and period $T$ for $N$ steps, we first load atoms into the 1068-site array and take an initial image. We then compute a sequence of linear phase gratings (where $k_x$ and $k_y$ are momentum-space coordinates, corresponding linearly to pixel indices, on the LC face):
\begin{align}
    h_x(d) &= k_xd \label{eq:xgrating} 
\end{align}
to add to the loading hologram pattern $\Phi_0$ :
\begin{align}
    \Phi_n &= \Phi_0\exp\left(ih_x(nd)\right). \label{eq:gratingrule}
\end{align}

The period of the grating controls the displacement of all tweezers in the array. Based on the optical specifications, including the wavelength and nominal focal lengths, we estimate a conversion factor of $\delta x = \SI{0.408}{\micro\meter/\textrm{(far-field px)}}$ between the computational grid ($1024 \times 1024$ far-field) and the atom plane. We then command the SLM to display $\Phi_n$ at time $Tn$ for $n \in \{0,1,\dots,N\}$. Finally, a second image measures survival for the chosen $d$, $T$, and $N$. The same can be done in the $y$ direction with a grating along $k_y$. The radial step-size threshold is fairly isotropic, with the same shape in all four directions (widths within ${\sim}20\%$, thresholds within ${\sim}7\%$).

\newcommand{\NA}{\text{NA}_N}

For axial movement ($+z$), we replace gratings with defocus kernels, in terms of $\rho$ which is the radius from the center of the SLM, normalized to $\pm 1$ at the edges of the LC face ($\sqrt{2}$ at the corners):
\begin{align}
    h_z(d) &= \frac{2\pi d}{\lambda}\left(\sqrt{1-\NA^2\rho^2}-1\right) + d \cgouy \label{eq:defocuskernelexact}\\
    &\approx -\frac{\pi d}{\lambda}\NA^2\rho^2 + d \cgouy \label{eq:defocuskernelzernike}
\end{align}
with $\Phi_n$ as in Eq.~\eqref{eq:gratingrule}, where $\lambda = \SI{532}{\nano\meter}$, and the Nyquist numerical aperture $\NA = \lambda / (2\delta x) = 0.65$ follows directly from the radial conversion factor $\delta x$ above ($\NA > \text{NA}_\text{obj}$ indicates that light from the outermost SLM pixels is clipped by the objective). The second term is a linear correction in step size that allows us to mitigate the Gouy phase of our optical tweezers~\pref{sm:superposition}.
In order to avoid imaging atoms at different planes, we utilize a ping-pong protocol: we perform $N/2$ steps outward and then $N/2$ back. We see significant asymmetry between $+z$ and $-z$ in both shape and threshold, even with the semi-symmetric ping-pong protocol (Fig.~\ref{fig:survivalmcatoms}). We sanity-check that atoms are actually leaving the initial plane by imaging without translating back, ensuring no atoms are detected.

To compare transport across directions and frame periods, we define $d_0$ as the step size at which survival over $n=50$ steps, normalized to the zero-step control, falls to $0.99^{n}$, obtained by linear interpolation between measured step sizes. The uncertainty on $d_0$ is the difference between linear interpolation and an erfc$^n$ fit, which exceeds the statistical error. The error on $v_\text{max} = d_0/T$ also includes the shot-to-shot spread of $T$ as realized by the SLM ($\leq 1\%$, $68\%$ CI). Calibration uncertainties common to all points (e.g. conversion to micrometers) are not included.

\section{Transport Model and Thresholds\label{sm:superposition}}

Here we model the light field during a hologram update and derive the critical step size beyond which atoms can no longer follow a moving tweezer.

Under the assumption that all pixels follow the same exponential settle curve as in Eq.~\eqref{eq:pixeltaudef}, and that the commanded LC phases are similar between frames, the light field at the atom plane during settling is approximately~\cite{kim2019gerchberg,lin2025ai}:
\begin{align}
E(\mb r, t) = e^{-t/\tau}E_0(\mb r) + \left(1-e^{-t/\tau}\right)E_1(\mb r) \label{eq:linsuperposition}
\end{align}
For updates faster than the settle time, the LC does not reach its commanded state before the next frame is commanded. The field therefore depends on several preceding frames. Starting from a fully settled field $E_0$, with successive fields $E_m$ commanded at times $mT$, the light field at the atoms under the superposition approximation is
\begin{align}
    E(\mathbf{r},t) = E_0(\mathbf{r})
    &+ \sum_{m=1}^{n} \left[ E_m(\mathbf{r}) - E_{m-1}(\mathbf{r}) \right]\left( 1 - e^{-(t - mT)/\tau} \right) \Theta(t - mT)
    \label{eq:linsupersum}
\end{align}
where $\Theta$ is a unit step function. This formula explains why phase alignment between tweezers is so crucial for atom survival: unmatched tweezers will interfere, causing lobing and reduced trap depth. 

Plugging our approximate tweezer shape into $E_m$ in $xy$ ($z$), and assuming perfect phase coherence between consecutive frames, recovers a critical distance $\dc^r$ ($\dc^z$). Below this distance, the trap continuously translates, modulating primarily in intensity. Above it, the old trap remains distinct from the new trap and fades as the new trap forms, producing release-and-recapture-like dynamics. If the tweezers are not phase aligned, this bifurcation occurs much below $\dc^r$ due to interference, causing significant atom loss.

We formally define bifurcation as a discontinuity of the local maximum of the tweezer intensity. This generally occurs as a splitting of one maximum into two or more maxima. When bifurcation occurs, the atom cannot follow the moving trap adiabatically, causing heating and loss. To quantify how close a transition comes to bifurcating without actually becoming discontinuous, we define
\begin{align}
    \kappa &= \min_{t\in[nT,(n+1)T]} \frac{-\partial_x^2 \ln |E(x,t)|\big|_{x=x_m(t)}}{-\partial_x^2 \ln |E(x,0)|\big|_{x=x_m(0)}} 
    \nonumber\\&
    = \min \left(\frac{w(0)}{w(t)}\right)^2,\label{eq:kappadef}
\end{align}
where $x$ is the coordinate along the transport direction, $x_m(t)$ is the position of the moving trap's intensity maximum, and $w(t)$ is the effective Gaussian trap waist at time $t$. The reference $t = 0$ refers to a fully-settled initial trap. A rigidly translated tweezer has $\kappa=1$, while $\kappa\to 0$ as the maximum flattens. We call a transition a near-bifurcation when $\kappa<0.1$.

For perfectly Gaussian spots, the zero-curvature condition gives $\dc^r \sim \sqrt{2}w_0 = \SI{0.74}{\micro\meter}$. Beyond this separation, two summed Gaussians develop two intensity maxima. With our calibrated spot shape, we find $\dc^r \approx \SI{0.78}{\micro\meter}$. At large period, this analytical expression for $\dc^r$ generalizes to sequences of multiple tweezers, since transitions are essentially independent (Eq.~\eqref{eq:linsupersum}); at low period, there is a very slight intrusion from the tails and sidelobes of earlier and later tweezers that are simultaneously visible, but this correction is very small. In simulation of the light field from superposition Eq.~\eqref{eq:linsupersum} across periods $T \geq T_\text{min}$, accounting for our calibrated tweezer shape, $\dc^r$ deviates from $\SI{0.78}{\micro\meter}$ by less than $3\%$.

However, axial transport is complicated by the nonlinear variation of the Gouy phase across the tweezer focus. To compensate for this, we approximate its arctangent dependence with a piston correction linear in step size (Eq.~\ref{eq:defocuskernelexact}), and fit $\cgouy = \SI{0.617}{\radian/\micro\meter}$ to maximize $d_0$ during axial transport. This fitted value agrees with the on-axis Gouy-phase slope $\lambda/(\pi w_0^2) \approx \SI{0.614}{\radian/\micro\meter}$ implied by the waist measured from the radial trap frequency~\pref{sm:trapcal}. For global defocus, this correction is applied to the entire hologram; for arbitrary movement, it can be applied individually through each tweezer’s phase.

With this linear correction, we can simulate Eq.~\eqref{eq:linsupersum} with our measured spot shape to determine $\dc^z$. Using our optimized $\cgouy$ value, we obtain $\dc^z = \SI{3.65}{\micro\meter}$, which is significantly larger than the $\sqrt{2}z_R = \SI{2.73}{\micro\meter}$ that would be predicted from a perfectly Gaussian spot with no Gouy phase. Importantly, this value $\dc^z$ still does not depend very much on $T \geq T_\text{min}$, drifting less than $1.5\%$. 

\section{Monte Carlo Simulation\label{sm:mc}}

The superposition model in \ref{sm:superposition} predicts a nearly period-independent geometric threshold, whereas the measured survival threshold $d_0$ decreases at shorter refresh periods (Fig.~\ref{fig:2}). To understand this difference, we simulate the atom-plane light field by modeling the settling of each LC pixel independently, then perform a classical Monte Carlo (MC) simulation of atomic motion in the resulting potential. Example trajectories are shown in  Fig.~\ref{fig:mcdemo}.

While MC survival numbers are imperfect due to LC settle curve inaccuracy (especially overestimation of settle times), spot shape inaccuracy, and not accounting for optical aberration, we recover $d_0$ values within 8\% of atomic measured values both radially and axially, as shown in Fig.~\ref{fig:survivalmcatoms}. The main mismatch between atom data and MC data is a larger site-to-site spread of survival in simulation versus a smaller spread in real atoms, an effect that is difficult to fully eradicate due to calibration limitations. Absolute temperature values also do not match well. However, ordering of the important observables (survival and heating) across periods and step sizes align between MC and atoms, so the conclusions made are qualitatively consistent, especially when combined with real atom data. 

We continue with our discussion on the effect of period on $v_\text{max}$. Even though $\dc$ is theoretically constant, we clearly measure difference in achievable step size between periods $\SI{0.7}{\milli\second}$ and $\SI{2.8}{\milli\second}$ (Fig.~\ref{fig:2}). The simplest possibility is reduced trap depth: at lower period, Eq.~\eqref{eq:linsupersum} predicts that more tweezers will be superposed at any one moment. Thus transient trap depth should go as
\begin{align}
    \min |E|^2 \propto \left(\frac{T}{d}\right)^2,
\end{align}
which decreases for very short periods as the beam intensity spreads across a larger area, meaning that atoms can easily evaporate out of the shallow light. To probe this, we sweep over laser power (Fig.~\ref{fig:trapdepth}) and find that the loss at a given step size is still much larger at lower period. It seems we are not limited by trap depth at our operating laser power---our \SI{418}{\micro\kelvin} traps, even dimmed during transport, are plenty deep enough for our initially \SI{7}{\micro\kelvin} atoms. Increasing laser power hardly changes $v_\text{max}$ and certainly does not close the period gap. MC simulation with varying initial trap powers confirms that trap depth does not significantly change $v_\text{max}$. Reduced trap depth is therefore not the primary cause.

In the multi-step transport during actual rearrangement, frames are refreshed consecutively at a steady period of \SI{0.7}{\milli\second}. This leads to another effect predicted by LC and MC simulation: a large portion of atom loss comes from the atoms exiting the traps of the initial frame. Since the first fully settled trap fades more slowly than the later transient traps, the second trap begins to fade out while the atom is still in the initial trap, making the direct jump necessary for atom survival much farther, all the way to the third or fourth trap. This effect, of a large loss at the second frame, is mirrored in both radial and axial atom data when we measure the loss at increasing step count, as in Fig.~\ref{fig:survivalsteps}. Even accounting for entry shock, however, the steady state per-step loss is still greater at shorter period than longer period at matched step size (equivalently, as shown in the figure, steady state loss matches at a higher step size for longer period).

Looking closer, simulation instead points to the breakdown of the coherent superposition approximation, Eq.~\eqref{eq:linsupersum}. In this short-period regime, the LC does not settle before each commanded frame, and pixels that must travel far in retardance (e.g. across the entire $2\pi$ modulation depth due to phase wrapping) make the realized field depend non-linearly on several preceding frames. This produces pseudo-random perturbations from superposition that soften the bifurcation threshold. Below the nominal $\dc^r$, bifurcations become more frequent, and near-bifurcations more frequent and more severe, so heating and loss set in earlier (Fig.~\ref{fig:foldrate}). Across radial and axial transport in MC simulation, nearly all lost atoms are first displaced from the trap during or immediately after a bifurcation or near-bifurcation, while atoms that experience neither are essentially never lost.

Note also that bifurcations separate more fully at lower period, as the commanded spot has more updates to move forward and away from the trailing spot, leaving atoms stranded more easily and making slight bifurcation more lethal. On this note, bifurcation does not immediately cause loss; atoms that survive near- or full bifurcations are often recaptured hotter, and are more likely to evaporate over the following steps. This leads to a small accumulation effect of loss, even at small step sizes, which is most prominently seen when we extend to hundreds or thousands of steps (Fig.~\ref{fig:2}i).

There are several features of axial transport that we cannot fully explain. First, we see significant $\pm z$ asymmetry (Fig.~\ref{fig:survivalmcatoms}). We see this effect in simulation, and isolate it to the non-single-exponential nature of the LC dynamics by ablation of complexity in our LC calibration. Second, axial heating (Fig.~\ref{fig:2}b) and survival versus number of steps (Fig.~\ref{fig:survivalsteps}) both imply an interaction between heating and loss other than the simple accumulation and evaporation we see radially. We hypothesize that axial transitions might selectively kill hotter atoms while leaving colder atoms untouched, which could lead to the sub-exponential survival curve we see over number of steps and the lower-than-expected measured temperatures versus step size.

We also use MC simulation to predict results beyond our physical limitations. With arbitrarily low period $T$ on our current SLM, we could approach $v_\text{max} \sim \SI{1.7}{\micro\meter / \milli\second}$ (${\sim}2 \times$ our $T_\text{min}$ speed). But what would happen if we had a faster SLM with the same proportional dynamics \cite{wei202610, bytyqi2026device, deters2026ultrafast, zhang2024scaled}? Simulating radial transport with proportionately sped-up LC dynamics and correspondingly shorter update periods, transport speed scales nearly linearly with settle speed up to MHz update rates. At these rates trap depth becomes the limiting factor, so faster LC response and deeper traps must be pursued together. At our current trap depth, a MHz SLM could already reach radial transport speeds on the order of ${\sim}\SI{0.1}{\meter/\second}$, comparable to those used in AOD-based rearrangement~\cite{pichard2024rearrangement,chiu2025continuous}.

\section{Rearrangement Pipeline\label{sm:rearrangement}}
\subsection{\label{sm:pathmatching}Path Matching}

In practice, both the initial array for atom loading and the final target array are optimized via the WGS algorithm to have nearly uniform trap depth. The loading array is designed to have enough extra sites such that loaded atoms will almost always fill the final array, and it is arranged in image-space to minimize average atomic movement path distance to the final array. After fluorescence imaging, we use a blocked version of the Hungarian algorithm \cite{lin2025ai} to assign near-optimal paths between loaded sites and target sites. We account for survival probability based on both movement distance and imaging fidelity (utilizing a slight excess of loaded atoms relative to target grid size) by assigning a cost function to each path based on both total movement distance and Gaussian-mixture probability of atom presence:
\begin{align}
C(x_i, x_f) &= -\log(P(x_i \rightarrow x_f)) \nonumber \\
&= -\log(p_\text{step}^n \cdot p_i) \nonumber\\
&= n\|x_f - x_i\|^2/(\beta n^2) - \log p_i \nonumber\\
&\propto  \|x_f - x_i\|^2 - \beta n \log p_i \label{eq:hungarianloss}
\end{align}
where in the first term we assume a very conservative (but tiebreaking) per step Gaussian survival probability, with variance $\beta/2$, and step number $n$. $p_i$ in the second term is the imaging fidelity for site $i$; this selects paths that most effectively minimize distance while disregarding empty or uncertain loaded atoms, protecting against sites with particularly bad imaging fidelity. In 3D, the path length can be generalized to a weighted Euclidean distance, where $xy$ distance weights ${\sim}(\dc^z/\dc^r)^2$ more than $z$ distance due to the difference in achievable step size. We optimize the second term by sweeping across $\beta$ values on our kagome lattice rearrangement (the larger loading array has smaller trap depths and therefore lower-quality imaging in our setup), settling on $\beta = \SI{50}{\micro\meter^2}$. The block size of the Hungarian algorithm can be tuned for the fastest possible result, leading to very little scaling with reasonable $N$. The path-matching precompute time is around \SI{9}{\milli\second} for ${\sim}2000$ atoms, only limited by CPU threading power. 

After calculating the atomic trajectories, we discretize each path into $n$ equally spaced intermediate positions, one per frame. We then generate a hologram for each frame to produce the target position and phase of every tweezer. As with position, we linearly interpolate the tweezer phase from the loading tweezer’s phase to the target tweezer’s phase, both of which are fixed from trap-depth uniformity optimization. This interpolation is straightforward in the $xy$ plane, but along $z$, we must account for the Gouy phase~\pref{sm:superposition}.

\subsection{\label{sm:2dinference} 2D Hologram Generation}

We mainly explore two methods for hologram generation. Following Lin et al., we train a 10-layer, 19k-parameter convolutional neural network (SLMnet) to approximate WGS in a single forward pass~\cite{lin2025ai}. We then finetune the model on realized far-field fidelity, including the coefficient of variation of tweezer intensities (CV) and the fraction of light near the target tweezers (hologram efficiency). SLMnet takes continuous, arbitrary tweezer positions as input and produces an atom plane array that can be inverse Fourier transformed to produce the desired SLM pattern. Intuitively, the CNN ``dirties up'' the input target, adding diffraction sidelobes and generating a near-realizable light field phase, so that a single unpadded inverse Fourier transform can produce a usable hologram. With CUDA graphing, kernel fusion, and compilation, SLMnet frame generation takes median $\SI{0.662}{\milli\second} < T_\text{min}$. By training SLMnet to reproduce the tweezer phases that WGS happens to produce, the model learns tweezer phase control (rms error ${\sim}\SI{0.15}{\radian}$).

We also test another proposed method for generating frames: warm-WGS~\cite{huang2026phase,zhang2026algorithm,kim2019gerchberg}.
Since consecutive frames are similar in both atom-plane target and hologram phases, WGS should require fewer iterations to converge for a particular frame’s target if seeded from the previous frame. Typical WGS does not guarantee phase control, but simply by replacing both the amplitudes and the phases near the tweezers each iteration, we can achieve both position and phase control in a few warm iterations. 
To achieve sub-$T_\text{min}$ inference with warm-WGS on our single RTX 5090 GPU, we zero-pad only to $2048\times 2048$, and we can afford to perform up to $3$ iterations in $\SI{0.649}{\milli\second} < T_\text{min}$. At $1024 \times 1024$ padding, we can afford $8$ iterations, but simulated fidelity and positional control are worse. See Fig.~\ref{fig:3}c. (We also measure rearrangement performance at $1024^2 \times 8$, which as expected does not beat performance at $2048^2 \times 3$).

Note that to characterize performance of our frame generation methods in simulation, we define simulated fidelity 
\begin{align}
    F_\text{sim}=\frac{\bigl|\sum_i a_i e^{i\delta\phi_i}\bigr|^2}{N\sum_i a_i^2}
\end{align}

with $a_i$ the realized amplitude of tweezer $i$ and $\delta\phi_i$ its phase error relative to the commanded phase; global phase and overall power are free. Two kinds of error lower this overlap: uneven tweezer intensities and phase deviations. We quantify the first by the coefficient of variation of the tweezer intensities $I_i=a_i^2$, $\mathrm{CV}_I=\sigma_I/\bar I$, which is the standard deviation divided by the mean across tweezers. For small errors the amplitude spread is half the intensity spread ($\mathrm{CV}_a\approx\mathrm{CV}_I/2$), so to first order
\begin{align}
    1-F_\text{sim}\approx\mathrm{CV}_I^2/4 +\mathrm{var}(\delta\phi).
\end{align}

\emph{Some technical details about warm-WGS}---Define $A = 1024$ as the SLM resolution and $P$ as the zero-padding resolution, in our case $2048$. We repeat the following $K=3$ times per frame, starting from the previous frame's $\varphi$ and $W_j$:
\vspace{-4pt}
\begin{alignat}{3}
&\text{1. Measure:}  &\quad m_j &= \Bigl|\sum_u w_j(u)\,F(u)\Bigr|, &\qquad F&=\mathcal F_P\!\left[B e^{i\varphi}\right] \label{eq:wgsmeasure}\\
&\text{2. Reweight:} & W_j &\leftarrow \frac{W_j\,(a_j/m_j)^{\gamma}}{\frac1N\sum_k W_k\,(a_k/m_k)^{\gamma}} && \label{eq:wgsweight}\\
&\text{3. Paint:}    & T(u) &= \sum_j W_j\,e^{i\theta_j}\,\tilde w_j(u) && \label{eq:wgspaint}\\
&\text{4. Update:}   & \varphi &\leftarrow \arg \mathcal F_P^{-1}[T]\big|_{A\times A} && \label{eq:wgsupdate}
\end{alignat}

Eq.~\eqref{eq:wgsmeasure} measures how bright tweezer $j$ is in the light field that the current hologram produces. That field is $F=\mathcal F_P[Be^{i\varphi}]$, the centered discrete Fourier transform of the SLM field, where $B$ is the illumination amplitude and $\varphi$ is the hologram phase. $F(u)$ is its complex amplitude at pixel $u$ of the $P\times P$ far-field grid. $w_j(u)$ is the weight of the tweezer amplitude at pixel $u$. To measure tweezer $j$, we compare this field with a picture of what an ideal tweezer looks like at that tweezer's position. A tweezer usually sits between pixels, so this picture spreads it over the four nearest pixels. Each of those pixels gets a real bilinear-interpolation weight, and the weight is larger the closer the pixel is to the tweezer. The weights are zero everywhere else and add up to one, $\sum_u w_j(u)=1$. The inner product $\sum_u w_j(u)F(u)$ therefore measures how closely the light around the tweezer matches the ideal one, and its modulus is $m_j$. The complex values are added before the modulus is taken. Light that reaches the four pixels with matching phases therefore adds up, and light with conflicting phases partly cancels.

Eq.~\eqref{eq:wgspaint} goes the other way: it writes the light field we want into a new target field $T(u)$. Each tweezer contributes a complex amplitude $W_j e^{i\theta_j}$, where $W_j$ is its equalization weight and $\theta_j$ is its commanded target phase. Its commanded target amplitude $a_j$ comes in later, in the weight update. This amplitude is not placed on a single pixel. It is spread over the tweezer's four-pixel patch according to a second picture, $\tilde w_j(u)$, which we call the painting kernel to set it apart from the measuring kernel $w_j$ of the first line. 
Where the patches of neighboring tweezers overlap, their contributions add as complex numbers, and $T(u)$ is zero outside all patches.

The two kernels answer different questions. The measuring kernel only reads light field at the atom plane that already exists, so comparing that light field with the ideal tweezer picture works directly. The painting kernel has to produce light. On its way to the far-field that light passes through the finite SLM aperture, which blurs it slightly; in Eq.~\eqref{eq:wgsupdate} this is the crop $|_{A\times A}$. A painting kernel equal to the ideal picture would therefore come out slightly smeared. In general $\tilde w_j$ must be a pre-corrected picture that turns into the ideal tweezer after this blurring. In 2D, we simplify by setting $\tilde w_j=w_j$, so painting uses exactly the same weights as measuring; holograms could be improved slightly with more accurate kernels. The later 3D version has more complicated kernels and so we use a separate, corrected $\tilde w_j$.

Eqs.~\eqref{eq:wgsweight} and~\eqref{eq:wgsupdate} close the loop by updating the two quantities that carry over to the next iteration. The weight update compares each tweezer's measured amplitude $m_j$ with its commanded amplitude $a_j$. A tweezer that comes out too dim ($m_j<a_j$) has its weight raised, and one that comes out too bright has its weight lowered. The exponent $\gamma$ sets how strongly each step corrects. The phase update converts the painted target field back to the SLM plane with the inverse transform $\mathcal F_P^{-1}$. It keeps only the phase of the result and crops it to the $A\times A$ SLM aperture, which gives the next hologram $\varphi$. One iteration therefore measures the current tweezers, paints a corrected target, and turns that target into a new hologram, and the next iteration measures the light that this hologram produces. Repeating the loop pushes dim tweezers up and bright ones down until every $m_j$ matches its $a_j$.

At the start of transport, every active trap gets the weight $W_j=1$, unfilled or unused traps get $W_j=0$, and $\varphi$ is seeded from the loading pattern. After that, both $\varphi$ and $W_j$ carry over from frame to frame. Each frame therefore starts from the previous frame's corrected hologram and weights rather than from scratch, and only needs a few iterations to follow the small change in tweezer positions. For simplicity, our implementation measures and paints at the new positions for every step; measuring at the previous frame's positions on the first iteration would slightly improve the uniformity of moving tweezers. 

Both methods, SLMnet and warm-WGS, produce frames faster than $T_\text{min}$; but under a pure Fourier transform simulation, warm-WGS scores better in intensity uniformity, phase control, and positional control (Fig.~\ref{fig:3}c). They achieve comparable hologram efficiency (${\sim}64\%$ of light lands close to target tweezer positions). WGS also naturally generalizes to a wide variety of array geometries, while SLMnet has to be trained according to its use case. An example is array pitch: while training over mid-arrangement patterns encourages generalization to irregular lattices, we must capture the full range of array spacings during training that we want to use in inference. See Fig.~\ref{fig:simgeneralization}.

We note that SLMnet, by construction, is parallelizable (with enough compute) across $n$ steps because its frame generations are completely independent. Our implementation of warm-WGS is not, since each frame is seeded from the previous one. However, we find that in simulation, we can achieve comparable hologram quality and computation time by initializing each frame from a superposition of target tweezers; the phase of each tweezer is determined by linear interpolation along the transport path~\pref{sm:pathmatching}. Importantly, this removes the dependence property of warm-WGS, allowing for parallelization (batching over frames) with enough compute. Combined with other improvements such as using the actual optimal $w_j$ and $\tilde w_j$ (as below in the 3D case) rather than a bilinear interpolation, even better simulated hologram fidelity is possible in comparable computational time. Note that a single inverse Fourier transform of the superposition initialization (i.e. 0-iteration WGS) would save significant computation time (${\sim}0.1$--$\SI{0.3}{\milli\second}$ on our GPU depending on padding); this could possibly generate high enough quality holograms for rearrangement while saving more computation time.

\subsection{\label{sm:3dinference} 3D Hologram Generation}

We can also apply both frame generation methods to 3 dimensions. 3D WGS normally forward-propagates the hologram to multiple depth layers each iteration, applying weights and correction to all layers and coherently superposing each corrected layer before the next iteration. We can bypass this computationally expensive multi-layer propagation, however, by using locally defocused splat targets for each tweezer. The 3D WGS algorithm then uses Eqs.~\eqref{eq:wgsmeasure}--\eqref{eq:wgsupdate} but with $\gamma = 0.4$ and with different measurement and painting target patterns $w_j$ and $\tilde w_j$. The kernels $w_j$ are complex, so we reinterpret the measurement step as an inner product:
\begin{align}
    m_j &= \frac{\left|\left\langle w_j, F \right\rangle \right|}{\|w_j\|^2}
    = \frac{1}{\sum_u |w_j(u)|^2}\left|\sum_{u}\overline{w_{j}(u)}\,F(u)\right|\label{eq:3dwgsmeasure}.
\end{align}
Parseval's theorem says that, up to a global constant and cropping, the inner product $\langle w_j, F\rangle$ equals the field amplitude at the tweezer's 3D focus $(x_j,y_j,z_j)$, although $F$ itself is only the field in the plane $z=0$. Particularly, one can use a cropped version of a defocus hologram Eq.~\eqref{eq:defocuskernelexact}, along with Eq.~\eqref{eq:xgrating} for radial placement, and set
\begin{align}
    L_j &= e^{i \left(h_z(z_j) + h_x(x_j) + h_y(y_j) \right)} \\
    w_j &= \mathcal{F}_P\mleft[L_j\mright] \big|_{S \times S}.
\end{align}
Note that we use the quadratic approximation kernel Eq.~\eqref{eq:defocuskernelzernike} in implementation; the exact spherical form Eq.~\eqref{eq:defocuskernelexact} gives better hologram quality in exchange for slightly more computation time. $L_j$ is the combined grating and defocus kernel for the perfect target tweezer, and $w_j$ is the discrete Fourier transform of $L_j$ at our padding resolution. The crop to dimension $S \times S$ reduces computational complexity, defining the advantage of this algorithm:  the kernel size $S$ is chosen as the smallest window that captures nearly all of the defocus kernel's energy, which grows linearly with the largest commanded absolute defocus $|z|_{\max}$. As one can imagine, a tweezer farther from the focal plane looks more spread out at $z=0$. The kernel size $S$ affects the computation time quadratically, so while multi-plane 3D WGS has time complexity $O(LP^2\log P)$, our 3D warm-WGS is $O(P^2\log P + N\cdot |z|_\text{max}^2)$, where $N$ is the number of tweezers and neither term depends on the number of distinct layers $L$. The kernel depends only on the tweezer's $z$-coordinate and its sub-pixel offset (the fractional part of its far-field position); we precompute kernels for a finite set of depths and sub-pixel offsets, and at run time each tweezer's kernel is looked up and shifted to its whole-pixel position. Thus we can arbitrarily place tweezers in $x,y$, and $z$ (within the range enabled by the kernel size). In practice, we use 2 iterations per frame for middle frames and 5 for the first frame. With these settings, in simulation, 3D warm-WGS reaches the same hologram fidelity as phase-controlled multi-layer WGS in significantly less computation time.

Importantly, painting with the cropped kernel itself, $\tilde w_j = w_j$, reproduces the defocus lens over the aperture only approximately, since cropping to $S\times S$ discards $w_j$'s outer tails. These tails carry the lens's high spatial frequencies, which come mostly from the steep defocus phase near the SLM aperture edge. Therefore, we instead paint with the $S \times S$ kernel that best matches the target SLM pattern for a single spot, $L_j$, i.e. the minimum norm solution of
\begin{align}
    \tilde w_j &= \argmin_{k \in \mathbb C^{S \times S}} \left\|\mathcal F_P^{-1}[k] - L_j\right\|_{A\times A}^2.
\end{align}
This is quadratic optimization: by linearity, we can write it in closed form $\tilde w_j = G^{+} w_j$, with $G^{+}$ the Moore--Penrose
pseudo-inverse of the aperture overlap matrix $G_{dd'} = \mathcal F_P[\mathds 1_{A\times A}](d-d')$. Intuitively, $G_{dd'}$ is the fraction of light painted at far-field pixel $d'$ that returns to pixel $d$ after passing through the $A\times A$ aperture; it depends only on the separation $d-d'$. So the same $G^{+}$ serves all depths and tweezers. This correction is necessary to make full use of the padding when $P > A$.

For the $\SI{10}{\micro\meter}$ maximum spacing in Fig.~\ref{fig:1}'s demos, we can generate frames in $\approx\SI{0.56}{\milli\second}$ by centering on the $\SI{5}{\micro\meter}$ midpoint between layers and post-applying a defocus kernel, rather than running the algorithm with $|z|_{\max} = \SI{10}{\micro\meter}$. 

Another option is to train a 3D SLMnet model in the same way as the 2D version is trained, but encoding input position, phase, and depth by using the defocused splat targets. We try this but achieve lower simulated fidelity with higher computation time than 3D warm-WGS; however, because kernel size is non-trivial in our 3D warm-WGS algorithm, scaling in the number of tweezers $N$ becomes significant much past our ${\sim}10^3$. Thus 3D SLMnet, which has the potential to scale much more slowly in $N$ and $|z|_\text{max}$, could be useful for very large 3D arrays.

\subsection{\label{sm:overdrive} Overdrive}
 
When a frame is commanded, and the SLM applies a voltage to LC backplane, each liquid crystal pixel immediately starts to relax from its instantaneous phase $\psi_0$ towards the commanded pixel phase we send to the SLM, $\psi_c$. Call the target pixel phase $\psi_1$. In general, the LC response for a single pixel goes as $\psi(t, \psi_0, \psi_c)$ where we enforce $\psi(0, \psi_0, \psi_c) = \psi_0$ and $\psi(t\to \infty, \psi_0, \psi_c) = \psi_c$. (Our LC calibration defines $\psi$ by Eq.~\eqref{eq:pixeltaudef} but with $\tau \mapsto \tau(\psi_c, \text{sign}(\psi_c - \psi_0))$, a time constant set by commanded phase and direction of the transition, interpolated from measured values.) For any finite $T$, especially for $T \lesssim \tau_0$, the next frame update occurs before the pixel has reached the commanded value. If we want each hologram to be displayed exactly on time, then, we would like to solve
\begin{align}
    \psi(T, \psi_0, \psi_c) &= \psi_1 \label{eq:overdrivecondition}
\end{align}
for $\psi_c$: to do so, we make the reasonable simplifying assumption that, for fixed $T$ and $\psi_0$, $\psi$ is continuous and strictly increasing in $\psi_c$. $\psi_c$ then is unique if it exists. Furthermore, if $\psi_1$ is between $\psi_\text{min} = \psi(T, \psi_0, 0)$ and $\psi_\text{max} = \psi(T, \psi_0, 2\pi)$, by the intermediate value theorem, there exists $\psi_c(T, \psi_0, \psi_1)$ that satisfies Eq.~\eqref{eq:overdrivecondition}. Commanding phase $\psi_c$ will result in the pixel having phase $\psi_1$ at the next frame update; the same logic then repeats for the following transition with initial phase $\psi_1$ and target phase $\psi_2$. This protocol is known as overdrive~\cite{thalhammer2013overdrive}.

It is not immediately clear how often we can achieve phase $\psi_1$ in time $T$. However, monotonicity and the reasonable assumption $\psi(t, \psi_0, \psi_0) = \psi_0$ (commanding the current instantaneous phase freezes the state of the pixel) give the inequality
\begin{align}
    \psi_\text{min}\leq \psi_0 \leq \psi_\text{max}.
\end{align} 
That is, depending on where $\psi_0$ sits in $[0,2\pi)$, how far $\psi_1$ is, and the value of $T$ relative to $\tau_0$, there is some space $[\psi_\text{min}, \psi_\text{max}]$ around $\psi_0$ where $\psi_1$ can sit and $\psi_c$ will exist. Our typical assumption of small hologram changes $|\psi_1 - \psi_0| \ll 2\pi$ suggests that, for a given pixel, there exist fairly low $T$ such that $\psi_c$ exists.

In practice, we relax the requirement of reaching the target hologram exactly in $T$: for each transition, we set a maximum clamping fraction $\alpha$ and then define $T^*$ as the smallest period such that Eq.~\eqref{eq:overdrivecondition} with $T\mapsto T^*$ has no solution for at most a fraction $\alpha$ of the pixels in the LC display. We take $T^* \leftarrow \max(T, T^*)$, and then we assign $\psi_c$ for pixels where $\psi_c(T^*, \psi_0, \psi_1)$ is well-defined and $\psi_c \in \{0,2\pi\}$ in the direction from $\psi_0$ to $\psi_1$ otherwise. Importantly, we keep track, using our LC calibration, of where we expect our pixels to be, and use this to seed $\psi_0$ for the next transition. The first transition is seeded from the settled SLM state, and the last frame, which is left to settle, is not overdriven. Operationally, we approximate the $T^\star$ construction by scaling down every pixel's overshoot past $\psi_1$ by a common factor, choosing the largest factor for which at most a fraction $\alpha$ of pixels fall outside $[0, 2\pi]$ (these are clamped).

Intuitively, we choose $\alpha$ large enough so that $T^*$ remains small, and small enough to control the number of pixels that lag behind the rest. $\alpha = 1$ (unconstrained overdrive) would mean $T^* = T$, so pixels with $\psi_c$ defined would reach their targets at the next update, and pixels without a $\psi_c$ would be commanded to $0$ or $2\pi$ and lag behind. $\alpha = 0$ would force a high enough $T^*$ such that $\psi_c$ exists for all pixels; inevitably, a few wrapping pixels that travel a phase difference $2\pi - \epsilon$ will dominate, forcing $T^*$ to grow far beyond $T$ and leaving speed on the table for the sake of outlier pixels ($T^* \sim \tau_0\ln(2\pi/\epsilon)$ for exponential relaxation). We heuristically choose $\alpha = 0.2$. Then overdrive is designed such that, within our calibration model, $\psi$ will converge at least as close to $\psi_1$ within the period $T$ each transition as it would have with no overdrive:
\begin{align}
    |\psi(T, \psi_0, \psi_c(T^*, \psi_0, \psi_1)) - \psi_1|&\leq |\psi(T, \psi_0, \psi_1) - \psi_1|.
\end{align}
We find that overdrive's better convergence yields faster transport by effectively accelerating the LC response (Fig.~\ref{fig:3}b); this effect is resistant to LC calibration error and LC pixel lag caused by clamping.

\section{Methods Clarifications\label{sm:figs}}

All error bars in main text figures are bootstrapped $68\%$ CIs, floored at binomial error, unless stated otherwise. All stated error bars in main text are at least as large as our statistical uncertainty, where some are rounded up for consistency and readability. We generally filter out shots with abnormally long frame generation times; abnormally low loading; or in which the real-time atom detection used for path planning disagreed with the final image analysis by more than ${\sim}5$--$10\%$ (typically dim-imaging).

(Fig.~\ref{fig:1}b,c) We first rearrange our 1068 site lattice into a checkerboard, and then we lift a subset of atoms. The $41$-site checkerboard regions share stationary boundary sites, but each region still lifts $16$ atoms into the second plane.

(Fig.~\ref{fig:1}b) For line lifting, we achieve average filling $92.9\%$ across $536$ sites using 196 frames at \SI{2.0}{\milli\second} period (slight frame time aliasing does not have a significant effect). This is dominated by loss in the static loading layer, likely caused by interference between moving and static tweezers; the lifted layer achieves $96.2\%$ filling on average. Measured over 44 of 60 kept shots. 

(Fig.~\ref{fig:1}c) For triangular lattice lifting, we achieve average filling $96.7\%$ across $536$ sites using 41 frames at \SI{2.0}{\milli\second} period. Measured over 28 of 30 kept shots.

(Fig.~\ref{fig:2}c) $d_{0}$ is obtained for each period by interpolating the 50-step survival curves of (a) and the analogous 80-step versions of (b), with survival threshold $0.99^{80} = 0.45$ in place of $0.99^{50} = 0.61$. Note that the axial data in (c) come from $n=80$ ping-pong steps (40 each way) rather than 50 (25 each way) as in (b), taken on a different day, at a slightly different focal plane, and in $+z$ first (whereas the (b) data is 50-step ping-pong, $-z$ first); the MC bands match accordingly. All numbers quoted in the main text come from (b) rather than this 80-step curve. The trend is consistent, including the roughly threefold increase of $v_\text{max}$ at $T_\text{min}$, and $d_0$ at $T=0.7$ and $2.8$ ms agrees between the two datasets to within 11\% and 5\%, respectively; the 80-step values lie systematically slightly higher (similar to the axial step-count measurements of Fig.~\ref{fig:survivalsteps}, where at $T_\text{min}$ the loss is concentrated in the first few steps, so the per-step survival rises with $n$, the opposite of accumulation; those data extend only to 25 steps). No error bars are included for atom $v_\text{max}$ values (they are generally small and do not affect the trend). The width of the MC bands represent $\pm 1$ standard deviation across 15 radial and 3--8 axial simulated sites from the $1068$ in the array. 

Furthermore, we display Monte Carlo simulation of atoms in the simulated SLM light field with matching protocol to the atom data~\pref{sm:mc}. The shaded bands represent $\pm 1$ standard deviation of $v_\text{max}$ over simulated trap sites (15 radial, 3--8 axial); i.e. site-to-site variation in the simulated array, not statistical uncertainty.

(Fig.~\ref{fig:2}d-h) Simulated light field is shown. Fig.~\ref{fig:mcdemo} shows a version of these images along with MC-simulated atoms.

(Fig.~\ref{fig:2}i) 50-step ping-pong (25 each way) is repeated to reach high step counts. We even extend $\SI{0.2}{\micro\meter}$ transit specifically out to $4000$ steps and still see no difference from control. The control and $d=\SI{0.2}{\micro\meter}$ curves are fit to memoryless exponentials; the $d=\SI{0.3}{\micro\meter}$ curve is fitted to the control curve times $\exp[-(c + An^b)]$ where $n$ is step count. We find $b = 3.06(21)$, a super-exponential survival decay that indicates accumulation, likely due to heating. Error bars are floored at quasi-binomial.

(Fig.~\ref{fig:3}a) These are representative, cropped single images at $n = 60$ steps. 

(Fig.~\ref{fig:3}b) Shots from all three methods are interleaved with no-movement control shots. Fidelity is normalized to a rolling average over time of no-movement survival to account for system drift and isolate transport loss.  Error bars are 68\% intervals from a bootstrap over shots that accounts for rare high-loss shots and imposes fidelity $\le 1$. Number of frames to 99\% fidelity comes from a joint fit of $F_{99}(n) = C[1-\epsilon(a/n)]^n$ to the plotted points, where $a$ is the lattice pitch, $\epsilon$ is an erfc-shaped per-step loss, each method has its own plateau $C$ and cliff position, and all three share one cliff width. Overdrive reaches every fidelity level from 97\% to 99.5\% in fewer frames than both other methods; warm-WGS and SLMnet are similar on the cliff and separate only near the plateau $n \gtrsim 54$.

(Fig.~\ref{fig:3}c) warm-WGS points are 1, 2, 3, 5, 10 and 20 iterations (left to right); SLMnet points are networks with 2, 8, 16, and 24 channels all trained with identical data and protocols (deployed has 16 channels). Measured on frames from 10 seeds of $28$-step checkerboard rearrangement. Note that production SLMnet, used e.g. in (b), is technically a different training seed and so differs slightly in performance from the production-size control in this graph. The comparison here is at equal training; results are qualitatively unchanged.

\newpage

\setcounter{figure}{0}
\renewcommand{\thefigure}{S\arabic{figure}}

\begin{figure*}[t]
    \centering
    \includegraphics[width=1\linewidth]{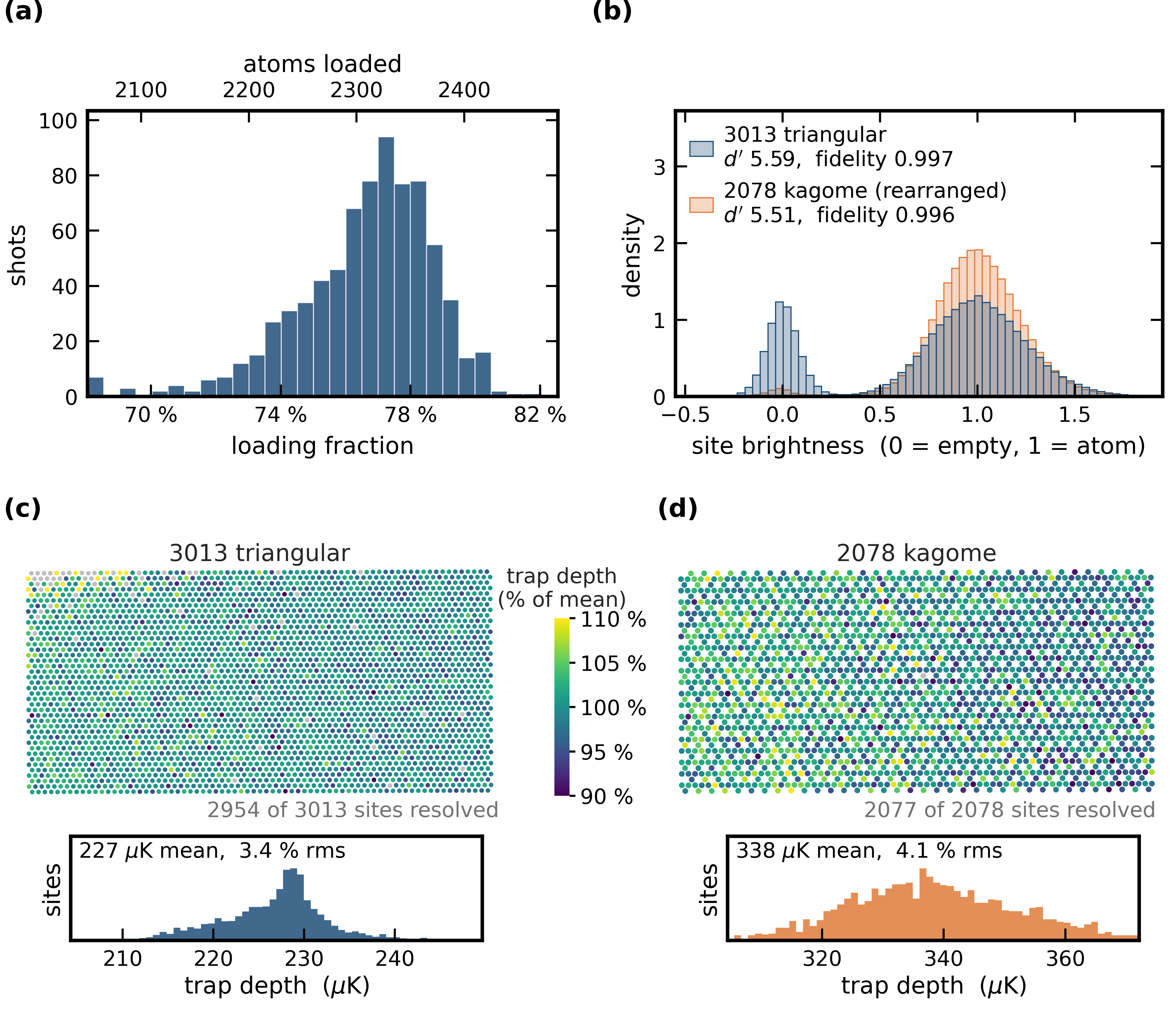}
        \caption{Large array loading and imaging. (a) Histogram showing number of atoms loaded into the 3013-site triangular lattice across 769 shots. 98\% of shots result in $>2078$ atoms (enough to fill the target array). The average fill is $2295$ atoms, or $76.2\%$ fill, using enhanced loading with blue-detuned LAC light. (b) Imaging histograms for the triangular 3013-site loading lattice and the kagome 2078-site target lattice, normalized to a $0$ to $1$ scale where $0$ represents no atom and $1$ represents an atom. The kagome density curve is post-rearrangement, so the bulk of sites are filled. The two arrays have mean imaging fidelities $99.7\%$ and $99.6\%$ respectively, with ${\sim}99\%$ imaging survival for both (approximately matching the survival of stationary atoms during rearrangement). $d'$ is the imaging discrimination between empty and filled sites, quantifying the effective signal-to-noise ratio of single-atom detection. (c) Trap depth site-map and histogram for the 3013-site triangular lattice after WGS uniformity feedback. Average trap depth \SI{227}{\micro\kelvin} with $3.4\%$ rms nonuniformity. (d) Trap depth site-map and histogram for the 2078-site kagome lattice after WGS uniformity feedback. Average trap depth \SI{338}{\micro\kelvin} with $4.1\%$ rms nonuniformity. }
    \label{fig:loadimage}
\end{figure*}

\begin{figure*}[t]
    \centering
    \includegraphics[width=0.4\linewidth]{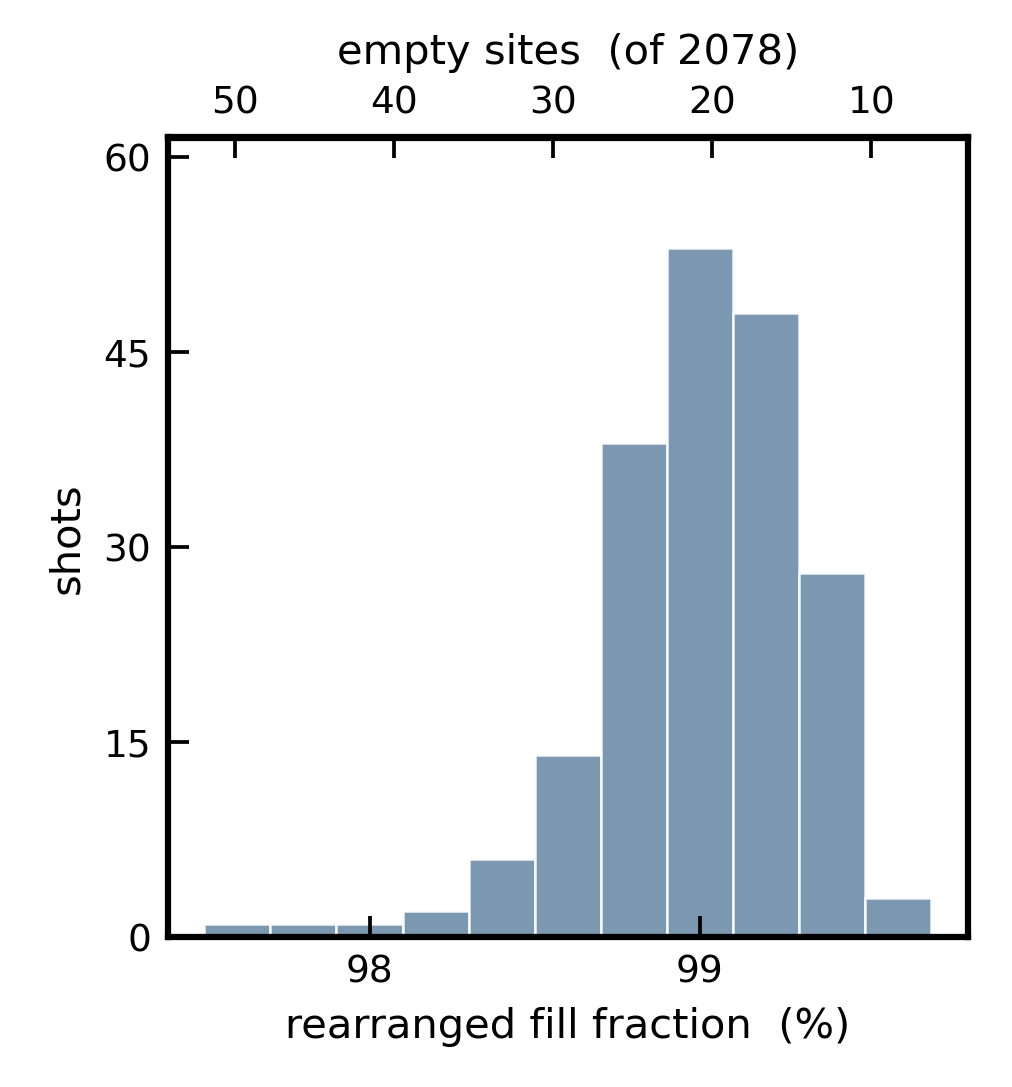}
    \caption{Large array histogram. Distribution of fill fractions resulting from rearrangement of a $3013$-site triangular lattice to a $2078$-site kagome lattice. Mean $99.01\%$ filling, or around $21$ defects. The best shot (Fig.~\ref{fig:1}d) has only $7$ defects; over $40$ frames, atoms move at a $99$th percentile speed of \SI{0.41}{\micro\meter / \milli\second}. Of $200$ shots, we remove $5$ with abnormally long computational time or low loading, leaving $195$ shots.  Rearrangement time: shots in this run take median time \SI{56.5}{\milli\second} from end to end (between images; \SI{56.3}{\milli\second} for the Fig.~\ref{fig:1}d shot), including \SI{8.9}{\milli\second} of path planning, \SI{28.1}{\milli\second} of transit time, and \SI{3.9}{\milli\second} of precompute buffer. The rest is unoptimized, miscellaneous GPU time.}
    \label{fig:largehistogram}
\end{figure*}

\begin{figure*}[t]
    \centering
    \includegraphics[width=0.8\linewidth]{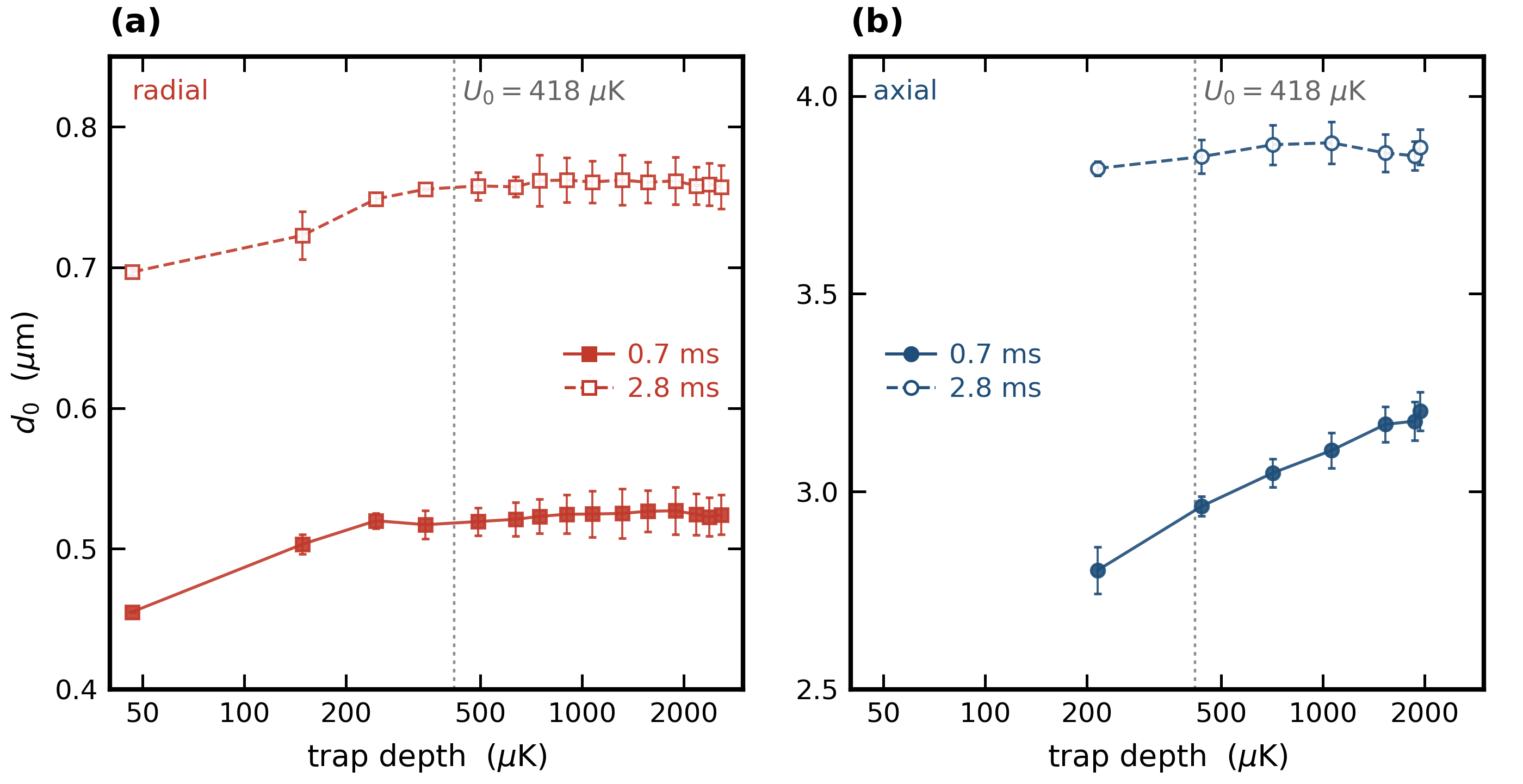}
    \caption{
    Effect of trap depth on $d_0$.
    Maximum movement distance $d_0$ versus log-scaled trap depth across periods and axes. Vertical dashed line in each graph represents our normal operating point $U_0 = \SI{418}{\micro\kelvin}$. Each $d_0$ value is linearly interpolated from survival versus step size curves as in Fig.~\ref{fig:2}c. Trap depth is controlled by laser power and measured as in~\ref{sm:trapcal}; a smaller array is used to increase trap depth range. (a) Radial $d_0$ versus trap depth at $T = 0.7,\SI{2.8}{\milli\second}$. A $d_0$ plateau is apparent around and above $U_0$. (b) Axial $d_0$ versus trap depth at $T = 0.7,\SI{2.8}{\milli\second}$. $d_0$ increases slightly above $U_0$ trap depth but seems to be slowing down. Both radially and axially, increasing laser power up to ${\sim}5U_0$ slightly increases $d_0$ but does not close the gap between the two periods.
    }
    \label{fig:trapdepth}
\end{figure*}

\begin{figure*}[t]
    \centering
    \includegraphics[width=0.8\linewidth]{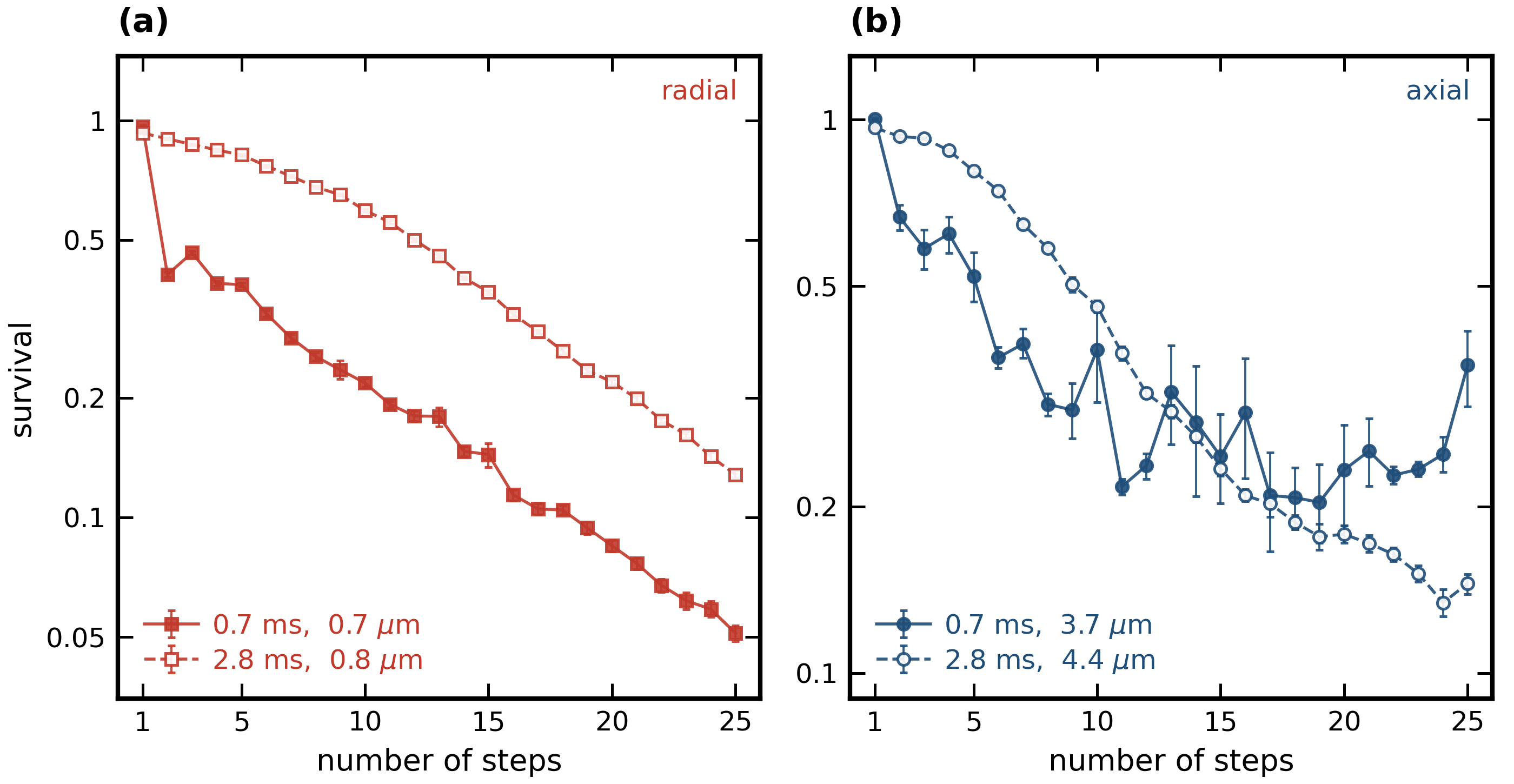}
    \caption{Loss rate during transport.
    Log-scaled survival vs. $n$, number of steps of transport. Step sizes are approximately chosen to produce a reasonable survival after 20-25 steps. A straight line indicates exponential, memoryless survival. (a) Radial ($+x$) transport. Entry shock is visible at $T = \SI{0.7}{\milli\second}$ as survival drops significantly during the second step; however, in steps 10-25, steady state loss rate at $T = \SI{0.7}{\milli\second}$, $d=\SI{0.7}{\micro\meter}$ approximately matches steady state loss rate at $T = \SI{2.8}{\milli\second}$, $d=\SI{0.8}{\micro\meter}$---on top of entry shock, there is still an impact of period on achievable step size. (b) Axial asymmetric ping-pong transport: atoms are transported outward in $-z$ with many $<\SI{1}{\micro\meter}$ steps, which are measured to be lossless and cause insignificant heating. Number of steps then tracks the number of steps in the return trip ($+z$) at the commanded step size $d$. For long period axial motion ($T = \SI{2.8}{\milli\second}$, $d = \SI{4.4}{\micro\meter}$), there is a super-exponential region and then a sub-exponential region, while for short period axial motion  ($T = \SI{0.7}{\milli\second}$, $d = \SI{3.7}{\micro\meter}$), the error bars are much larger and loss rate seems to fluctuate and decrease to near-zero on average across steps 10-25.}
    \label{fig:survivalsteps}
\end{figure*}

\begin{figure*}[t]
    \centering
    \includegraphics[width=0.8\linewidth]{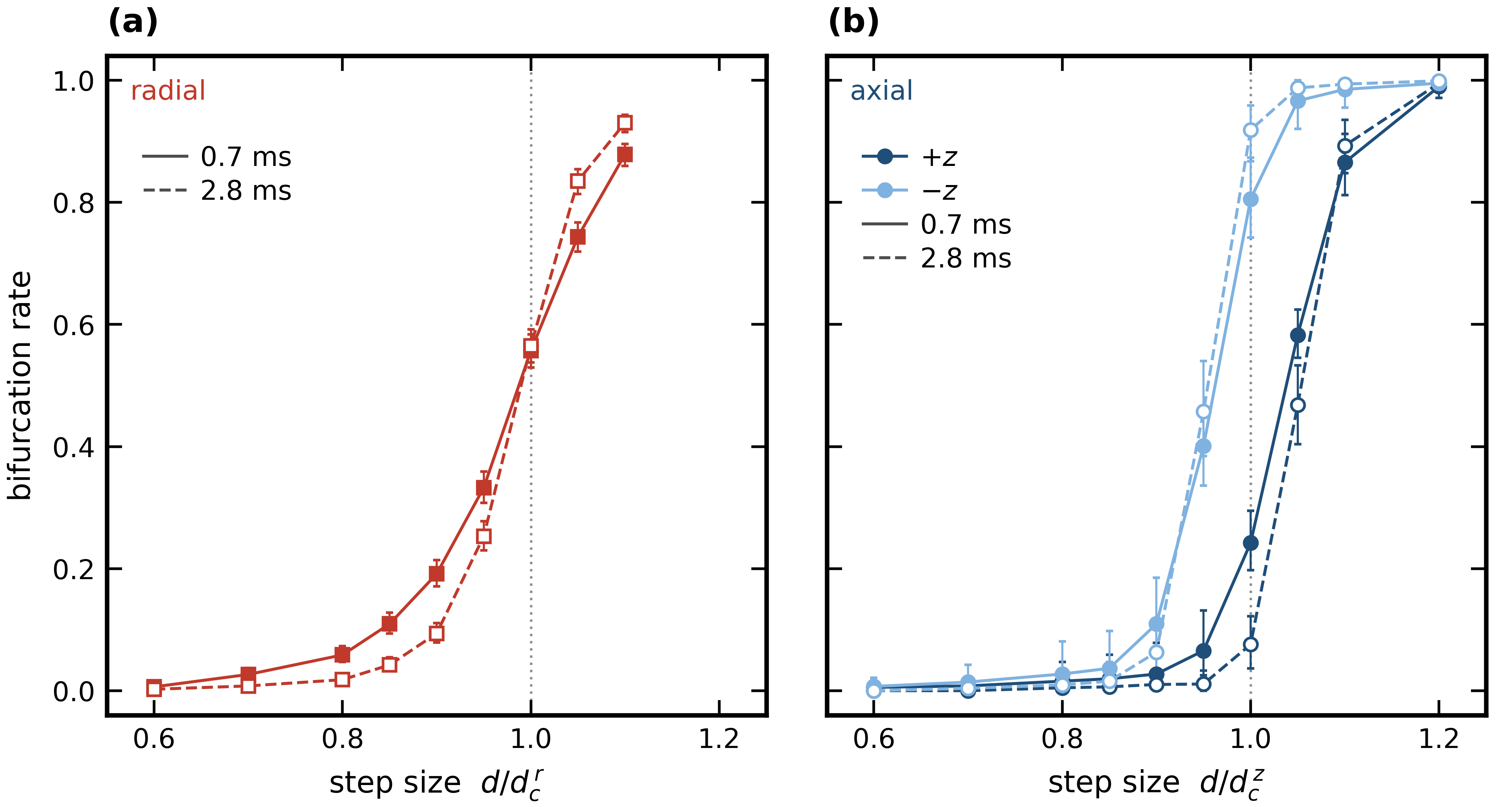}
    \caption{
    Bifurcation rate versus step size across periods and axes. 
    Fraction of transitions in which the simulated tweezer bifurcates, (a) radially along $+x$ and (b) axially along $+z$ (dark) and $-z$ (light), at frame periods of \SI{0.7}{\milli\second} (solid, filled) and \SI{2.8}{\milli\second} (dashed, open).
    The radial (axial) step $d$ is given in units of $\dc^{r}(\dc^z)$. 
    Each point pools $1280$ steady-state transitions simulated with our calibrated LC response; error bars are \SI{95}{\percent} Wilson intervals.
    Radially, faster pacing raises the bifurcation rate below $\dc^r$ while the curves cross at $d = \dc^r$; axially the $\pm z$ asymmetry is apparent, and there is a slight bifurcation rate change with period.}
    \label{fig:foldrate}
\end{figure*}

\begin{figure*}[t]
    \centering
    \includegraphics[width=0.9\linewidth]{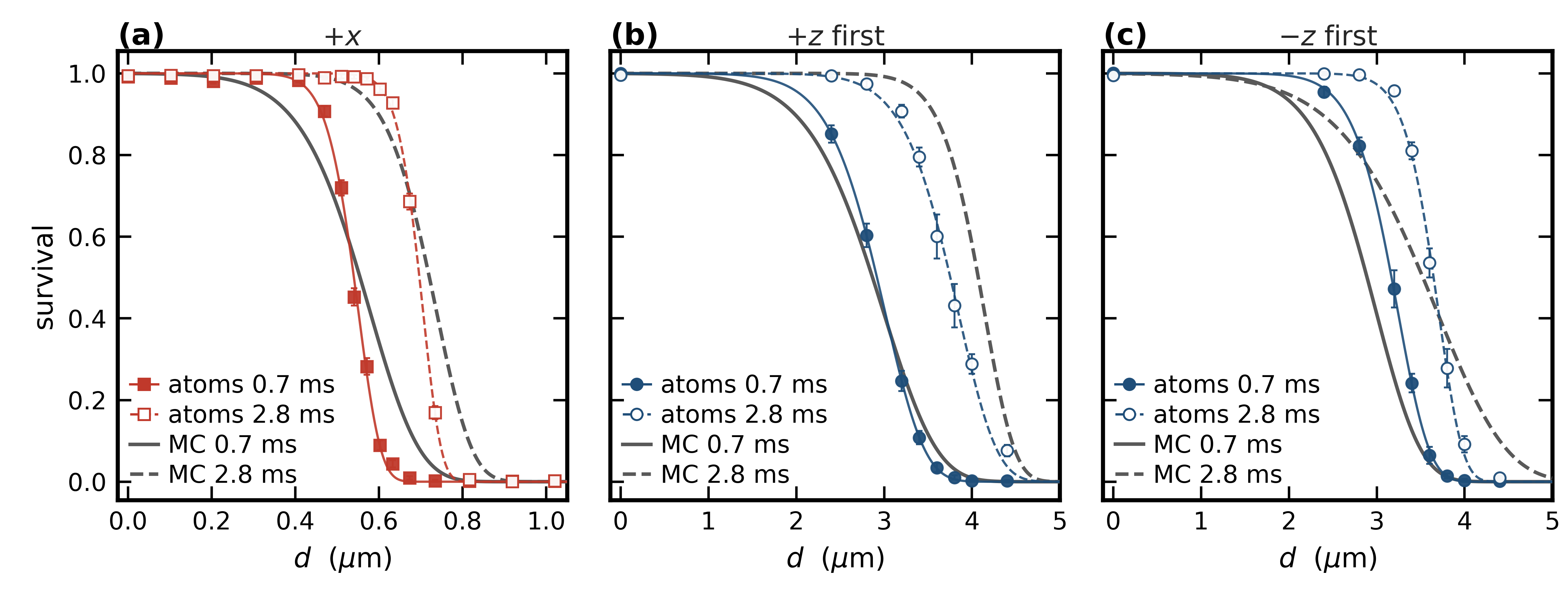}
    \caption{
    Survival versus step size $d$: real atoms and MC atoms.
    (a) Radial $+x$ 50-step transport.
    (b),~(c) Axial 80-step ping-pong transport, $+z$ first (b) or $-z$ first (c).
    All survival data is normalized to the zero-step-size plateau.
    Filled markers and solid lines: SLM frame period \SI{0.7}{\milli\second}; open markers and dashed lines: \SI{2.8}{\milli\second}.
    Thin colored lines are fits to the atoms, grey lines are fits to the MC, both of the form $S = \left[\tfrac{1}{2}\,\mathrm{erfc}\!\left[(d - d_*)/\sqrt{2}\sigma\right]\right]^{n}$, $n$ the number of steps.
    }
    \label{fig:survivalmcatoms}
\end{figure*}

\begin{figure*}[t]
    \centering
    \includegraphics[width=0.4\linewidth]{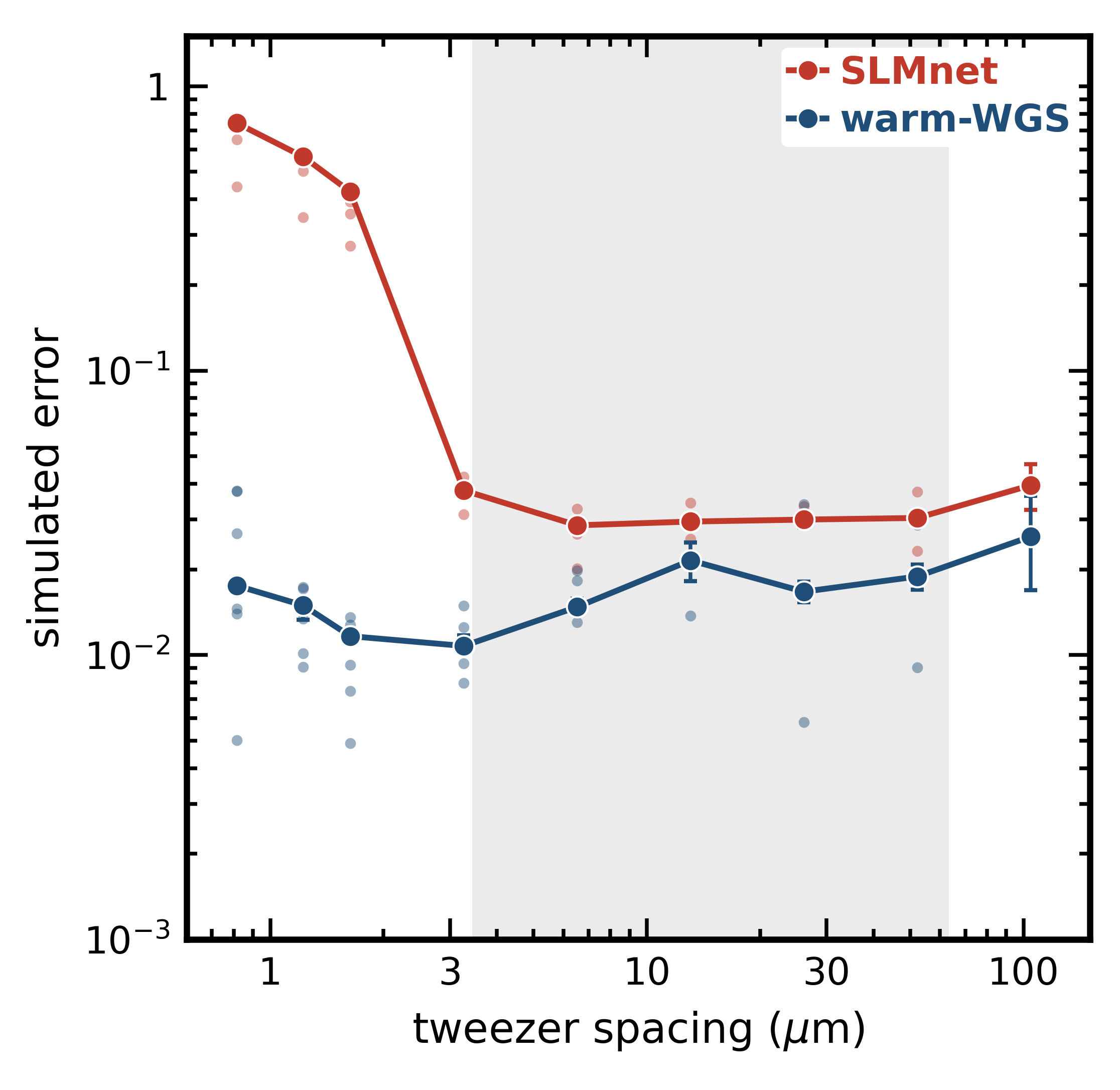}
    \caption{
    Generalization of the hologram generators across tweezer spacing.
    Hologram error $1-F_\text{sim}$ of the deployed SLMnet (red) and of warm-WGS ($2048 \times 3$, blue) versus tweezer spacing. Each test is a simulated rearrangement (random half-filled lattice $\rightarrow$ compact $n\times n$ array at the same pitch, 12 frames) for $n\in\{2,\dots,90\}$ and spacings of \SIrange{0.8}{104}{\micro\meter}, wherever the array fits the field of view (42 geometries).
    Faint points: individual geometries, averaged over ten random loadings. Line: median over array sizes at each spacing. Error bars are 68\% bootstrap CIs over loadings. The grey band spans the 0.5--99.5 percentile of the nearest-neighbour spacing of SLMnet's training frames.
    On spacings below its training distribution, SLMnet's error rises by more than an order of magnitude while warm-WGS is unaffected.
    }
    \label{fig:simgeneralization}
\end{figure*}

\begin{figure*}[t]
    \centering
    \includegraphics[width=\linewidth]{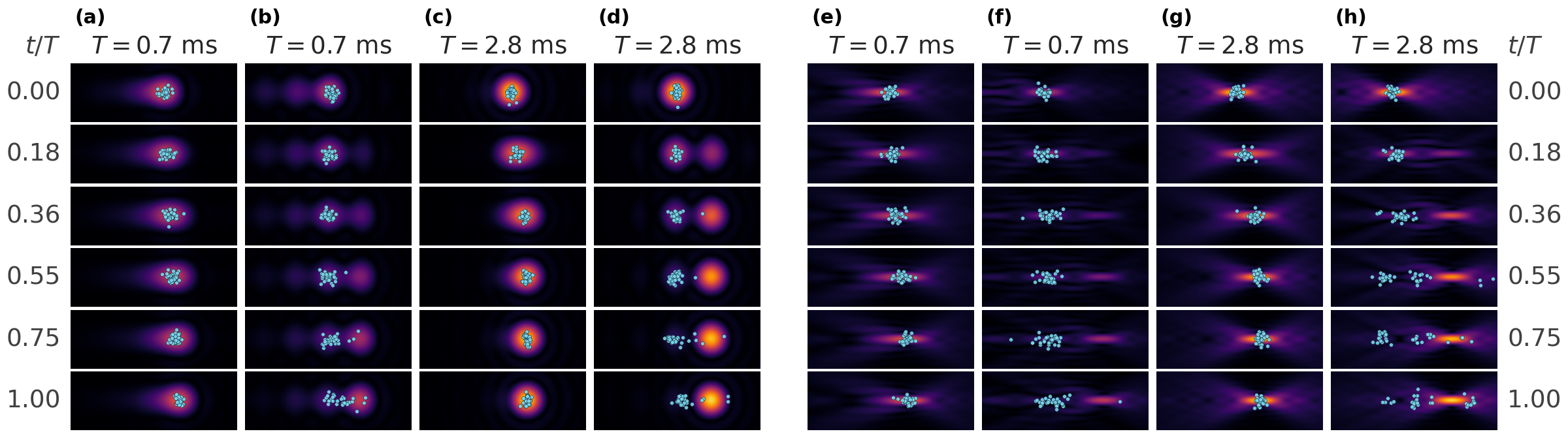}
    \caption{Monte Carlo demonstration. Freeze-frames of Monte-Carlo-simulated atoms being transported by a simulated light field~\pref{sm:mc}. Atom temperature is shown as a color gradient. Atoms follow continuously moving tweezers (a,c,e,g); bifurcation strands atoms in a previous frame's trap (b,d,f,h). 
    Steps and periods: (a) \SI{0.40}{\micro\meter}, \SI{0.7}{\milli\second}; (b) \SI{0.84}{\micro\meter}, \SI{0.7}{\milli\second}; (c) \SI{0.44}{\micro\meter}, \SI{2.8}{\milli\second}; (d) \SI{0.96}{\micro\meter}, \SI{2.8}{\milli\second} (radial); (e) \SI{1.60}{\micro\meter}, \SI{0.7}{\milli\second}; (f) \SI{4.80}{\micro\meter}, \SI{0.7}{\milli\second}; (g) \SI{2.00}{\micro\meter}, \SI{2.8}{\milli\second}; (h) \SI{4.80}{\micro\meter}, \SI{2.8}{\milli\second} (axial). Each strip shows $|E|$ during one hologram update at $t/T = 0$, 0.18, 0.36, 0.55, 0.75 and 1.00 (top to bottom), with 30 of 600 Monte Carlo atoms overlaid.
    }
    \label{fig:mcdemo}
\end{figure*}

\clearpage

\renewcommand{\addcontentsline}[3]{}%
\bibliography{slm}